\documentclass{aa}
 
 \usepackage[varg]{txfonts}
 
\usepackage[]{hyphenat}
\usepackage[]{microtype}
\usepackage{amssymb}
\usepackage{float}
\usepackage{bm}
\usepackage{epsfig}
\usepackage{soul}
\usepackage{subcaption}
\usepackage{multirow,balance}
\usepackage{amsfonts}
\usepackage{amsmath}
\usepackage{aas_macros}
\usepackage{graphicx}
\usepackage{natbib}
\usepackage{color}
\usepackage{listings}
\usepackage{hyperref}  
\usepackage{cases}
\usepackage{tikz}
\usetikzlibrary{decorations.pathreplacing}

\hypersetup{dvips, colorlinks=true, linkcolor=cyan, citecolor=black, filecolor=black, urlcolor=black}

\definecolor{grey}{rgb}{0.75,0.75,0.75}
\definecolor{Orange}{rgb}{1.0,0.5,0.15}
\definecolor{brown}{rgb}{0.7,0.25,0.0}
\definecolor{pink}{rgb}{1.0,0.5,0.5}
\definecolor{darkerred}{rgb}{0.8,0,0}
\definecolor{darkerblue}{rgb}{0,0,0.8}
\definecolor{Blue}{rgb}{0,0.08,0.65}
\definecolor{Red}{rgb}{0.65,0.08,0.05}
\definecolor{Green}{rgb}{0.15,0.45,0.25}

\begin{document}

\setcounter{tocdepth}{3}

\title
{
Secular diffusion in discrete self-gravitating tepid discs\\
I: analytic solution in the tightly wound limit
}

\author{J.~B. Fouvry
\inst{\ref{inst1}}
\and
C. Pichon
\inst{\ref{inst1},\ref{inst2}}
\and
P.~H. Chavanis
\inst{\ref{inst3}}
}

\institute{
Institut d'Astrophysique de Paris and UPMC, CNRS (UMR 7095), 98 bis Boulevard Arago, 75014, Paris, France\\
\email{fouvry@iap.fr; pichon@iap.fr}\label{inst1}
\and
Institute of Astronomy \& KICC, University of Cambridge, Madingley Road, Cambridge, CB3 0HA, United Kingdom
\label{inst2}
\and
Laboratoire de Physique Th\'eorique (IRSAMC), CNRS and UPS, Univ. de Toulouse, F-31062 Toulouse, France\\
\email{chavanis@irsamc.ups-tlse.fr}\label{inst3}
}

\date{Received \today /
Accepted --}

\abstract{
The secular evolution of an infinitely thin tepid isolated galactic disc  made of a finite number of particles
 is described using the  inhomogeneous Balescu-Lenard equation. 
Assuming that only tightly wound transient spirals are present in the   disc, a WKB approximation
provides a simple and tractable quadrature for the corresponding drift and diffusion coefficients.
It provides insight into the physical processes at work during the secular diffusion of a self-gravitating discrete disc and makes quantitative predictions
on the initial variations of the distribution function in action space. 
\\
When applied to the secular evolution of an isolated stationary self-gravitating Mestel disc, this formalism predicts {\sl initially} the importance of the corotation resonance in the inner regions of the disc leading to a regime involving radial migration and heating. 
It   predicts in particular the formation of a ``ridge like'' feature  in action space,
 in agreement with simulations, but over-estimates the timescale involved in its appearance. 
Swing amplification is likely to resolve this discrepancy.

In astrophysics, the inhomogeneous Balescu-Lenard equation and its WKB limit may also describe the secular diffusion of giant molecular clouds in galactic discs, the secular migration  and segregation of planetesimals in proto-planetary discs,  or even the long-term evolution of  population of stars within the Galactic center.

}
\keywords{Galaxies: evolution - Galaxies: kinematics and dynamics - Galaxies: spiral - Diffusion - Gravitation}

\maketitle

\section{Introduction}
\label{sec:introduction}

Understanding the dynamical  evolution of galactic discs over cosmic times is a long-standing endeavour.
Self-gravitating discs are cold  dynamical systems, for which rotation represents an important reservoir of free energy.  Fluctuations of 
the potential 
 induced by discrete (possibly distant) encounters may be strongly amplified, 
 while resonances tend to confine and localise their dissipation: such small stimuli can lead to discs spontaneous evolution to distinct equilibria.
 Quantifying the relative importance of this intrinsically driven evolution w.r.t. that driven by the environment is of renewed interest 
 now that their cosmological environment   is   firmly established in the context of the $\Lambda$CDM paradigm. 
 The effect of intrinsic susceptibility on secular timescales can be addressed in the context of kinetic theory,
  which takes explicitly into account such  interactions.

The kinetic theory of stellar systems is an old, yet fundamental, topic in
astrophysics.\footnote{For an historical account of the development of kinetic
theories in astrophysics, plasma physics, and for systems with long-range
interactions, see the introductions of the references
\cite{Chavanis2013b,Chavanis2013}.} 
It was initiated by \cite{Jeans1929} and 
\cite{Chandrasekhar1942}  in the context of ${ 3D }$ stellar systems such as
elliptical galaxies and globular clusters. The kinetic theory of Coulombian
plasmas was developed in parallel by  \cite{Landau1936} and
\cite{Vlasov1938}. When encounters are neglected between the particles
(stars or electric charges), one gets  a purely mean
field equation \citep{Jeans1915,Vlasov1938} called the collisionless Boltzmann
equation, or  the Vlasov equation. When encounters
are taken into account, one gets a kinetic equation that includes a collision
term. In
early works, the collision term was obtained by assuming that a particle
experiences a succession of independent two-body encounters with the other
particles. The corresponding kinetic equation can be derived either from the
Boltzmann equation by considering a limit of weak deflections \citep{Landau1936},
or directly from the general form of the Fokker-Planck equation by evaluating
the diffusion and drift coefficients in a binary collisions approximation
\citep{Chandrasekhar1949,Rosenbluth1957}.  In the case of neutral plasmas, the
system is spatially homogeneous, so the distribution function depends only on
the velocity, hence the name kinetic theory. 
By contrast, stellar systems are  spatially inhomogeneous, so the
distribution function depends on position and velocity. In early works on
stellar dynamics, spatial inhomogeneity was taken into account in the advection
term (Vlasov) but the collisional term was calculated by making a local
approximation, as if the system were homogeneous. In ${ 3D }$, the
collisional term displays a logarithmic divergence at large
scales \citep{Jeans1929,Landau1936,Chandrasekhar1942}. In the case
of plasmas, this divergence is due to the neglect of collective effects that are
responsible for Debye shielding. \cite{Landau1936} phenomenologically
introduced a cut-off at the Debye length to regularize the divergence.

Later on, \cite{Balescu1960} and
\cite{Lenard1960} developed a rigorous kinetic theory of plasmas, taking
collective effects into account, and obtained a kinetic equation, the so-called
Balescu-Lenard equation, that does not present any divergence at large scales.
The Debye shielding is taken naturally into account in their treatment through
the dielectric function (that is absent from the Landau equation). In the case
of stellar systems, the divergence at large scales is solved by the spatial
inhomogeneity of the system and its finite extent. One can
phenomenologically
introduce a cut-off at the Jeans scale \citep{Weinberg1993}, \textit{i.e.} at the system's size, which would correspond to the
 analogue of the Debye length in plasma physics, but this \textit{ad hoc}
treatment is not fully satisfactory. Furthermore, it cannot be applied to cold (centrifugally supported) 
stellar discs where spatial inhomogeneity is more crucial than in ${ 3D }$.

A
more fruitful procedure is to write the
kinetic equation with angle-action variables that are the appropriate variables
to describe spatially inhomogeneous multi-periodic systems. When collective effects are
neglected, one obtains the inhomogeneous Landau equation
\citep{Chavanis2007,Chavanis2013}. When collective effects are accounted for, one
gets the inhomogeneous Balescu-Lenard equation \citep{Heyvaerts2010,Chavanis2012}.
For self-gravitating systems, where the interaction is attractive (instead of
being repulsive as in Coulombian plasmas), collective effects are responsible
for an  anti-shielding which tends to  increase the effective mass of the
stars, hence reducing the relaxation time.  The Balescu-Lenard equation is valid
at the order ${ 1/N }$ in an expansion of the dynamics in terms of this
small parameter, where ${ N \!\gg\! 1 }$ is the number of stars. Therefore, it takes
finite-$N$ effects into account 
 and describes the evolution of
the system on a timescale of the order ${ N t_D }$, where $t_D$ is the dynamical
time. For times ${ t \!\ll\!  N t_{D} }$, or for ${ N \!\rightarrow\! +\infty }$, it reduces to the
Vlasov equation which ignores distant encounters between stars. Although the
kinetic theory was initially developed for ${ 3D }$ stellar systems, the final form of
the inhomogeneous Balescu-Lenard equation also applies to stellar discs such as
those considered in this paper.

Indeed, the Balescu-Lenard non-linear equation 
 accounts for self-driven orbital secular diffusion of a gravitating system induced by the intrinsic shot noise due to its discreteness 
 and the corresponding long range correlations.
 Even though this kinetic equation was first written down more than fifty years ago, it has
hardly ever been applied in its prime context, but only in various limits
where it reduces to simpler kinetic equations, as discussed above.

In this paper, we will focus on solving explicitly  such an equation describing the self-gravitating response of a tepid thin disc to its own  stochastic  fluctuating potential induced by
its finite number of components. In this cool regime, the self-gravity of the disc can be tracked via a local WKB-like response,
which in turn allows us to simplify the \textit{a priori} $2D$ formalism to an effective (non degenerate) $1D$ formalism.
We will  compare the prediction of the WKB limit to a numerical experiment presented in the literature,
 and discuss its diagnosis power and possible limitations.

The paper is organized as follows. Section~\ref{sec:inhomogeneousBL} briefly presents the content of the inhomogeneous Balescu-Lenard equation.
Section~\ref{sec:WKBlimit} focuses on razor thin axisymmetric galactic discs within the WKB approximation.
Section~\ref{sec:application} investigates the formation of a narrow resonant ridge in an isolated self-gravitating Mestel disc. Finally, section~\ref{sec:conclusion} wraps up.  Appendix~\ref{sec:derivation} provides  a short sketch of the derivation of the  Balescu-Lenard equation.
Appendix~\ref{sec:noselfgravity} considers the inhomogeneous Balescu-Lenard equation without collective effects.
Appendix~\ref{sec:comp} compares it to other similar kinetic equations, and in particular its Fokker-Planck limit.

\section{The inhomogeneous Balescu-Lenard equation}
\label{sec:inhomogeneousBL}

\begin{figure}
\begin{center}
\epsfig{file=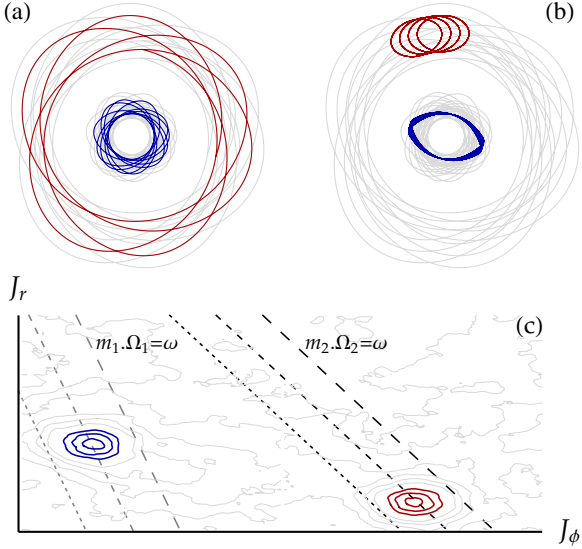,angle=-00,width=0.45\textwidth}
\caption{\small{{\sl Top}:~(a) a set of two resonant orbits in the inertial frame ; (b) in the rotating frame in which they are resonant -- here through ILR-COR coupling.
{\sl Bottom}:~(c)
Fluctuations of the distribution function in action-space caused by finite$-N$ effects showing overdensities for the blue and red orbits. The dashed lines correspond to $3$ contour levels of the intrinsic frequency ${ \omega \!=\! \bm{m} \!\cdot\! \bm{\Omega} }$ respectively associated with the resonance vector $\bm{m}_{1}$ (gray lines) and $\bm{m}_{2}$ (black lines). The two sets of orbits satisfy the resonant condition ${ \bm{m}_{1} \!\cdot\! \bm{\Omega}_{1} \!=\! \bm{m}_{2} \!\cdot\! \bm{\Omega}_{2} }$, and therefore lead to a secular diffusion of the orbital structure of the disc according to equation~\eqref{initial_BL}. Note that the resonant orbits need not be caught in the same resonance ${ (\bm{m}_{1} \!\neq\! \bm{m}_{2}) }$, be close in position space nor in action space.
}}
\label{plot_BL_resonances}
\end{center}
\end{figure}

We consider a system made of $N$ particles. We suppose that the gravitational
background, associated to the Hamiltonian $H_{0}$, is stationary and integrable,
so that we may always remap the physical phase-space coordinates ${
(\bm{x} , \bm{v}) }$ to the angle-actions coordinates ${ (\bm{\theta} , \bm{J})
}$~\citep{Goldstein,born1960mechanics,BinneyTremaine2008}. We also introduce the
intrisinc frequencies of the system $\bm{\Omega}$ defined as
\begin{equation}
\bm{\Omega} (\bm{J}) = \dot{\bm{\theta}} = \frac{\partial H_{0}}{\partial \bm{J}} \, .
\label{definition_Omega}
\end{equation}
Along the unperturbed trajectories, the angles $\bm{\theta}$ are ${ 2
\pi-}$periodic evolving with the frequency $\bm{\Omega}$, whereas the actions
$\bm{J}$ are conserved. To describe the long-term evolution of such a system,
one assumes that there are two decoupled timescales: a short dynamical timescale
and a secular timescale of collisional evolution. We assume that the system is
always in a virialized stable state (\textit{i.e.} is a stable stationary
solution of the Vlasov equation), so that the distribution function can be
written as a quasi-stationary distribution ${ F \!=\! F (\bm{J} , t) }$.
This is a function of the actions only that slowly evolves in
time due to stellar encounters (finite-$N$ effects).\footnote{In this paper, we are
not interested in the initial complex mechanism of \textit{violent relaxation}
\citep{LyndenBell1967}, during which the system gets virialized, since we intend
to describe the long-term evolution of an already and continuously virialized
system.}
From~\cite{Heyvaerts2010} and~\cite{Chavanis2012} (see also Appendix~\ref{sec:derivation} for a short sketch of the derivation), the secular evolution, induced by collisional finite-$N$ effects, of such a quasi-stationary distribution function ${ F (\bm{J},t) }$ is given by the inhomogeneous Balescu-Lenard equation which reads
\begin{align}
\frac{\partial F}{\partial t} = \pi (2 \pi)^{d}  \frac{\partial }{\partial \bm{J}_{1}} \!\cdot\! \bigg[ & \sum_{\bm{m}_{1} , \bm{m}_{2}} \!\! \!\bm{m}_{1} \!\! \int \!\!  \mathrm{d} \bm{J}_{2} \, \frac{ \delta_{\rm D} (\bm{m}_{1} \!\cdot\! \bm{\Omega}_{1} \!-\! \bm{m}_{2} \!\cdot\! \bm{\Omega}_{2})}{| \mathcal{D}_{\bm{m}_{1} , \bm{m}_{2}} (\bm{J}_{1} , \bm{J}_{2} , \bm{m}_{1} \!\cdot\! \bm{\Omega}_{1}) |^{2}} \nonumber \\ \hskip -0.25cm
& \left( \bm{m}_{1} \!\cdot\! \frac{\partial }{\partial \bm{J}_{1}} \!-\! \bm{m}_{2} \!\cdot\! \frac{\partial }{\partial \bm{J}_{2}} \right) F (\bm{J}_{1} , t) \, F (\bm{J}_{2} , t) \bigg] \, ,
\label{initial_BL}
\end{align}
where $d$ is the dimension of the physical space, and where we used the shortened notation ${ \bm{\Omega}_{i} \!=\! \bm{\Omega} (\bm{J}_{i}) }$. The r.h.s of equation~\eqref{initial_BL} is the Balescu-Lenard operator which encompasses the secular diffusion due to collisional effects,
see figure~\ref{plot_BL_resonances}. Because it is the divergence of a flux, this writing ensures that the total number of stars is exactly conserved during the secular diffusion. The Dirac delta ${ \delta_{\rm D} (\bm{m}_{1} \!\cdot\! \bm{\Omega}_{1} \!-\! \bm{m}_{2} \!\cdot\! \bm{\Omega}_{2}) }$ is the sharp resonance condition. One must note that this condition allows to describe non-trivial gravitational interactions. Indeed, it can cause \textit{non-local} resonances by coupling different regions of action-space $\bm{J}_{1}$ and $\bm{J}_{2}$. Even for \textit{local} resonances (\textit{i.e.} ${ \bm{J}_{1} \!=\! \bm{J}_{2} }$), it can allow for non-trivial coupling of oscillations, as soon as $\bm{m}_{1}$ and $\bm{m}_{2}$ have non-zero components. The coefficients ${ 1/ |\mathcal{D}_{\bm{m}_{1} , \bm{m}_{2}} (\bm{J}_{1} , \bm{J}_{2} , \omega)|^{2} }$ represent the \textit{dressed} susceptibilities of the system, for which collective effects have been taken into account\footnote{In the secular timescale limit, the amplification through the  propagation of waves between resonances is assumed to 
be instantaneous, see Appendix~\ref{sec:derivation}.}. To deal with the resolution of the \textit{non-local} Poisson equation, following Kalnajs' matrix method~\citep{Kalnajs2}, one has to introduce a complete biorthonormal basis of potentials and densities ${ \psi^{(p)} (\bm{x}) }$ and ${ \rho^{(p)} (\bm{x}) }$ such that
\begin{equation}
\Delta \psi^{(p)} = 4 \pi G \rho^{(p)} \, , \quad
\int \!\! \mathrm{d} \bm{x} \, [ \psi^{(p)} (\bm{x}) ]^{*} \, \rho^{(q)} (\bm{x}) = - \, \delta_{p}^{q} \, .
\label{definition_basis}
\end{equation}
Thanks to this basis, the susceptibility coefficients are given by
\begin{equation}
\frac{1}{\mathcal{D}_{\bm{m}_{1} , \bm{m}_{2}} (\bm{J}_{1} , \bm{J}_{2} , \omega)} =  \sum_{p , q}  \psi_{\bm{m}_{1}}^{(p)} \!(\bm{J}_{1}) \, [ \mathbf{I} \!-\! \widehat{\mathbf{M}} (\omega) ]^{-1}_{p q} \, [ \psi_{\bm{m}_{2}}^{(q)} \!(\bm{J}_{2}) ]^{*} \, ,
\label{definition_1/D}
\end{equation}
where $\widehat{\mathbf{M}}$ is the response matrix given by
\begin{equation}
\widehat{\mathbf{M}}_{pq} (\omega) = (2 \pi)^{d} \! \sum_{\bm{m}} \!\! \int \!\! \mathrm{d} \bm{J} \, \frac{\bm{m} \!\cdot\! \partial F / \partial \bm{J} }{\omega \!-\! \bm{m} \!\cdot\! \bm{\Omega}} [\psi_{\bm{m}}^{(p)} \!(\bm{J})]^{*} \psi_{\bm{m}}^{(q)} \!(\bm{J}) \, .
\label{Fourier_M}
\end{equation}
In this expression, ${ \psi_{\bm{m}}^{(p)} (\bm{J}) }$ corresponds to the Fourier transform in angles of the basis elements ${ \psi^{(p)} (\bm{x}) }$, where we used the convention that the Fourier transform of a function ${ X (\bm{\theta} , \bm{J}) }$ is given by
\begin{equation}
\begin{cases}
\displaystyle X (\bm{\theta} , \bm{J}) = \!\! \sum_{\bm{m} \in \mathbb{Z}^{d}} X_{\bm{m}} (\bm{J}) \, e^{i \bm{m} \cdot \bm{\theta}} \, ,
\\
\displaystyle X_{\bm{m}} (\bm{J}) = \frac{1}{(2 \pi)^{d}} \!\! \int \!\! \mathrm{d} \bm{\theta} \, X (\bm{\theta} , \bm{J}) \, e^{- i \bm{m} \cdot \bm{\theta}} \, .
\end{cases}
\label{definition_Fourier_angles}
\end{equation}
Note that equation~\eqref{initial_BL} is therefore a non-linear function of $F$, both explicitly via the products of $F$ on the r.h.s., describing the effects of the binary collisions (as in the Boltzmann or Landau equations), but also
implicitly via equation~\eqref{Fourier_M}, which encompasses the collective effects and is specific to the Balescu-Lenard equation.
Note also importantly that -- in  contrast to its counterpart in plasma physics --
 equation~\eqref{initial_BL} does {\sl not} assume that {\sl local} encounters drive secular evolution; resonant interactions of correlated 
possibly distant  dressed orbits are sourcing long term orbital distortions.

\subsection{Content of the diffusion equation}

One may also rewrite the Balescu-Lenard equation~\eqref{initial_BL} as an anisotropic Fokker-Planck equation introducing the associated drift and diffusion coefficients. It then reads
\begin{equation}
  \frac{\partial F}{\partial t} =  \sum_{\bm{m}_{1}} \frac{\partial }{\partial \bm{J}_{1}} \cdot  \left[  \bm{m}_{1} \left( A_{\bm{m}_{1}} ( \bm{J}_{1} ) \, F ( \bm{J}_{1} ) +  D_{\bm{m}_{1}}  ( \bm{J}_{1} ) \, \bm{m}_{1}  \cdot  \frac{\partial F}{\partial \bm{J}_{1}}  \right)  \right] ,
\label{initial_BL_rewrite}
\end{equation}
where ${ A_{\bm{m}_{1}} (\bm{J}_{1}) }$ and ${ D_{\bm{m}_{1}} (\bm{J}_{1}) }$ are respectively the anisotropic drift and diffusion coefficients associated to a given resonance $\bm{m}_{1}$, \textit{i.e.} to a given Fourier mode $\bm{m}_{1}$ in angles. They both secularly depend on the distribution function $F$, but this dependence has not been explicitly written out to shorten the notations. The drift coefficients are given by
\begin{equation}
\! A_{\bm{m}_{1}} (\bm{J}) \! = \! - \pi (2 \pi)^{d} \!\!  \sum_{\bm{m}_{2}}  \!\!  \int  \!\!  \mathrm{d} \bm{J}_{2}  \frac{\delta_{\rm D} (\bm{m}_{1}  \!\cdot\!  \bm{\Omega}_{1}  \!-\!  \bm{m}_{2}  \!\cdot\!  \bm{\Omega}_{2})}{|\mathcal{D}_{\bm{m}_{1} , \bm{m}_{2}} (\bm{J}_{1} , \bm{J}_{2},\bm{m}_{1}  \!\cdot\!  \bm{\Omega}_{1}) |^{2}} \bm{m}_{2}  \!\cdot\!  \frac{\partial F}{\partial \bm{J}_{2}} \, ,
\label{initial_drift}
\end{equation}
while the diffusion coefficients are given by
\begin{equation}
D_{\bm{m}_{1}} (\bm{J}_{1}) \! =\! \pi (2 \pi)^{d} \!\! \sum_{\bm{m}_{2}} \!\! \int \!\! \mathrm{d} \bm{J}_{2} \frac{\delta_{\rm D} (\bm{m}_{1} \!\cdot\! \bm{\Omega}_{1} \!-\! \bm{m}_{2} \!\cdot\! \bm{\Omega}_{2})}{|\mathcal{D}_{\bm{m}_{1} , \bm{m}_{2}} (\bm{J}_{1} , \bm{J}_{2},\bm{m}_{1} \!\cdot\! \bm{\Omega}_{1}) |^{2}} F (\bm{J}_{2}) \, .
\label{initial_diff}
\end{equation}
The rewriting from equation~\eqref{initial_BL_rewrite} allows us to discuss some important properties of such anisotropic diffusion equation. We introduce the total flux, $\bm{\mathcal{F}}_{\rm tot}$, associated with this diffusion, which reads
\begin{equation}
\bm{\mathcal{F}}_{\rm tot} = \sum_{\bm{m}} \bm{m} \left(\! A_{\bm{m}} (\bm{J}) \, F (\bm{J}) + D_{\bm{m}} (\bm{J}) \, \bm{m} \!\cdot\! \frac{\partial F}{\partial \bm{J}} \!\right) \, .
\label{definition_Ftot}
\end{equation}
As a consequence, the Balescu-Lenard diffusion equation given by the expressions~\eqref{initial_BL} and~\eqref{initial_BL_rewrite} takes the shortened form
\begin{equation}
\frac{\partial F}{\partial t} = \text{div} \left( \bm{\mathcal{F}}_{\rm tot} \right) \, .
\label{BL_divergence_flux}
\end{equation}
We then define as ${ M (t) }$ the mass contained in a volume $\mathcal{V}$ of the action-space at time $t$, so that we have
\begin{equation}
M (t) = \!\! \int_{\mathcal{V}} \!\!\! \mathrm{d} \bm{J} \, F (\bm{J} , t) \, .
\label{definition_Mt}
\end{equation}
Thanks to the divergence theorem, the variation of mass in $\mathcal{V}$ due to secular diffusion corresponds to the flux of particles through the boundary $\mathcal{S}$ of this volume so that
\begin{equation}
\!\!\frac{\mathrm{d} M}{\mathrm{d} t} \!=\!\!  \int_{\mathcal{S}} \!\! \bm{\mathcal{F}}_{\!\rm tot}  \cdot \mathrm{d} \bm{S}   \!=\! \!
\sum_{\bm{m}} \!\! \int_{\mathcal{S}} \!\!\! \mathrm{d} S (\bm{m} \!\cdot\! \bm{n}) \left[\! A_{\bm{m}} (\!\bm{J}) F (\!\bm{J}) \!+\! D_{\bm{m}} (\!\bm{J}) \, \bm{m} \!\cdot\! \frac{\partial F}{\partial \bm{J}} \!\right] ,
\label{flux_through_S}
\end{equation}
where $\bm{n}$ is the exterior pointing normal vector. One can note in equation~\eqref{flux_through_S} that the contribution from a given resonance $\bm{m}$ takes the form of a preferential diffusion in the direction $\bm{m}$. This diffusion is therefore anisotropic because it is maximum for ${ \bm{n} \!\propto\! \bm{m} }$ and equal to $0$ for ${ \bm{n} \!\cdot\! \bm{m} \!=\! 0 }$. To emphasize the anisotropy of the diffusion, one may use the formalism of the \textit{slow} and \textit{fast} actions~\citep{Lynden1979,Lynden1996}. For simplicity, we consider the ${ 2D }$ case. For a given resonance ${ \bm{m} \!=\! ( m_{1} , m_{2} ) }$, we consider the change of coordinates
\begin{equation}
J_{\bm{m}}^{s} = \frac{\bm{J} \!\cdot\! \bm{m}}{|\bm{m}|} \;\;\; ; \;\;\; J_{\bm{m}}^{f} = \frac{\bm{J} \!\cdot\! \bm{m}^{\perp}}{|\bm{m}|} \, ,
\label{fast_slow_actions}
\end{equation}
where ${ J_{\bm{m}}^{s} }$ and ${ J_{\bm{m}}^{f} }$ are respectively the slow and fast actions associated to the resonance $\bm{m}$. Here ${ \bm{m}^{\perp} }$ corresponds to the direction perpendicular to the resonance so that ${ \bm{m}^{\perp} \!=\! (m_{2} , - m_{1}) }$, and ${ |\bm{m}| \!=\!\! \sqrt{\bm{m} \!\cdot\! \bm{m}} }$. Thanks to the chain rule, for any function ${ X (\bm{J}) }$, one has
\begin{equation}
\bm{m} \!\cdot\! \frac{\partial X}{\partial \bm{J}} = |\bm{m}| \, \frac{\partial X}{\partial J_{\bm{m}}^{s}} \bigg|_{J_{\bm{m}}^{f} = cst.} \, .
\label{chain_rule}
\end{equation}
Introducing the natural vector basis elements ${ \bm{e}_{\bm{m}}^{s} \!=\! \bm{m} / |\bm{m}| }$ and ${ \bm{e}_{\bm{m}}^{f} \!=\! \bm{m}^{\perp} / | \bm{m} | }$ associated with this change of coordinates, the diffusion flux ${ \bm{\mathcal{F}}_{\bm{m}} }$ associated with a resonance $\bm{m}$ takes the form
\begin{equation}
\bm{\mathcal{F}}_{\bm{m}} (J_{\bm{m}}^{s} , J_{\bm{m}}^{f}) = |\bm{m}| \left[ A_{\bm{m}}(\bm{J}) F (\bm{J}) \!+\! |\bm{m}| D_{\bm{m}} (\bm{J}) \frac{\partial F}{\partial J_{\bm{m}}^{s}} \right] \bm{e}_{\bm{m}}^{s} \, .
\label{Flux_m_fast_slow}
\end{equation}
Such a rewriting illustrates the fact that as soon as only one resonance $\bm{m}$ dominates the secular evolution, the diffusion flux will be aligned with this resonance. Hence one will observe a narrow mono-dimensional diffusion in the preferential ${J_{\bm{m}}^{s}-}$direction. During this diffusion, particles will conserve their fast action ${ J_{\bm{m}}^{f} }$, which can therefore be seen as an adiabatic invariant, whereas their slow action $J_{\bm{m}}^{s}$ gets to change. This strong anistropy in the diffusion is an essential property of the Balescu-Lenard equation~\eqref{initial_BL_rewrite}.

\section{Thin tepid discs and their WKB limit}
\label{sec:WKBlimit}

When implementing the inhomogeneous Balescu-Lenard equation, one encounters two main difficulties. The first one is the explicit construction  of a mapping ${ (\bm{x} , \bm{v}) \!\mapsto\! (\bm{\theta} , \bm{J}) }$ since the Balescu-Lenard drift and diffusion coefficients must be computed using  angle-actions coordinates. The second difficulty arises from the \textit{non-locality} of Poisson's equation which requires to introduce potential basis elements $\psi^{(p)}$ as in equation~\eqref{definition_basis}. One can then compute the response matrix from equation~\eqref{Fourier_M}, which must subsequently be inverted. The following step is to compute the drift and diffusion coefficients from equations~\eqref{initial_drift} and~\eqref{initial_diff} which requires to explicitly deal with the resonance constraint ${ \bm{m}_{1} \!\cdot\! \bm{\Omega}_{1} \!-\! \bm{m}_{2} \!\cdot\! \bm{\Omega}_{2} \!=\! 0 }$. In the case of a ${ 2D }$ axisymmetric disc, one may implement a WKB approximation~\citep{WKB,Toomre1964,Kalnajs1965,Lin1966,Palmer1989} which assumes that the diffusion of the system is made of tightly wound spirals. Such an assumption has two main consequences. First of all, Poisson's equation becomes \textit{local}, resulting in a diagonal response matrix. Moreover, it also entails that all the resonances becomes exactly local, allowing an explicit calculation of the resonant constraint ${ \delta_{\rm D} (\bm{m}_{1} \!\cdot\! \bm{\Omega}_{1} \!-\! \bm{m}_{2} \!\cdot\! \bm{\Omega}_{2} ) }$. We now detail these two elements: epicyclic approximation and WKB assumption.

\subsection{Epicyclic approximation}

For a sufficiently cold disc (\textit{i.e.} a disc where the radial excursions of the stars are small), one can explicitly build up a mapping ${ (\bm{x} , \bm{v}) \!\mapsto\! (\bm{\theta} , \bm{J}) }$ thanks to the epicyclic approximation. We introduce the polar coordinates ${ (R , \phi) }$ to describe the infinitely thin galactic disc, and introduce their associated momenta ${ (p_{R} , p_{\phi}) }$. As the disc at equilibrium is axisymmetric, the stationary Hamiltonian of the system reads
\begin{equation}
H_{0} (R,\phi,p_{R},p_{\phi}) = \frac{1}{2} \!\left[ p_{R}^{2} \!+\! \frac{p_{\phi}^{2}}{  R^{2} }\right] \!+ \psi_{0}(R) \, ,
\label{definition_Hamiltonian_disc}
\end{equation}
where $\psi_{0}$ is the stationary background potential within the disc. Because $\psi_{0}$ is axisymmetric, it does not depend on $\phi$, so that $p_{\phi}$ is a conserved quantity. We may then define the first action of the system, the angular momentum $J_{\phi}$, as
\begin{equation}
J_{\phi} = \frac{1}{2 \pi} \!\! \oint \! d \phi \, p_{\phi} = p_{\phi} = R^{2} \dot{\phi} \, .
\label{definition_Jphi}
\end{equation}
For a given value of $J_{\phi}$, the equation of evolution of $R$ is then given by
\begin{equation}
\ddot{R} = - \frac{\partial \psi_{\rm eff}}{\partial R} \, ,
\label{equation_evolution_R}
\end{equation}
where $\psi_{\rm eff}$ is an effective potential defined as
\begin{equation}
\psi_{\rm eff} (R) = \psi_{0} (R) \!+\! \frac{J_{\phi}^{2}}{2 R^{2}} \, .
\label{definition_psi_eff}
\end{equation}
The heart of the epicyclic approximation is to assume that small radial excursions can be approximated as harmonic librations. For a given value of $J_{\phi}$, we implicitly introduce the guiding radius $R_{g}$ as
\begin{equation}
0 = \frac{\partial \psi_{\rm eff}}{\partial R} \bigg|_{R_{g}} \!\!\!\!\! = \, \frac{\partial \psi_{0}}{\partial R} \bigg|_{R_{g}} \!\!\!\!\!- \frac{J_{\phi}^{2}}{R_{g}^{3}} \, .
\label{definition_Rg}
\end{equation}
Here ${ R_{g} (J_{\phi}) }$ is the radius for which stars with an angular momentum of $J_{\phi}$ are on exactly circular orbits. It is important to note that the mapping between $R_{g}$ and $J_{\phi}$ is bijective and unambiguous (up to the sign of $J_{\phi}$). We may then define the two frequencies of evolution: ${ \Omega_{\phi} (R_{g}) }$ the azimuthal frequency and ${ \kappa (R_{g}) }$ the epicyclic frequency as follows
\begin{equation}
\begin{cases}
\displaystyle \Omega_{\phi}^{2} (R_{g}) = \frac{1}{R_{g}} \frac{\partial \psi_{0}}{\partial R} \bigg|_{R_{g}} \!\!\!\! = \frac{J_{\phi}^{2}}{R_{g}^{4}} \, ,
\\
\displaystyle \kappa^{2} (R_{g}) = \frac{\partial^{2} \psi_{\rm eff}}{\partial R^{2}} \bigg|_{R_{g}} \!\!\!\! = \frac{\partial^{2} \psi_{0}}{\partial R^{2}} \bigg|_{R_{g}} \!\!\!\! + 3 \frac{J_{\phi}^{2}}{R_{g}^{4}} \, .
\end{cases}
\label{definition_Omega_kappa}
\end{equation}
As the radial oscillations are supposed to be small, one may perform a Taylor expansion at first order of the evolution equation~\eqref{equation_evolution_R} in the neighborhood of the minimum ${ R \!=\! R_{g} }$ so that $R$ satisfies the differential equation ${ \ddot{R} \!=\! - \kappa^{2} (R \!-\! R_{g}) }$. Hence one can note that in this limit the evolution of the radius of a star is the one of a harmonic oscillator centered on $R_{g}$. Up to an initial phase, one has therefore ${ R(t) \!=\! R_{g} \!+\! A \cos(\kappa t) }$, where $A$ is the amplitude of the radial oscillations. The associated radial action $J_{r}$ is then given by
\begin{equation}
J_{r} = \frac{1}{2 \pi} \!\! \oint \! \mathrm{d} R \, p_{R} = \frac{1}{2} \kappa A^{2} \, ,
\label{definition_Jr}
\end{equation}
For ${ J_{r} \!=\! 0 }$, the orbit is circular. Within the epicyclic approximation, the frequencies of motion along the action-torii, introduced in equation~\eqref{definition_Omega} are given by ${ \bm{\Omega} (\bm{J}) \!=\! (\Omega_{\phi} (J_{\phi}) , \kappa (J_{\phi})) }$. An important dynamical consequence of this approximation is that these two frequencies are only function of $J_{\phi}$ and do not depend on $J_{r}$, so that the resonance constraint ${ \bm{m}_{1} \!\cdot\! \bm{\Omega}_{1} \!-\! \bm{m}_{2} \!\cdot\! \bm{\Omega}_{2} \!=\! 0 }$ becomes simpler. Finally, one can explicitly construct the mapping between ${ (R,\phi,p_{R},p_{\phi}) }$ and ${ (\theta_{R} , \theta_{\phi} , J_{r} , J_{\phi}) }$ \citep{LyndenBell1972,Palmer1994,BinneyTremaine2008}, which takes at first order the form
\begin{equation}
\begin{cases}
\displaystyle R = R_{g} \!+\! A \cos (\theta_{R}) \, ,
\\
\displaystyle \phi = \theta_{\phi} \!-\! \frac{2 \Omega_{\phi}}{\kappa} \frac{A}{R_{g}} \sin (\theta_{R}) \, .
\end{cases}
\label{mapping_action}
\end{equation}
Thanks to this mapping and the definitions of the actions from equations~\eqref{definition_Jphi} and~\eqref{definition_Jr}, the epicyclic approximation allows us to build up an explicit mapping between the physical phase-space coordinates and the angle-actions ones.

Finally, throughout our calculation, we will assume that the stationary distribution function of the disc is a Schwarzschild distribution function (or \textit{locally isothermal} DF) given by
\begin{equation}
F (R_{g} , J_{r}) = \frac{\Omega_{\phi} (R_{g}) \, \Sigma (R_{g})}{\pi \, \kappa (R_{g}) \, \sigma_{r}^{2} (R_{g})} \exp \!\left[\! - \frac{\kappa (R_{g}) \, J_{r}}{\sigma_{r}^{2} (R_{g})} \!\right] \, ,
\label{definition_DF_Schwarzschild}
\end{equation}
where ${ \Sigma (R_{g}) }$ is the surface density of the disc and ${ \sigma_{r}^{2} (R_{g}) }$, which varies within the disc, represents the radial velocity dispersion of the stars at a given radius. Increasing values of $\sigma_{r}^{2}$ correspond to hotter discs that  are therefore more stable.

\subsection{The WKB basis}

As we are considering a ${ 2D }$ case, the potential basis elements $\psi^{(p)}$ introduced in equation~\eqref{definition_basis} must be written as ${ \psi^{(p)} (R , \phi) }$ in the disc polar coordinates and must be orthonormal to the associated surface density ${ \Sigma^{(p)} (R , \phi) }$. Using a WKB approximation amounts to building up \textit{local} basis elements thanks to which the response matrix will become diagonal.

\subsubsection{Definition of the basis elements}

\begin{figure}
\begin{center}
\begin{tikzpicture}
\draw [->] [thick] (-0.5,0) -- (7.5,0) ; \draw (7.5,0) node[font = \small, right]{$R$} ;
\draw [->] [thick] (-0.3,-2) -- (-0.3,2) ; \draw (-0.3,2) node[font = \small, above]{$\psi$} ;
\draw [dashed , smooth] plot [domain=0:4] ( \x,{exp(- (\x-2)*(\x-2)*2)} ) ; \draw [dashed , smooth] plot [domain=0:4] ( \x,{-exp(- (\x-2)*(\x-2)*2)} ) ;
\draw [thick, smooth, samples = 100] plot [domain=0:4] ( \x,{1.005 * exp(- (\x-2)*(\x-2)*2) * cos(1200*(\x-2))} );
\draw [dashed] (2,0) -- (2,-1.3) ; \draw [thin] (2,-0.05) -- (2,0.05) ; \draw (2.05,-1.32) node[font = \small, below]{$R_{0}^{p}$} ;
\draw [|->] [thin] (2,-1.3) -- (3,-1.3) ; \draw (2.5,-1.35) node[font = \small, above]{$\sigma$} ;
\draw [|->] [thin] (2,1.1) -- (2.35,1.1) ; \draw (2.175,1.1) node[font = \small , above]{$ 1 \!/\! k_{r}^{p}$} ;
\draw [dashed , smooth] plot [domain=3.5:7.5] ( \x,{exp(- (\x-5.5)*(\x-5.5)*2)} ) ; \draw [dashed , smooth] plot [domain=3.5:7.5] ( \x,{-exp(- (\x-5.5)*(\x-5.5)*2)} ) ;
\draw [thick, smooth, samples = 200] plot [domain=3.5:7.5] ( \x,{1.005 * exp(- (\x-5.5)*(\x-5.5)*2) * cos(2300*(\x-5.5))} );
\draw [dashed] (5.5,0) -- (5.5,-1.3) ; \draw [thin] (5.5,-0.05) -- (5.5,0.05) ; \draw (5.55,-1.32) node[font = \small, below]{$R_{0}^{q}$} ;
\draw [|->] [thin] (5.5,-1.3) -- (6.5,-1.3) ; \draw (6.0,-1.35) node[font = \small, above]{$\sigma$} ;
\draw [|->] [thin] (5.5,1.1) -- (5.68,1.1) ; \draw (5.675,1.1) node[font = \small , above]{$ 1 \!/\! k_{r}^{q}$} ;
\end{tikzpicture}
\caption{\small{Two WKB basis elements. Each Gaussian is centered around a radius $R_{0}$. The typical extension of the Gaussian is given by the decoupling scale $\sigma$, and they are modulated at the radial frequency $k_{r}$.
}}
\label{plot_WKB_basis_radial}
\end{center}
\end{figure}
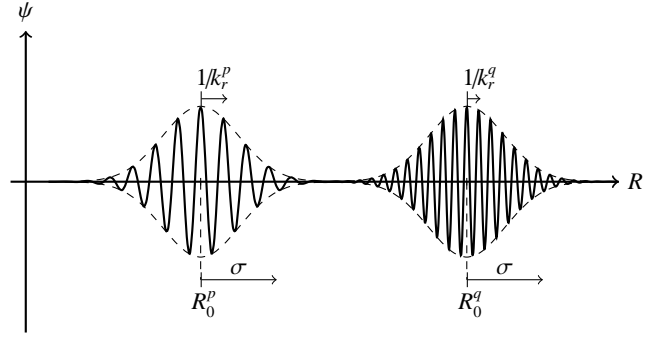

\begin{figure}
\begin{center}
\epsfig{file=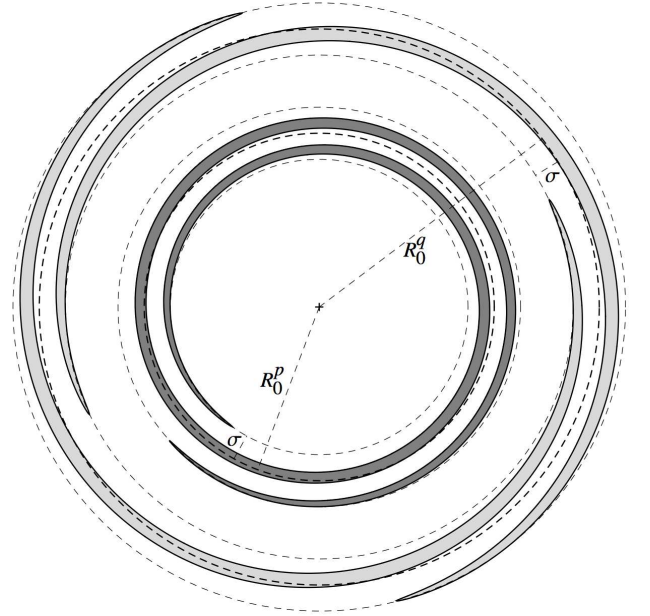,angle=-00,width=0.45\textwidth}
\caption{\small{Two WKB basis elements in the polar ${ (R,\phi)-}$plane. Each basis element is located around a central radius $R_{0}$, on a region of size $\sigma$. The winding of the spirals is governed by the radial frequency $k_{r}$. The number of azimuthal patterns is given by the index $k_{\phi}$, \textit{e.g.} ${ k_{\phi} \!=\! 1 }$ for the interior dark gray element, whereas ${ k_{\phi} \!=\! 2 }$ for the exterior light gray one.
}}
\label{plot_WKB_basis_polar}
\end{center}
\end{figure}

We introduce the basis elements
\begin{equation}
\psi^{[k_{\phi} , k_{r} , R_{0}]} (R , \phi) = \mathcal{A} \, e^{i (k_{\phi} \phi + k_{r} R)} \, \mathcal{B}_{R_{0}} (R) \, ,
\label{definition_WKB_basis}
\end{equation}
where the \textit{window} function ${ \mathcal{B}_{R_{0}} (R) }$ is defined as
\begin{equation}
\mathcal{B}_{R_{0}} (R) = \frac{1}{(\pi \sigma^{2})^{1\!/\!4}} \, \exp \!\left[\! - \frac{(R \!-\! R_{0})^{2}}{2 \sigma^{2}} \!\right] \, .
\label{definition_window_B}
\end{equation}
The basis elements are indexed by three numbers: $k_{\phi}$ is an azimuthal number which parametrizes the angular component of the basis elements, $R_{0}$ is the radius position in the disc around which the Gaussian window $\mathcal{B}_{R_{0}}$ is centered, and $k_{r}$ is the radial frequency of the basis element. We also introduced an additional parameter $\sigma$ of scale-separation, which will ensure the biorthogonality of the basis elements, as detailed later on. Finally, $\mathcal{A}$ is an amplitude which will be tuned in order to normalize correctly the basis elements. Thanks to  a somewhat unsual normalization of $\mathcal{B}_{R_{0}}$, we will ensure that $\mathcal{A}$ is independent of $\sigma$. Figure~\ref{plot_WKB_basis_radial} illustrates the radial dependence of the basis elements.
Figure~\ref{plot_WKB_basis_polar} illustrates the shape of these basis elements in the polar ${ (R , \phi)-}$plane.
The next steps will be to ensure that these WKB basis elements have all the properties required to allow for the computation of the dressed susceptibility coefficients introduced in equation~\eqref{definition_1/D}. Therefore, we will successively compute the associated surface density elements ${ \Sigma^{[k_{\phi} , k_{r} , R_{0}]} }$, ensure the biorthogonality of the basis elements and their correct normalization, and finally compute the Fourier transform in angles of the basis elements.

\subsubsection{Associated surface densities}

In order to ensure the biorthogonality of the basis,  we will first build up the surface densities associated to the potential elements introduced in equation~\eqref{definition_WKB_basis}. We extend the WKB potential in the ${z-}$direction using the Ansatz
\begin{equation}
\psi^{[k_{\phi} , k_{r} , R_{0}]} (R , \phi ,z) = \mathcal{A} \, e^{i (k_{\phi} \phi + k_{r} R)} \mathcal{B}_{R_{0}} (R) \, Z (z) \, .
\label{ansatz_WKB_potential}
\end{equation}
Poisson's equation in vacuum ${ \Delta \psi^{[k_{\phi} , k_{r} , R_{0}]} \!=\! 0 }$ leads to
\begin{align}
\frac{Z''}{Z} = k_{r}^{2} \bigg[ 1 \!-\! \frac{i}{k_{r} R} \!+\! 2 i \frac{R \!-\! R_{0}}{\sigma^{2}} \frac{1}{k_{r}} \!+\! \frac{R \!-\! R_{0}}{R} \frac{1}{(\sigma k_{r})^{2}}  \nonumber
\\
 + \frac{1}{(\sigma k_{r})^{2}} \!+\! \frac{k_{\phi}^{2}}{(k_{r} R)^{2}} \!-\! \bigg[ \frac{R \!-\! R_{0}}{\sigma^{2}}\frac{1}{k_{r}} \bigg]^{2} \bigg] \, . \label{Poisson_psi_vacuum}
\end{align}
We now explicitly introduce our WKB assumptions. We assume that the spirals are tightly wound so that
\begin{equation}
k_{r} \, R \gg 1 \, .
\label{WKB_assumption_I}
\end{equation}
Introducing the typical size of the system $R_{\rm sys}$, we also additionally suppose that we have
\begin{equation}
k_{r} \sigma \gg \frac{R_{\rm sys}}{\sigma} \, .
\label{WKB_assumption_II}
\end{equation}
Assuming that $k_{\phi}$ is of the order of  unity, equation~\eqref{Poisson_psi_vacuum} becomes
\begin{equation}
\frac{Z''}{Z} = k_{r}^{2} \, .
\label{Poissons_psi_vacuum_short}
\end{equation}
Hence within the WKB limit, the \textit{extended} potential from equation~\eqref{ansatz_WKB_potential} takes the form
\begin{equation}
\psi^{[k_{\phi} , k_{r} , R_{0}]} (R , \phi , z) = \psi^{[k_{\phi} , k_{r} , R_{0}]} (R , \phi) \, e^{- |k_{r}| z} \, ,
\label{ansatz_WKB_potential_final}
\end{equation}
where we ensured that for ${ z \!\to\! \pm\infty }$ the potential tends to $0$, therefore introducing a discontinuity for ${ \partial \psi / \partial z }$ in ${ z \!=\! 0 }$. Thanks to Gauss theorem, the associated surface density satisfies
\begin{equation}
\Sigma (R , \phi) = \frac{1}{4 \pi G} \left[\lim\limits_{z \to 0^{+}} \frac{\partial \psi}{\partial z} \,  - \lim\limits_{z \to 0^{-}} \frac{\partial \psi}{\partial z} \right] \, ,
\label{Poisson_equation_2D}
\end{equation}
so that we have
\begin{equation}
\Sigma^{[k_{\phi} , k_{r} , R_{0}]} (R , \phi) = - \frac{|k_{r}|}{2 \pi G} \, \psi^{[k_{\phi} , k_{r} , R_{0}]} (R , \phi) \, .
\label{WKB_surface_density}
\end{equation}

\subsubsection{Biorthogonality condition and normalization}

One must now ensure that the basis elements introduced in equations~\eqref{definition_WKB_basis} and~\eqref{WKB_surface_density} form a biorthogonal basis as assumed in equation~\eqref{definition_basis}. Indeed, it has to satisfy the property
\begin{equation}
\delta_{k_{\phi}^{p}}^{k_{\phi}^{q}} \delta_{k_{r}^{p}}^{k_{r}^{q}} \delta_{R_{0}^{p}}^{R_{0}^{q}} = - \!\! \int \!\! \mathrm{d} R \, R \, \mathrm{d} \phi \, \!\left[\! \psi^{[k_{\phi}^{p} , k_{r}^{p} , R_{0}^{p}]}  \!\right]^{*} \, \Sigma^{[k_{\phi}^{q} , k_{r}^{q} , R_{0}^{q}]}  \, .
\label{biorthogonality_condition}
\end{equation}
The r.h.s of this expression takes the form
\begin{align}
\frac{|k_{r}^{q}|}{2 \pi G} & \mathcal{A}_{p} \mathcal{A}_{q} \frac{1}{\sqrt{\pi \sigma^{2}}} \!\! \int \!\! \mathrm{d} \phi \, e^{i (k_{\phi}^{p} - k_{\phi}^{q}) \phi} \nonumber
\\
& \times \!\! \int \!\! \mathrm{d} R \, R \, e^{i (k_{r}^{p} - k_{r}^{q}) R} \exp \!\left[\! - \frac{(R \!-\! R_{0}^{p})^{2}}{2 \sigma^{2}} \!\right] \, \exp \!\left[\! - \frac{(R \!-\! R_{0}^{q})^{2}}{2 \sigma^{2}} \!\right] \, .  \label{calculation_biorthogonality_I}
\end{align}
The integration on $\phi$ is straightforward and is equal to ${ 2 \pi \delta_{k_{\phi}^{p}}^{k_{\phi}^{q}} }$. In order to perform the integration on $R$, we have to introduce additional  assumptions to ensure the biorthogonality of the basis. The peaks of the Gaussians in equation~\eqref{calculation_biorthogonality_I} can be considered as separated if ${ \Delta R_{0} \!=\! R_{0}^{p} \!-\! R_{0}^{q} }$ satisfies the condition
\begin{equation}
\Delta R_{0} \gg \sigma \;\;\; \text{if} \;\;\; R_{0}^{p} \neq R_{0}^{q} \, .
\label{step_distance_R0}
\end{equation}
Under this assumption\footnote{This is an assumption that one might want to lift partially to account for weakly non local effects.}, the term from equation~\eqref{calculation_biorthogonality_I} can be assumed to be non-zero only for ${ R_{0}^{p} \!=\! R_{0}^{q} }$. The r.h.s of equation~\eqref{biorthogonality_condition} then takes the form
\begin{equation}
\delta_{k_{\phi}^{p}}^{k_{\phi}^{q}} \delta_{R_{0}^{p}}^{R_{0}^{q}} \frac{|k_{r}^{q}|}{G} \mathcal{A}_{p} \mathcal{A}_{q}  \frac{1}{\sqrt{\pi \sigma^{2}}} \!\! \int \!\! \mathrm{d} R \, R \, e^{i (k_{r}^{p} - k_{r}^{q}) R} \exp \!\left[\! - \frac{(R \!-\! R_{0}^{p})^{2}}{\sigma^{2}} \!\right] \, .
\label{calculation_biorthogonality_II}
\end{equation}
The integration on $R$ takes the form of a radial Fourier transform of a Gaussian of spread ${ \sigma }$ at the frequency ${ \Delta k_{r} \!=\! k_{r}^{p} \!-\! k_{r}^{q} }$. It is therefore proportional to ${ \exp [- (\Delta k_{r})^{2} / (4/\sigma)^{2}] }$. Hence we will suppose that the frequency spread ${ \Delta k_{r} }$ satisfies
\begin{equation}
\Delta k_{r} \gg \frac{1}{\sigma} \;\;\; \text{if} \;\;\; k_{r}^{p} \neq k_{r}^{q} \, .
\label{step_distance_kr}
\end{equation}
Under this assumption, the term from equation~\eqref{calculation_biorthogonality_II} is non-zero only for ${ k_{r}^{p} \!=\! k_{r}^{q} }$. In order to have a biorthogonal basis, one must therefore consider a spread $\sigma$, central radii $R_{0}$, and radial frequencies $k_{r}$ such that
\begin{equation}
\Delta R_{0} \gg \sigma \gg \frac{1}{\Delta k_{r}} \, .
\label{step_distances_sum_up}
\end{equation}
With these constraints, one must necessarily have ${ k_{\phi}^{p} \!=\! k_{\phi}^{q} }$, ${ k_{r}^{p} \!=\! k_{r}^{q} }$ and ${ R_{0}^{p} \!=\! R_{0}^{q} }$ in order to have a non negligible term in equation~\eqref{biorthogonality_condition}. It then only remains to explicitly estimate the amplitude $\mathcal{A}$ of the basis elements. Equation~\eqref{biorthogonality_condition} gives
\begin{equation}
\mathcal{A}^{2} \frac{|k_{r}|}{G} \frac{1}{\sqrt{\pi \sigma^{2}}} \!\! \int \!\! \mathrm{d} R \, R \, \exp \!\left[\! - \frac{(R \!-\! R_{0})^{2}}{\sigma^{2}} \!\right] = 1 \, .
\label{calculation_normalization}
\end{equation}
Thanks to the WKB assumption from equation~\eqref{step_distances_sum_up}, the integration can be straightforwardly computed and leads to
\begin{equation}
\mathcal{A} = \sqrt{\frac{G}{|k_{r}| \, R_{0}}} \, .
\label{expression_amplitude_WKB}
\end{equation}

\subsubsection{Fourier transform in angles}

In order to estimate the susceptibility coefficients and the response matrix from equations~\eqref{definition_1/D} and~\eqref{Fourier_M}, one has to be able to calculate ${ \psi_{\bm{m}}^{(p)} (\bm{J}) }$ for the WKB basis elements. Thanks to the explicit mapping from equation~\eqref{mapping_action}, we have to compute
\begin{align}
\psi_{\bm{m}}^{[k_{\phi} , k_{r} , R_{0}]} (\bm{J}) = \frac{\mathcal{A} \, e^{i k_{r} R_{g}}}{(2 \pi)^{2}} \!\! \int \!\! \mathrm{d} \theta_{\phi} \!\! \int \!\! \mathrm{d} \theta_{R} \, e^{- i m_{\phi} \theta_{\phi}} e^{- i m_{r} \theta_{R}} e^{i k_{\phi} \theta_{\phi}} \nonumber
\\
\times \, e^{i [ k_{r} A \cos (\theta_{R}) - k_{\phi} \!\frac{2 \Omega_{\phi}}{\kappa} \frac{A}{R_{g}} \sin (\theta_{R}) ] } \mathcal{B}_{R_{0}} \!(R_{g} \!+\! A \cos (\theta_{R})) \, . \label{calculation_psi_m}
\end{align}
The integration on $\theta_{\phi}$ is straighforward and equal to ${ 2 \pi \,\delta_{m_{\phi}}^{k_{\phi}} }$. Regarding the dependence on $\theta_{R}$ in the complex exponential, we may write
\begin{equation}
k_{r} A \cos (\theta_{R}) \!-\! k_{\phi} \frac{2 \Omega_{\phi}}{\kappa} \frac{A}{R_{g}} \sin (\theta_{R}) = H_{k_{\phi}} (k_{r}) \sin (\theta_{R} \!+\! \theta_{R}^{0}) \, ,
\label{dependence_theta_R}
\end{equation}
where the amplitude ${ H_{k_{\phi}} (k_{r}) }$ and the phase shift $\theta_{R}^{0}$ are given by
\begin{equation}
H_{k_{\phi}} (k_{r}) = A \, | k_{r}|  \sqrt{1 \!+\! \left[ \frac{\Omega_{\phi}}{\kappa} \frac{2 k_{\phi}}{k_{r} R_{g}} \right]^{2}} \; ; \; \theta_{R}^{0} = \tan^{-1} \!\!\left[\! - \frac{\kappa}{\Omega_{\phi}} \frac{k_{r} R_{g}}{2 k_{\phi}} \!\right] \, .
\label{expression_H_theta0}
\end{equation}
For typical galaxies, we have ${ 1/2 \!\leq\! \Omega_{\phi} / \kappa \!\leq\! 1}$ \citep{BinneyTremaine2008}. Assuming that the azimuthal number $k_{\phi}$ is of the order of unity, one can use the WKB hypothesis introduced in equation~\eqref{WKB_assumption_I}, so that equations~\eqref{expression_H_theta0} can be approximated as
\begin{equation}
H_{k_{\phi}} (k_{r}) \simeq A \, |k_{r}| \simeq \sqrt{\frac{2 J_{r}}{\kappa}} \, |k_{r}| \;\;\; ; \;\;\; \theta_{R}^{0} \simeq - \frac{\pi}{2} \, .
\label{approximation_H_theta0}
\end{equation}
Because we have assumed that the disc is tepid, the radial oscillations are small so that ${ A \!\ll\! R_{g} }$. We may then get rid of the dependances on $A$ in ${ \mathcal{B}_{R_{0}} (R_{g} \!+\! A \cos (\theta_{R})) }$ by replacing it with ${ \mathcal{B}_{R_{0}} (R_{g}) }$, so that the only remaining dependence on $A$ will be in the complex exponentials. To be able to explicitly perform the remaining integration on $\theta_{R}$ in equation~\eqref{calculation_psi_m}, we introduce the Bessel functions ${ \mathcal{J}_{\ell} }$ of the first kind which satisfy the relation
\begin{equation}
e^{i z \sin (\theta)} = \sum_{\ell \in \mathbb{Z}} \mathcal{J}_{\ell} [z] \, e^{i \ell \theta} \, .
\label{summation_property_Bessel}
\end{equation}
We then finally obtain the expression of the Fourier transform in angles of the WKB basis elements which reads
\begin{equation}
\psi_{\bm{m}}^{[k_{\phi} , k_{r} , R_{0}]} (\bm{J}) = \delta_{m_{\phi}}^{k_{\phi}} e^{i k_{r} R_{g}} e^{i m_{r} \theta_{R}^{0}} \mathcal{A} \, \mathcal{J}_{m_{r}} [H_{m_{\phi}} (k_{r})] \, \mathcal{B}_{R_{0}} (R_{g}) \, .
\label{Fourier_WKB}
\end{equation}

\subsection{Estimation of the response matrix}

Thanks to the WKB basis introduced in equation~\eqref{definition_WKB_basis}, one can now explicitly compute the response matrix from equation~\eqref{Fourier_M}. Indeed, we use the expression~\eqref{Fourier_WKB} of the Fourier transform of the basis elements, and after simplification of the phase-shift terms thanks to the approximation from equation~\eqref{approximation_H_theta0}, one has to evaluate
\begin{align}
 & \widehat{\mathbf{M}}_{[k_{\phi}^{p} , k_{r}^{p} , R_{0}^{p}] , [k_{\phi}^{q} , k_{r}^{q} , R_{0}^{q}]} (\omega) = \nonumber
\\
& \;\;\;\;\;\;\;\;\;\; (2 \pi)^{2} \sum_{\bm{m}} \!\! \int \!\! \mathrm{d} \bm{J} \, \frac{\bm{m} \!\cdot\! \partial F / \partial \bm{J}}{\omega \!-\! \bm{m} \!\cdot\! \bm{\Omega}} \delta_{m_{\phi}}^{k_{\phi}^{p}} \delta_{m_{\phi}}^{k_{\phi}^{q}} e^{i (k_{r}^{q} - k_{r}^{p}) R_{g}} \mathcal{A}_{p} \mathcal{A}_{q} \nonumber
\\
& \;\;\;\;\;\;\;\;\;\; \times \, \mathcal{J}_{m_{r}} \!\left[\!\! \sqrt{\tfrac{2 J_{r}}{\kappa}} k_{r}^{p} \!\right] \mathcal{J}_{m_{r}} \!\left[\!\! \sqrt{\tfrac{2 J_{r}}{\kappa}} k_{r}^{q} \!\right] \, \mathcal{B}_{R_{0}^{p}} (R_{g}) \, \mathcal{B}_{R_{0}^{q}} (R_{g}) \, .  \label{calculation_M_I}
\end{align}
The first step of the calculation is to show that in the WKB limit, the response matrix becomes diagonal. One should note that the previous expression~\eqref{calculation_M_I} is similar to equation~\eqref{calculation_biorthogonality_I}, where we discussed the biorthogonality of the WKB basis. In equation~\eqref{calculation_M_I}, the azimuthal Kronecker symbols ensures that ${ k_{\phi}^{p} \!=\! k_{\phi}^{q} }$. Moreover, because of our WKB assumptions from equation~\eqref{step_distances_sum_up} on the step distances of the basis elements, the product of the two radial Gaussians in $R_{g}$ imposes that ${ R_{0}^{p} \!=\! R_{0}^{q} }$ in order to have a non-zero contribution. In order to shorten the notations, we temporarily introduce the function ${ h (R_{g}) }$ defined as
\begin{equation}
h(R_{g}) = \left| \frac{\mathrm{d} J_{\phi}}{\mathrm{d} R_{g}} \right| \frac{\bm{m} \!\cdot\! \partial F / \partial \bm{J}}{\omega \!-\! \bm{m} \!\cdot\! \bm{\Omega}} \mathcal{A}_{p} \mathcal{A}_{q} \mathcal{J}_{m_{r}} \!\left[\!\! \sqrt{\tfrac{2 J_{r}}{\kappa}} k_{r}^{p} \!\right] \mathcal{J}_{m_{r}} \!\left[\!\! \sqrt{\tfrac{2 J_{r}}{\kappa}} k_{r}^{q} \!\right] \, ,
\label{definition_h} \nonumber
\end{equation}
which encompasses all the additional radial dependences appearing in equation~\eqref{calculation_M_I}. Thanks to the change of variables ${ J_{\phi} \!\mapsto\! R_{g} }$, the integral on $J_{\phi}$ which has to be evaluated in equation~\eqref{calculation_M_I}, when estimated for ${ R_{0}^{p} \!=\! R_{0}^{q} }$, is qualitatively of the form
\begin{equation}
\int \!\! \mathrm{d} R_{g} \, h (R_{g}) \, e^{i R_{g} (k_{r}^{q} - k_{r}^{p})} \exp \!\left[\! - \frac{(R_{g} \!-\! R_{0}^{p})^{2}}{\sigma^{2}} \!\right] \, .
\label{shape_integration_h_I}
\end{equation}
This expression corresponds to a radial Fourier transform $\mathcal{F}$ at the frequency ${ \Delta k_{r} }$. It can be rewritten as a convolution of two radial Fourier transforms so that it becomes
\begin{equation}
\eqref{shape_integration_h_I} \propto \int \!\! \mathrm{d} k' \, \mathcal{F} [h] (k') \, \exp \!\left[\! - \frac{(\Delta k_{r} \!-\! k')^{2}}{4 / \sigma^{2}} \!\right] \, ,
\label{shape_integration_h_II}
\end{equation}
where ${ \Delta k_{r} \!=\! k_{r}^{p} \!-\! k_{r}^{q} }$. We now rely on the WKB assumption from equation~\eqref{step_distance_kr}. If one has ${ \Delta k_{r} \!\neq\! 0 }$, because of the Gaussian from equation~\eqref{shape_integration_h_II}, the contribution from ${ \mathcal{F} [h] }$ will come from the region ${ k' \!\sim\! \Delta k_{r} \!\gg\! 1 / \sigma }$. We assume that the function $h$ is such that its Fourier Transform is limited to the frequency region ${ |k'| \!\lesssim\! 1 / \sigma }$. This is consistent with assuming that the properties of the disc are   radially slowly varying, and this implies that non-zero contributions to the response matrix can only be obtained when ${ \Delta k_{r} \!=\! k_{r}^{p} \!-\! k_{r}^{q} \!=\! 0 }$. Therefore, we have shown that within our WKB formalism, the response matrix from equation~\eqref{Fourier_M} is diagonal.

In order to shorten the notations, we will denote the matrix eigenvalues as
\begin{equation}
\lambda_{[k_{\phi} , k_{r} , R_{0}]} (\omega) = \widehat{\mathbf{M}}_{[k_{\phi} , k_{r} , R_{0}] , [k_{\phi} , k_{r} , R_{0}]} (\omega) \, .
\label{definition_lambda_short}
\end{equation}
For these diagonal coefficients, the last step is to explicitly compute the integrals over $J_{\phi}$ and $J_{r}$ in equation~\eqref{calculation_M_I} to obtain the expression of the response matrix eigenvalues. We now detail this calculation. Thanks to our scale-decoupling approach, we may replace the radial Gaussian from equation~\eqref{shape_integration_h_I} by a Dirac delta ${ \delta_{\rm D} (R_{g} \!-\! R_{0}^{p}) }$ while paying a careful attention to the correct normalization of the Gaussian. Hence we have to evaluate
\begin{align}
& \lambda_{[k_{\phi} , k_{r} , R_{0}]} (\omega) = \nonumber
\\
& (2 \pi)^{2} \mathcal{A}^{2} \! \left| \frac{\mathrm{d} J_{\phi}}{\mathrm{d} R_{g}} \right|_{R_{0}} \!\! \sum_{\bm{m}}  \delta_{k_{\phi}}^{m_{\phi}} \!\!\! \int \!\!\! \mathrm{d} J_{r}  \frac{\bm{m} \!\cdot\! \partial F / \partial \bm{J}}{\omega \!-\! m_{\phi} \Omega_{\phi} \!-\! m_{r} \kappa} \mathcal{J}_{m_{r}}^{2} \!\left[\!\! \sqrt{\tfrac{2 J_{r}}{\kappa}} k_{r} \!\right] \, . \label{calculation_M_II}
\end{align}
Because of the presence of the azimuthal Kronecker symbol, we may drop the sum on $m_{\phi}$. The intrinsic frequencies from equations~\eqref{definition_Omega_kappa} allow us to compute
\begin{equation}
\left| \frac{\mathrm{d} J_{\phi}}{\mathrm{d} R_{g}} \right|_{R_{0}} \!\! = \frac{R_{0} \kappa^{2}}{2 \Omega_{\phi}} \, .
\label{dJphi_dRg}
\end{equation}
Moreover, we assume that the galactic disc is tepid so that ${ | \partial F / \partial J_{\phi} | \!\ll\! |\partial F / \partial J_{r} | }$. We may then only keep the term corresponding to a gradient with respect to the radial action $J_{r}$. Thanks to the expression of the Schwarzschild distribution function from equation~\eqref{definition_DF_Schwarzschild} and the expression of the basis ampitude from equation~\eqref{expression_amplitude_WKB}, equation~\eqref{calculation_M_II} becomes after some simple algebra
\begin{align}
 \lambda_{[k_{\phi} , k_{r} , R_{0}]} (\omega) & =  \frac{2 \pi G \Sigma |k_{r}|}{\kappa^{2}} \frac{\kappa^{4}}{k_{r}^{2} \sigma_{r}^{4}} \nonumber
\\
& \times \, \sum_{m_{r}} \!\! \int \!\! \mathrm{d} J_{r} \, \frac{- m_{r} \exp [ - \kappa J_{r} / \sigma_{r}^{2} ]}{\omega \!-\! k_{\phi} \Omega_{\phi} \!-\! m_{r} \kappa} \mathcal{J}_{m_{r}}^{2} \!\left[\!\! \sqrt{\tfrac{2 J_{r}}{\kappa}} k_{r} \!\right] \, . \label{calculation_M_III}
\end{align}
We may now use the following integration formula (see formula (6.615) from \cite{Gradshteyn2007})
\begin{equation}
\int_{0}^{+ \infty} \!\!\!\!\!\!\! \mathrm{d} J_{r} \, e^{- \alpha J_{r}} \mathcal{J}_{m_{r}}^{2} \!\left[ \beta \sqrt{J_{r}} \right] = \frac{e^{- \beta^{2} / 2 \alpha}}{\alpha} \mathcal{I}_{m_{r}} \!\!\left[\! \frac{\beta^{2}}{2 \alpha} \!\right] \, ,
\label{integration_formula_Bessel}
\end{equation}
where ${ \alpha > 0}$, ${ \beta >0 }$, ${ m_{r} \in \mathbb{Z} }$ and $\mathcal{I}_{m_{r}}$ are the modified Bessel functions of the first kind. We apply this formula with ${ \alpha \!=\! \kappa / \sigma_{r}^{2} }$ and ${ \beta \!= \!\!\sqrt{2 k_{r}^{2} / \kappa} }$. We also introduce the notation
\begin{equation}
\chi = \frac{\sigma_{r}^{2} \, k_{r}^{2}}{\kappa^{2}} \, ,
\label{definition_chi}
\end{equation}
so that equation~\eqref{calculation_M_III} becomes
\begin{equation}
\lambda_{[k_{\phi} , k_{r} , R_{0}]} (\omega) = \frac{2 \pi G \Sigma |k_{r}|}{\kappa^{2}} \frac{\kappa}{\chi} \sum_{m_{r}} \frac{- m_{r} \, e^{- \chi} \, \mathcal{I}_{m_{r}} [\chi] }{\omega \!-\! k_{\phi} \Omega_{\phi} \!-\! m_{r} \kappa}  \, .
\label{calculation_M_IV}
\end{equation}
We now define the dimensionless shifted frequency $s$ as
\begin{equation}
s = \frac{\omega \!-\! k_{\phi} \Omega_{\phi}}{\kappa} \, .
\label{definition_s}
\end{equation}
Because we have ${ \mathcal{I}_{- m_{r}} [\chi] \!=\! \mathcal{I}_{m_{r}} [\chi]}$, we may rewrite equation~\eqref{calculation_M_IV} using the reduction factor \citep{Kalnajs1965,Lin1966} defined as
\begin{equation}
\mathcal{F} (s , \chi) = 2 \, (1 \!-\! s^{2}) \frac{e^{-\chi}}{\chi} \!\! \sum_{m_{r} = 1}^{+ \infty} \! \frac{\mathcal{I}_{m_{r}} [\chi]}{1 \!-\! [s / m_{r}]^{2}} \, .
\label{definition_reduction_F}
\end{equation}
As a conclusion, we obtain that within our WKB formalism the response matrix ${ \widehat{\mathbf{M}} }$ becomes diagonal and in the limit of tepid discs reads
\begin{equation}
\widehat{\mathbf{M}}_{[k_{\phi}^{p} , k_{r}^{p} , R_{0}^{p}] , [k_{\phi}^{q} , k_{r}^{q} , R_{0}^{q}]} = \delta_{k_{\phi}^{p}}^{k_{\phi}^{q}} \delta_{k_{r}^{p}}^{k_{r}^{q}} \delta_{R_{0}^{p}}^{R_{0}^{q}} \frac{2 \pi G \Sigma |k_{r}| }{\kappa^{2} (1 \!-\! s^{2})} \mathcal{F} (s ,\chi) \, ,
\label{diagonal_M_tepid}
\end{equation}
This  eigenvalue recovered using  the WKB basis introduced in equation~\eqref{definition_WKB_basis} is in full agreement with the seminal results from~\cite{Kalnajs1965} and~\cite{Lin1966}. In order to handle the singularity of the eigenvalue appearing for ${ s \!=\! n \!\in\! \mathbb{Z} }$, one adds a small imaginary part to the frequency of evaluation, so that ${ s \!=\! n \!+\! i \eta }$. Indeed, as long as $\eta$ is small compared to the imaginary part of the least damped mode of the disc, adding this complex part makes a negligible contribution to the expression of ${ \text{Re} (\lambda)}$.

\subsection{Estimation of the susceptibility coefficients}

One can now estimate the dressed susceptibility coefficients from equation~\eqref{definition_1/D}. In order to shorten the notations, we will write the WKB basis elements introduced in equation~\eqref{definition_WKB_basis} as
\begin{equation}
\psi^{(p)} = \psi^{[k_{\phi}^{p} , k_{r}^{p} , R_{0}^{p}]} \, .
\label{short_notation_WKB_basis}
\end{equation}
We have shown previously in equation~\eqref{diagonal_M_tepid} that within our WKB basis, the response matrix ${ \widehat{\mathbf{M}} }$ is diagonal. Its eigenvalues will be noted as $\lambda_{p}$ so that we have ${ \widehat{\mathbf{M}}_{pq} \!=\! \delta_{p}^{q} \lambda_{p} }$. Hence the expression~\eqref{definition_1/D} of the susceptibility coefficients takes the form
\begin{equation}
\frac{1}{\mathcal{D}_{\bm{m}_{1} , \bm{m}_{2}} (\bm{J}_{1} , \bm{J}_{2} , \omega)} = \sum_{p} \psi_{\bm{m}_{1}}^{(p)} \!(\bm{J}_{1}) \,  \!\left[\! \frac{1}{1 \!-\! \lambda_{p} (\omega)} \!\right]\! \, \psi_{\bm{m}_{2}}^{(p) *} \!(\bm{J}_{2}) \, .
\label{initial_1/D_WKB} \nonumber
\end{equation}
Using the expression of the Fourier transformed basis elements obtained in equation~\eqref{Fourier_WKB}, we obtain
\begin{align}
& \frac{1}{ \mathcal{D}_{\bm{m}_{1} , \bm{m}_{2}} (\bm{J}_{1} , \bm{J}_{2} , \omega)}  = \!\!\!\! \sum_{k_{\phi}^{p} , k_{r}^{p} , R_{0}^{p}} \!\!\!\! \delta_{m_{1}^{\phi}}^{k_{\phi}^{p}} \delta_{m_{2}^{\phi}}^{k_{\phi}^{p}} \frac{G}{k_{r}^{p} R_{0}^{p}} \frac{1}{1 \!-\! \lambda_{p}} \nonumber
\\
& \;\;\;\;\;\;\; \times \, \mathcal{J}_{m_{1}^{r}} \!\!\left[\!\! \sqrt{\!\tfrac{2 J_{r}^{1}}{\kappa_{1}}} k_{r}^{p} \!\right] \mathcal{J}_{m_{2}^{r}} \!\!\left[\!\! \sqrt{\!\tfrac{2 J_{r}^{2}}{\kappa_{2}}} k_{r}^{p}\!\right] e^{i k_{r}^{p} (R_{1} - R_{2})} e^{i \theta_{R}^{0 p }  (m_{1}^{r} - m_{2}^{r})} \nonumber
\\
& \;\;\;\;\;\;\; \times \, \frac{1}{\sqrt{\pi \sigma^{2}}} \exp \!\left[\! - \frac{(R_{1} \!-\! R_{0}^{p})^{2}}{2 \sigma^{2}} \!\right] \exp \!\left[\! - \frac{(R_{2} \!-\! R_{0}^{p})^{2}}{2 \sigma^{2}} \!\right] \, ,\label{calculation_1/D_I}
\end{align}
where we used the shortened notations ${ \kappa_{i} \!=\! \kappa (\bm{J}_{i}) }$, and ${ R_{i} \!=\! R_{g} (\bm{J}_{i}) }$. We also used the approximation introduced in equation~\eqref{approximation_H_theta0} for the values at which the Bessel functions have to be evaluated. Thanks to the Kronecker symbols in $m_{\phi}$ and $k_{\phi}^{p}$, we necessarily have
\begin{equation}
m_{1}^{\phi} = m_{2}^{\phi} = k_{\phi}^{p} \, ,
\label{equality_mphi_kphi}
\end{equation}
so that the sum on $k_{\phi}^{p}$ can be dropped.

\subsection{Restriction on the loci of resonance }

Before proceeding with the evaluation of the susceptibility coefficients obtained in equation~\eqref{calculation_1/D_I}, let us first emphasize a crucial consequence  of the WKB basis from equation~\eqref{definition_WKB_basis}, which is the restriction to only exactly local resonances. One can note that the expressions~\eqref{initial_drift} and~\eqref{initial_diff} of the drift and diffusion coefficients all involve an integration over the mute variable $\bm{J}_{2}$. For a given value of $\bm{J}_{1}$, $\bm{m}_{1}$ and $\bm{m}_{2}$, this integration should be seen as a \textit{scan} of the entire action-space, searching for resonant region where the constraint ${ \bm{m}_{1} \!\cdot\! \bm{\Omega}_{1} \!-\! \bm{m}_{2} \!\cdot\! \bm{\Omega}_{2} \!=\! 0 }$ is satisfied. We first recall the rule for the composition of a Dirac delta and a function which reads
\begin{equation}
\delta_{\rm D} (f (x)) = \sum_{y \in Z_{f}} \frac{\delta_{\rm D} (x \!-\! y)}{|f'(y)|} \, ,
\label{composition_Dirac_function}
\end{equation}
where ${ Z_{f} \!=\! \{ y \,|\, f(y) \!=\! 0 \} }$, and we have supposed that all the poles of $f$ are simple. As  noted in equation~\eqref{definition_Omega_kappa}, within the epicyclic approximation, the intrinsic frequencies ${ \bm{\Omega} \!=\! (\Omega_{\phi} , \kappa) }$ only depend on ${ R_{g} \!=\! R_{g} (J_{\phi}) }$ and are independent of $J_{r}$. Hence, the resonance condition ${ \bm{m}_{1} \!\cdot\! \bm{\Omega}_{1} \!-\! \bm{m}_{2} \!\cdot\! \bm{\Omega}_{2} \!=\! 0 }$ only depends on $J_{\phi}^{2}$ and is independent of $J_{r}^{2}$. Hence if we consider fixed $\bm{J}_{1}$, $\bm{m}_{1}$ and $\bm{m}_{2}$, the resonant Dirac delta which has to be studied takes the form
\begin{equation}
\delta_{\rm D} (\bm{m}_{1} \!\cdot\! \bm{\Omega}_{1} \!-\! \bm{m}_{2} \!\cdot\! \bm{\Omega}_{2}) = \!\!\! \sum_{R_{2}^{r} \,|\, f (R_{2}^{r}) = 0} \frac{\delta_{\rm D} (R_{2} \!-\! R_{2}^{r})}{\left| \frac{\partial }{\partial R} [\bm{m}_{2} \!\cdot\! \bm{\Omega}] \right|_{R_{2}^{r}}} \, ,
\label{rewriting_Dirac_delta}
\end{equation}
where the resonance condition ${ f (R_{2}^{r}) = 0 }$ is given by
\begin{equation}
f (R_{2}^{r}) = \bm{m}_{1} \!\cdot\! \bm{\Omega} (R_{1}) \!-\! \bm{m}_{2} \!\cdot\! \bm{\Omega} (R_{2}^{r}) \, .
\label{definition_f_R2r}
\end{equation}
The radii $R_{2}^{r}$ therefore correspond to the \textit{resonant radii} for which the resonance condition is satisfied. When writing equation~\eqref{rewriting_Dirac_delta}, we have assumed that the zeros of the resonance function are simple, which corresponds to the assumption that for any resonant radius $R_{2}^{r}$, we have ${ f'(R_{2}^{r}) \!\neq\! 0 }$. Assuming that ${ m_{2}^{\phi} \neq 0 }$, this condition can be rewritten as
\begin{equation}
\frac{\partial \Omega_{\phi}}{\partial \kappa} \bigg|_{R_{2}^{r}} \!\!\! \neq - \frac{m_{2}^{r}}{m_{2}^{\phi}} \, .
\label{condition_simple_poles}
\end{equation}
Resonance poles are therefore simple as long as the rates of change of the two intrinsic frequencies are not in a rational ratio. One must note that the Keplerian case for which ${ \kappa \!=\! \Omega_{\phi} }$ and the harmonic case for which ${ \kappa = 2 \Omega_{\phi} }$ are in this sense degenerate. It can lead to resonant poles of higher multiplicity and would therefore require a more involved evaluation of the Balescu-Lenard collision operator. In what follows we assume that the potential is not degenerate.

Let us now use the properties of the WKB basis to restrict the range of  resonant radii, $R_{2}^{r}$. The expression~\eqref{calculation_1/D_I} of the susceptibility coefficients,  thanks to the two Gaussians, imposes that the relevant resonant radius $R_{2}^{r}$ must necessarily be \textit{close} to $R_{1}$. As noted in equation~\eqref{equality_mphi_kphi}, in order to have a non-zero susceptibility, one also has to satisfy the constraint ${ m_{1}^{\phi} = m_{2}^{\phi} }$. The resonant condition which has to be satisfied is therefore given by
\begin{equation}
m_{1}^{\phi} \Omega_{\phi} (R_{1}) \!+\! m_{1}^{r} \kappa (R_{1}) = m_{1}^{\phi} \Omega_{\phi} (R_{2}^{r}) \!+\! m_{2}^{r} \kappa (R_{2}^{r}) \, ,
\label{statement_resonance}
\end{equation}
where the distance ${ \Delta R \!=\! R_{1} \!-\! R_{2}^{r} }$ is such that ${ |\Delta R| \!\leq\! \text{(few)} \, \sigma }$. Because the scale-decoupling parameter $\sigma$ is supposed to be small compared to the size of the system, we may approximate the previous resonant condition as
\begin{equation}
\bigg[ m_{2}^{\phi} \frac{\partial \Omega_{\phi}}{\partial R} \!+\! m_{2}^{r} \frac{\partial \kappa}{\partial R} \bigg] \, \Delta R = \bigg[ m_{1}^{r} \!-\! m_{2}^{r} \bigg] \, \kappa (R_{1}) \, .
\label{approximate_resonance_condition}
\end{equation}
On the l.h.s of equation~\eqref{approximate_resonance_condition}, the term within bracket is non-zero, because we assumed in equation~\eqref{condition_simple_poles} that the resonant poles are simple. Moreover, ${ \Delta R }$ is small, because of our scale-decoupling approach. The r.h.s of equation~\eqref{approximate_resonance_condition} is discrete: it is either zero or at least of the order of ${ \kappa (R_{1}) }$.
Because the l.h.s is necessarily small, we must have
\begin{equation}
\displaystyle R_{2}^{r} = R_{1} \, ,\quad
\displaystyle m_{2}^{r} = m_{1}^{r} \, .
\label{restriction_local_resonances}
\end{equation}
This result is a crucial consequence  of our WKB tightly wound spiral assumption, because it implies that
only  {\sl local} resonances are allowed. 
In particular this implies that the WKB limit does not allow for distant orbits to resonate (through \textit{e.g.} propagation of swing amplified wave packets, see below).
Then the sum ${ \sum_{R_{2}^{r}} }$ from equation~\eqref{rewriting_Dirac_delta} can be limited to the evaluation in ${ R_{2}^{r} \!=\! R_{1} }$. Hence within this WKB limit, the susceptibility coefficients from equation~\eqref{calculation_1/D_I} have to be evaluated only for ${ \bm{m}_{2} \!=\! \bm{m}_{1} }$ and ${ R_{2} \!=\! R_{1} }$, so that we have to deal with the expression
\begin{align}
&\frac{1}{\mathcal{D}_{\bm{m}_{1} , \bm{m}_{1}} (R_{1} , J_{r}^{1} , R_{1} , J_{r}^{2} , \omega)} =   \!\! \sum_{k_{r}^{p} , R_{0}^{p}} \! \frac{G}{k_{r}^{p} R_{0}^{p}} \frac{1}{1 \!-\! \lambda_{p}} \nonumber
\\
& \;\;  \times \, \mathcal{J}_{m_{1}^{r}} \!\left[\!\! \sqrt{\tfrac{2 J_{r}^{1}}{\kappa_{1}}} k_{r}^{p} \!\right] \mathcal{J}_{m_{1}^{r}} \!\left[\!\! \sqrt{\tfrac{2 J_{r}^{2}}{\kappa_{1}}} k_{r}^{p} \!\right]
\frac{1}{\sqrt{\pi \sigma^{2}}} \exp \!\left[\! - \frac{(R_{1} \!-\! R_{0}^{p})^{2}}{\sigma^{2}} \!\right] \, . \label{calculation_1/D_II}
\end{align}

\subsection{Asymptotic continuous limit}

One can note that in equation~\eqref{calculation_1/D_II} the susceptibility coefficients are still expressed as a discrete sum on the basis index $k_{r}^{p}$ and $R_{0}^{p}$. Our next step  is  to replace these sums by continuous integrals. The discrete basis elements are separated by the step distances ${ \Delta R_{0} }$ and ${ \Delta k_{r} }$, which must satisfy the WKB hypothesis detailed in equation~\eqref{step_distances_sum_up}. We use the Riemann sum formula ${ \sum \! f (x) \Delta x \!\simeq\! \! \int \! \mathrm{d} x \, f(x) }$, where ${ \Delta x }$ controls the distance between the basis elements. This transformation is a subtle stage of the calculation, because one has to consider step distances ${ \Delta R_{0} }$ and ${ \Delta k_{r} }$, which have to simultaneously be \textit{large} to comply with the WKB assumption from equation~\eqref{step_distances_sum_up} and \textit{small} to allow the use of the Riemann sum formula. As we are going to transform both the sums on $k_{r}^{p}$ and ${ R_{0}^{p} }$, the exact value of the susceptibility coefficients will depend on our choice for ${ \Delta R_{0} \, \Delta k_{r} }$. One has to consider the case
\begin{equation}
\Delta R_{0} \, \Delta k_{r} = 2 \pi \, .
\label{critical_sampling}
\end{equation}
This sampling corresponds to a critical sampling condition \citep{Gabor1946,Daubechies1990} \citep[See also][]{FouvryPichonPrunet2015}. Equation~\eqref{calculation_1/D_II} then takes the form
\begin{align}
& \frac{1}{\mathcal{D}_{\bm{m}_{1} , \bm{m}_{1}} (R_{1} , J_{r}^{1} , R_{1} , J_{r}^{2} , \omega) } = \frac{G}{2 \pi} \!\! \int \!\! \mathrm{d} k_{r} \, \mathrm{d} R_{0} \, \frac{1}{k_{r} R_{0}} \frac{1}{1 \!-\! \lambda_{k_{r}} (R_{0} , \omega)} \nonumber
\\
& \;\;\;\;\;\;\; \times \, \mathcal{J}_{m_{1}^{r}} \!\left[\!\! \sqrt{\tfrac{2 J_{r}^{1}}{\kappa_{1}}} k_{r} \!\right] \mathcal{J}_{m_{1}^{r}} \!\left[\!\! \sqrt{\tfrac{2 J_{r}^{2}}{\kappa_{1}}} k_{r} \!\right] \frac{1}{\sqrt{\pi \sigma^{2}}} \exp \!\left[\! - \frac{(R_{1} \!-\! R_{0})^{2}}{\sigma^{2}} \!\right] \, . \label{calculation_1/D_II_bis}
\end{align}
One can now assume that the radial Gaussian present in equation~\eqref{calculation_1/D_II_bis} is sufficiently peaked. Because it is correctly normalized, we may in this limit replace it by ${ \delta_{\rm D} (R_{1} \!-\! R_{0}) }$. The integration on ${ R_{0} }$ can then be immediately performed to give
\begin{align}
& \frac{1}{\mathcal{D}_{\bm{m}_{1} , \bm{m}_{1}} (R_{1} , J_{r}^{1} , R_{1} , J_{r}^{2} , \omega)} = \frac{1}{2 \pi} \frac{G}{R_{1}}  \nonumber
\\
& \;\;\;\;\; \times \, \int \!\! \mathrm{d} k_{r} \, \frac{1}{k_{r}} \frac{1}{1 \!-\! \lambda_{k_{r}} (R_{1} , \omega)} \mathcal{J}_{m_{1}^{r}} \!\left[\!\! \sqrt{\tfrac{2 J_{r}^{1}}{\kappa_{1}}} k_{r} \!\right]  \mathcal{J}_{m_{1}^{r}} \!\left[\!\! \sqrt{\tfrac{2 J_{r}^{2}}{\kappa_{1}}} k_{r} \!\right] \, ,  \label{calculation_1/D_III}
\end{align}
where only ${ \lambda_{k_{r}} }$ depends on the frequency of evaluation $\omega$. One may note that in equation~\eqref{calculation_1/D_III}, all the dependencies in $\sigma$ have disappeared, so that the value of the susceptibility coefficients is independent of the precise choice of the WKB basis.The square of the susceptibility coefficients which is required to estimate the drift and diffusion coefficients from equation~\eqref{initial_drift} and~\eqref{initial_diff} is therefore given by
\begin{align}
&\left| \frac{1}{\mathcal{D}_{\bm{m}_{1} , \bm{m}_{1}} (R_{1} , J_{r}^{1} , R_{1} , J_{r}^{2} , \omega)} \right|^{2} =  \frac{1}{4 \pi^{2}} \frac{G^{2}}{R_{1}^{2}}  \nonumber
\\
& \;\;\;\;\; \times \, \left\{ \!\! \int_{1\!/\!\sigma_{k}}^{+ \infty} \!\!\!\!\!\!\!\! \mathrm{d} k_{r} \, \frac{1}{k_{r}} \frac{1}{1 \!-\! \lambda_{k_{r}} (R_{1} , \omega)} \mathcal{J}_{m_{1}^{r}} \!\left[\!\! \sqrt{\tfrac{2 J_{r}^{1}}{\kappa_{1}}} k_{r} \!\right] \! \mathcal{J}_{m_{1}^{r}} \!\left[\!\! \sqrt{\tfrac{2 J_{r}^{2}}{\kappa_{1}}} k_{r} \!\right] \right\}^{2} , \label{calculation_1/D^2_I}
\end{align}
where we introduced a cut-off at ${ 1/\sigma_{k} }$ for the integration on $k_{r}$. This bound is justified by the WKB constraint from equation~\eqref{step_distances_sum_up}, which imposes that the probed radial frequency region is bounded from below. It is also important to note that these susceptibility coefficients should be evaluated at ${ R_{2} \!=\! R_{1} }$, since we proved in equation~\eqref{restriction_local_resonances} that, consistently with our WKB approximation, exactly local resonances are the only ones which have to be considered.

At this stage, there is an arbitration to make between two possible behaviors depending on the physical  properties of the underlying disc. First of all, if the amplification function ${ k_{r} \!\mapsto\! 1/(1 \!-\! \lambda_{k_{r}}) }$ is asymptotically  a sharp function reaching a maximum value $\lambda_{\rm max}$ for ${ k_{r} \!=\! k_{\rm max} }$, one can assume that the susceptibility coefficients are dominated by the contribution from the peak in ${\lambda_{k_{r}}}$. In this situation, we can perform an approximation of the \textit{small denominators}. The second possible behavior arises if the function ${ k_{r} \!\mapsto\! 1 / (1 \!-\! \lambda_{k_{r}}) }$ is asymptotically flat, so that there is no characteristic scale of \textit{blow-up} of the amplification eigenvalues. In such a situation, the susceptibility coefficients are mostly dominated by the behavior at the boundaries  of  integration from equation~\eqref{calculation_1/D^2_I} where ${ k_{r} \!\to\! 1/ \sigma_{k} }$. The detailed response structure of the self-gravitating disc then does not play a significant role.

We place ourselves within the approximation of the small denominators, assuming that the biggest contribution to the susceptibility coefficients comes from waves which yield the largest $\lambda_{k_{r}}$. Therefore, one has to suppose that the function ${ k_{r} \!\mapsto\! 1 /(1 \!-\! \lambda_{k_{r}}) }$ is a sharp function reaching a maximum value ${ \lambda_{\rm max} (R_{1} , \omega) }$, for ${ k_{r} \!=\! k_{\rm max} (R_{1} , \omega) }$, with a characteristic spread ${ \Delta k_{\lambda} (R_{1} , \omega)}$. The expression~\eqref{calculation_1/D^2_I} then becomes
\begin{align}
& \left| \frac{1}{\mathcal{D}_{\bm{m}_{1} , \bm{m}_{1}} (R_{1} , J_{r}^{1} , R_{1} , J_{r}^{2} , \omega) } \right|^{2} = \frac{1}{4 \pi^{2}} \frac{G^{2}}{R_{1}^{2}} \frac{(\Delta k_{\lambda})^{2}}{k_{\rm max}^{2}}  \nonumber
\\
& \;\;\;\;\;\;\;\;\;\; \times \, \left[ \frac{1}{1 \!-\! \lambda_{\rm max} } \right]^{2} \mathcal{J}_{m_{1}^{r}}^{2} \!\left[\!\! \sqrt{\tfrac{2 J_{r}^{1}}{\kappa_{1}}} k_{\rm max} \!\right] \mathcal{J}_{m_{1}^{r}}^{2} \!\left[\!\! \sqrt{\tfrac{2 J_{r}^{2}}{\kappa_{1}}} k_{\rm max} \!\right] \, .  \label{expression_1/D^2_ASD}
\end{align}
While still focusing on the contribution to the susceptibility coefficients due to the waves with the largest amplification ${ \lambda (k_{r}) }$, one can improve the approximation of the small denominators from equation~\eqref{expression_1/D^2_ASD}. Indeed, starting from equation~\eqref{calculation_1/D^2_I}, one can instead perform the $k_{r}-$integration for ${ k_{r} \!\in\! [k_{\rm inf}\,;\, k_{\rm sup}] }$, where these bounds are given by ${ \lambda (k_{\rm inf/sup}) \!=\! \lambda_{\rm max} / 2 }$. This approach is numerically more demanding but does not alter the principal conclusions drawn in this paper, while allowing a more precise determination of the secular flux structure. All the calculations presented in section~\ref{sec:application} were performed with this improved approximation.
Finally, in Appendix~\ref{sec:noselfgravity}, we detail how this same WKB formalism may be applied to the inhomogeneous Balescu-Lenard equation without collective effects \citep{Chavanis2013}.

\subsection{Estimation of the drift and diffusion coefficients}

The drift and diffusion coefficients are given by equations~\eqref{initial_drift} and~\eqref{initial_diff}. Within the WKB approximation, we have shown in equations~\eqref{equality_mphi_kphi} and~\eqref{restriction_local_resonances} that the susceptibility coefficients have to be evaluated only for ${ \bm{m}_{1} \!=\! \bm{m}_{2} }$, so that the sum on $\bm{m}_{2}$ in the expressions of the drift and diffusion coefficients may be dropped. As the resonances are exactly local, using the formula from equation~\eqref{rewriting_Dirac_delta} adds a prefactor of the form ${ 1/ |\partial (\bm{m}_{1} \!\cdot\! \bm{\Omega}) / \partial J_{\phi}| }$, so that the drift coefficients from equation~\eqref{initial_drift} become
\begin{align}
A_{\bm{m}_{1}} (\bm{J}_{1}) = & - \frac{4 \pi^{3}}{\left| \frac{\partial }{\partial J_{\phi}} [\bm{m}_{1} \!\cdot\! \bm{\Omega}_{1}] \right|_{J_{\phi}^{1}}}  \nonumber
\\
& \times \, \!\! \int \!\! \mathrm{d} J_{r}^{2} \, \frac{\bm{m}_{1} \!\cdot\! \partial F / \partial \bm{J} (J_{\phi}^{1} , J_{r}^{2}) }{|\mathcal{D}_{\bm{m}_{1} , \bm{m}_{1}} (J_{\phi}^{1} , J_{r}^{1} , J_{\phi}^{1} , J_{r}^{2} , \bm{m}_{1} \!\cdot\! \bm{\Omega}_{1}) |^{2}} \, . \label{WKB_drift}
\end{align}
Similarly, the diffusion coefficients are given by
\begin{align}
D_{\bm{m}_{1}} (\bm{J}_{1}) = & \frac{4 \pi^{3}}{\left| \frac{\partial }{\partial J_{\phi}} [\bm{m}_{1} \!\cdot\! \bm{\Omega}_{1}] \right|_{J_{\phi}^{1}}} \nonumber
\\
& \times \, \!\! \int \!\! \mathrm{d} J_{r}^{2} \, \frac{F (J_{\phi}^{1} , J_{r}^{2})}{|\mathcal{D}_{\bm{m}_{1} , \bm{m}_{1}} (J_{\phi}^{1} , J_{r}^{1} , J_{\phi}^{1} , J_{r}^{2} , \bm{m}_{1} \!\cdot\! \bm{\Omega}_{1}) |^{2}} \, . \label{WKB_diff}
\end{align}
In both equations~\eqref{WKB_drift} and~\eqref{WKB_diff}, the susceptibility coefficients are given by equation~\eqref{calculation_1/D^2_I} or, within the approximation of the small denominators, by equation~\eqref{expression_1/D^2_ASD}. 

Such simple expressions of the drift and diffusion coefficients along with the required expression of the susceptibility coefficients constitute a main result of this paper.
Note importantly  that this WKB formalism is self contained and is obtained without any \textit{ad hoc} assumptions or fittings.
Except for the explicit recovery of the amplification eigenvalues from equation~\eqref{diagonal_M_tepid}, the calculations presented previously are not limited to the Schwarzchild distribution function from equation~\eqref{definition_DF_Schwarzschild}. Indeed, the drift and diffusion coefficients from equations~\eqref{WKB_drift} and~\eqref{WKB_diff} are valid for any tepid disc, as long as the epicyclic angles-actions mapping from equation~\eqref{mapping_action} may be used.
In Appendix~\ref{sec:explicitappendix}, we show how the drift and diffusion coefficients from equations~\eqref{WKB_drift} and~\eqref{WKB_diff} can be explicitly computed, when one considers a Schwarzschild distribution function as in equation~\eqref{definition_DF_Schwarzschild} and when the susceptibility coefficients are estimated thanks to the approximation of the small denominators from equation~\eqref{expression_1/D^2_ASD}. Finally, in Appendix~\ref{sec:comp}, we compare this ${ 2D }$ WKB Balescu-Lenard equation with other similar kinetic equations.

\section{Application}
\label{sec:application}

Let us now illustrate how this WKB approximation of the inhomogeneous Balescu-Lenard equation can be implemented to recover some results obtained via well-crafted numerical simulations. Indeed, \cite{Sellwood2012} (hereafter S12) using careful numerical simulations studied the long-term evolution of an isolated stable and stationary Mestel disc \citep{Mestel1963} sampled by pointwise particles. When evolved for hundreds of dynamical times, such a disc would secularly diffuse in action-space through the spontaneous generation of transient spiral structures. Figure 7 of S12 shows for instance the late time  formation of resonant ridges along very specific resonant directions. Such features are possible signatures of secular evolution, which results in a long-term aperiodic evolution of a self-gravitating system, during which small resonant and cumulative effects can add up in a coherent way.
These small effects, which are then amplified through the self-gravity of the system originate from \textit{finite-$N$} effects. 
Indeed, the distribution function of the system is made of a finite number $N$ of pointwise particles. Even with a perfect numerical integrator, the system would necessarily undergo \textit{encounters} during which orbits \textit{feel} the discreteness of the joint DF
through its two point correlation. 
Note  that these interactions need not be local but assume that the potential fluctuations
are resonant so as to build up a secular evolution of the system.
This effect, which is still present even in the absence of any numerical noise, is the effect captured by the Balescu-Lenard equation (see figure~\ref{plot_BL_resonances}).

\subsection{The disc model}

\begin{figure}
\begin{center}
\epsfig{file=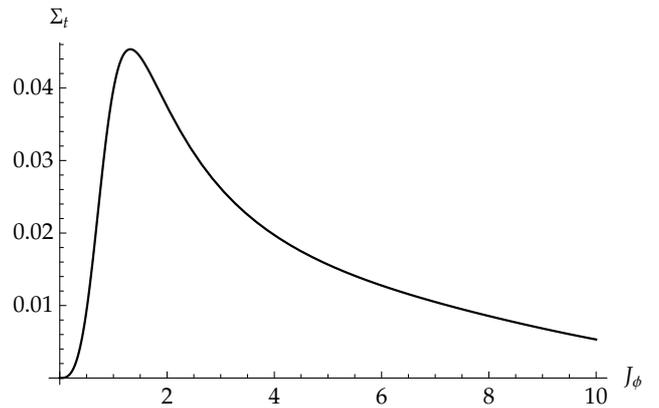,angle=-00,width=0.45\textwidth}
\caption{\small{Surface density $\Sigma_{\rm t}$ of the tapered Mestel disc. The unit system has been chosen so that $V_{0} \!=\! G \!=\! R_{i} \!=\! 1$.
Because of the tapering functions, the self-gravity of the disc is turned off in the inner and outer regions.
}}
\label{figSigmaDF}
\end{center}
\end{figure}
The disc considered by S12 is an infinitely thin Mestel disc, for which the circular speed ${ v_{\phi} }$ is a constant $V_{0}$ independent of the radius. 
Such a model represents fairly well the observed rotation curve of real galaxies.
The stationary background potential $\psi_{\rm M}$ of such a disc and its associated surface density $\Sigma_{\rm M}$ are given by
\begin{equation}
\psi_{\rm M} (R) = V_{0}^{2} \ln \!\left[\! \frac{R}{R_{i}} \!\right] \;\;\; ; \;\;\; \Sigma_{\rm M} (R) = \frac{V_{0}^{2}}{2 \pi G R} \, ,
\label{potential_Sigma_Mestel}
\end{equation}
where $R_{i}$ is a scale parameter. Because of this scale invariance, the relationship between the angular momentum $J_{\phi}$ and the guiding radius $R_{g}$ is straightforwardly given by
\begin{equation}
J_{\phi} = R_{g} \, V_{0} \, .
\label{link_Jphi_Rg_Mestel}
\end{equation}
Within the epicyclic approximation, the intrinsic frequencies $\Omega_{\phi}$ and $\kappa$ can be computed thanks to equations~\eqref{definition_Omega_kappa} and read
\begin{equation}
\Omega_{\phi} (J_{\phi}) = \frac{V_{0}^{2}}{J_{\phi}} \;\;\; ; \;\;\; \kappa (J_{\phi}) = \sqrt{2} \,\,  \Omega_{\phi} (J_{\phi}) \, .
\label{intrinsic_frequencies_Mestel}
\end{equation}
\begin{figure}[!htbp]
\begin{center}
\epsfig{file=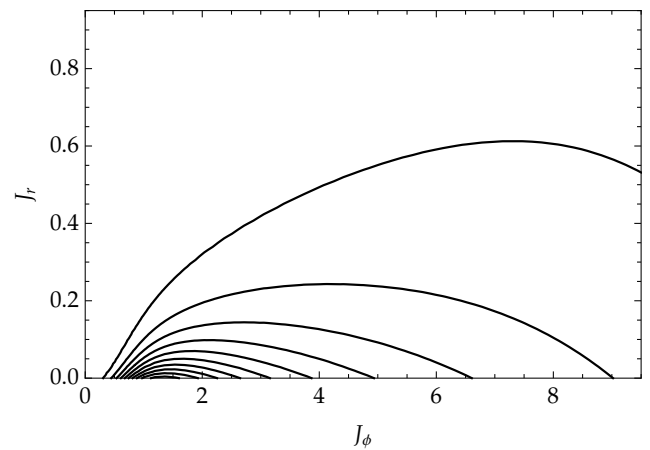,angle=-00,width=0.45\textwidth}
\caption{\small{Contours of the initial distribution function in action-space $(J_{\phi},J_{r})$, within the epicyclic approximation. The contours are spaced linearly between 95\% and 5\% of the distribution function maximum.
}}
\label{figcontourDF}
\end{center}
\end{figure}
We note that  ${ \kappa / \Omega_{\phi} \!=\! \sqrt{2} }$, so that the Mestel disc
could be seen as an intermediate case between the Keplerian case for which ${
\kappa / \Omega_{\phi} \!=\! 1 }$ and the harmonic case for which ${ \kappa /
\Omega_{\phi} \!=\! 2 }$. The ratio of the intrinsic frequencies is a important
parameter for the system since it will determine the location of the resonances
and a constant ratio may introduce dynamical degeneracies. This
is the case for the Keplerian and harmonic discs for which $ \kappa /
\Omega_{\phi}$ is a  rational number, as discussed below
equation~\eqref{condition_simple_poles}. By contrast, for the Mestel disc, the
non-rational ratio  $ \kappa / \Omega_{\phi} \!=\! \sqrt{2}$ ensures that the
potential is non-degenerate. Using the epicyclic approximation, the
DF considered by S12 takes, as in equation~\eqref{definition_DF_Schwarzschild},
the form of a Schwarzschild DF, where the intrinsic frequencies are given by
equation~\eqref{intrinsic_frequencies_Mestel}, the velocity dispersion
$\sigma_{r}$ is constant throughout the entire disc, and the surface density is
given by $\Sigma_{\rm t}$, \textit{i.e.} the \textit{active} surface density of
the disc. Indeed, in order to accommodate  the central singularity and the
infinite extent of the Mestel disc, one introduces tapering functions ${ T_{\rm
inner} }$ and ${ T_{\rm outer} }$ to damp out  the inner and outer regions,
which read
\begin{equation}
\begin{cases}
\displaystyle T_{\rm inner} (J_{\phi}) = \frac{J_{\phi}^{\nu}}{(R_{i} V_{0})^{\nu} \!+\! J_{\phi}^{\nu}} \, ,
\\
\displaystyle T_{\rm outer} (J_{\phi}) = \bigg[ 1 \!+\! \bigg[ \frac{J_{\phi}}{R_{0} V_{0}} \bigg]^{\mu} \bigg]^{-1} \, ,
\end{cases}
\label{definition_tapering_functions}
\end{equation}
where $\nu$ and $\mu$ control the sharpness of the two tapers and $R_{0}$ is an additional scale parameter.
\begin{figure}
\begin{center}
\epsfig{file=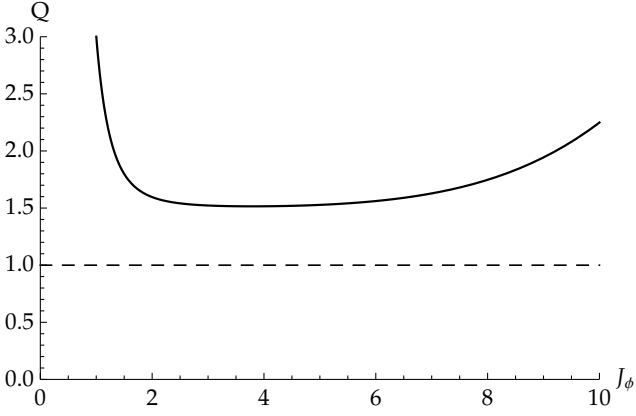,angle=-00,width=0.45\textwidth}
\caption{\small{Dependence of the local $Q$ Toomre parameter with the angular momentun. It is scale invariant except in the inner/outer regions because of the presence of the tapering functions $T_{\rm inner}$ and $T_{\rm outer}$. The unit system has been chosen so that $V_{0} \!=\! G \!=\! R_{i} \!=\! 1$.
}}
\label{figQDF}
\end{center}
\end{figure}
The two tapers are physically motivated by the presence of a bulge and an outer truncation for the disc.
Moreover, in order to reduce the susceptibility of the disc, we also suppose that only a fraction $\xi$ of the disc is self-gravitating, with ${ 0 \!\leq\! \xi \!\leq\! 1 }$, so that the rest of the gravitational field is provided by the static halo. As a conclusion, the active surface density $\Sigma_{\rm t}$ is given by
\begin{equation}
\Sigma_{\rm t} (J_{\phi}) = \xi \, \Sigma_{\rm M} (J_{\phi}) \, T_{\rm inner} (J_{\phi}) \, T_{\rm outer} (J_{\phi}) \, ,
\label{active_surface_density}
\end{equation}
where $\Sigma_{\rm M}$ is the surface density of the Mestel disc from equation~\eqref{potential_Sigma_Mestel}. We place ourselves in the same units system as in S12, so that we have ${ V_{0} \!=\! G \!=\! R_{i} \!=\! 1 }$. The other numerical factors are given by ${ \sigma_{r} \!=\! 0.284 }$, ${ \nu \!=\! 4 }$, ${ \mu \!=\! 5 }$, ${ \xi \!=\! 0.5 }$ and ${ R_{0} \!=\! 11.5 }$. The shape of the active surface density is illustrated in figure~\ref{figSigmaDF}.
The initial contours of the Schwarzschild DF from equation~\eqref{definition_DF_Schwarzschild} are shown in figure~\ref{figcontourDF}.
For such an almost scale invariant disc, the local Toomre parameter, $Q$~\citep{Toomre1964}
\begin{equation}
Q (J_{\phi}) = \frac{\sigma_{r} \, \kappa (J_{\phi})}{3.36 \, G \, \Sigma_{\rm t} (J_{\phi}) } \, ,
\label{definition_Q_Toomre}
\end{equation}
which for ${ Q \!>\! 1}$ ensures the stability of the disc with respect to local axisymmetric disturbances, becomes almost independent of the radius, especially in the intermediate regions of the disc.
As illustrated in figure~\ref{figQDF}, ${ Q \!\simeq\! 1.5 }$ between the tapers and increases strongly in the tapered regions.

The expression~\eqref{definition_Ftot} of the secular diffusion flux requires to sum on all the resonances $\bm{m}$.  S12 restricted pertubations forces to ${ m_{\phi} \!=\! 2 }$, so that we may impose this same restriction on the considered azimuthal number $m_{\phi}$. Throughout our numerical calculations, we will restrict ourselves to only three different resonances which are: the inner Lindblad resonance (ILR) corresponding to ${ (m_{r}^{\rm ILR} , m_{\phi}^{\rm ILR}) \!=\! (-1,2) }$, the outer Lindblad resonance (OLR) given by ${ (m_{r}^{\rm OLR} , m_{\phi}^{\rm OLR}) \!=\! (1,2) }$ and the corotation resonance (COR) for which ${ (m_{r}^{\rm COR} , m_{\phi}^{\rm COR}) \!=\! (0,2) }$. Moreover, all the calculations in the upcoming sections have also been performed while taking into account the contributions from the resonances with ${ m_{r} \!=\! \pm 2 }$, which were checked to be subdominant. Being able to perform such a restriction to the relevant resonances appearing in the secular flux $ \bm{\mathcal{F}}_{\rm tot} $ from equation~\eqref{definition_Ftot} is an important step of the calculation.

Returning to the fast and slow coordinates from equation~\eqref{fast_slow_actions}, note that the diffusion associated to the COR resonance amounts to diffusion along the ${ J_{\phi}-}$axis. Such diffusion brings stars from one quasi-circular orbit to another of a different radius and is called radial migration. Conversely, the diffusion associated to the ILR and OLR resonances exhibits a non-zero diffusion component in the ${J_{r}-}$direction. It therefore increases the velocity dispersion within the disc so as to \textit{heat} it, while either decreasing (ILR) or increasing (OLR) star's angular momentum.

\subsection{Disc amplification}

\begin{figure}
\begin{center}
\epsfig{file=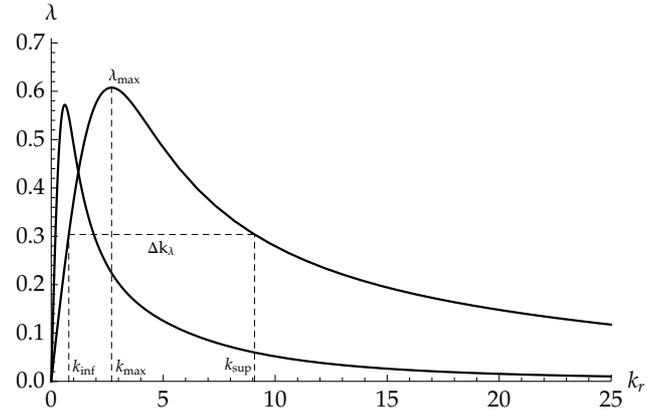,angle=-00,width=0.45\textwidth}
\caption{\small{Variations of the response matrix eigenvalues $\lambda$ with the WKB-frequency $k_{r}$, for $\bm{m} \!=\! \bm{m}_{\rm COR}$ and two values of $J_{\phi}$. The curve that peaks at large $k_{r}$ is for the smaller value of $J_{\phi}$.}}
\label{figLambdakr}
\end{center}
\end{figure}

\begin{figure}
\begin{center}
\epsfig{file=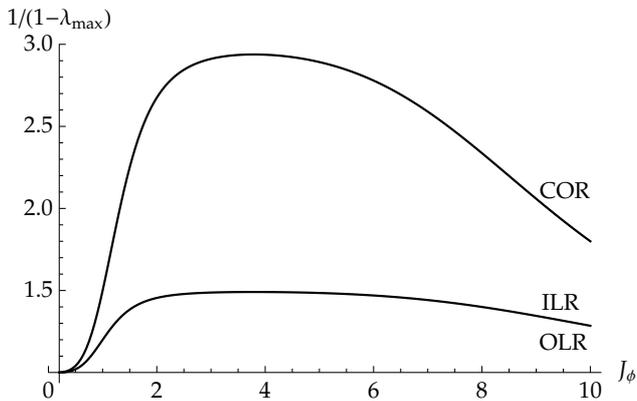,angle=-00,width=0.45\textwidth}
\caption{\small{Dependence of the amplification factor ${ {1}/{(1 \!-\! \lambda_{\rm max})} }$ with the position $J_{\phi}$ in the disc. The amplification associated to the COR is always larger than the one associated to the ILR and OLR.}}
\label{figLambdaJphi}
\end{center}
\end{figure}

One may now study the behavior of the amplification eigenvalues $\lambda_{k_{r}}$ given by equation~\eqref{diagonal_M_tepid}, thanks to which one can perform the improved approximation of the small denominators. For a given resonance $\bm{m}$ and angular momentum $J_{\phi}$, the amplification function ${ k_{r} \!\mapsto\! \lambda_{k_{r}} }$ is presented in figure~\ref{figLambdakr}.
As equation~\eqref{diagonal_M_tepid} only depend on ${ s^{2} }$, the ILR and OLR resonances always have the same response matrix eigenvalues. One can also note that the eigenvalues ${ \lambda (k_{r}) }$ are maximum for a frequency ${ k_{\rm max} (J_{\phi}) }$, where ${ \lambda (k_{r}) \!=\! \lambda_{\rm max} }$, in a region whose size is given by the width at half maximum ${ \Delta k_{\lambda} }$. Because of the scale-invariance property of the Mestel disc, it is straighforward to show that ${ \Delta k_{\lambda} \!\propto\! 1/ J_{\phi} }$, ${ k_{\rm max} \!\propto\! 1 / J_{\phi} }$ and ${ k_{\rm inf/sup} \!\propto 1 / J_{\phi} }$. One can then consider the behavior of the amplification factor ${ 1/ (1 \!-\! \lambda_{\rm max}) }$, which encodes the strength of the self-gravitating amplification, as shown in figure~\ref{figLambdaJphi}.
Note that the COR resonance is always more amplified than the ILR and OLR resonances, but the maximum amplification (${ \sim\! 3} $ for the COR and ${ \sim\! 1.5 }$ for the ILR and OLR) remains sufficiently small, so that the susceptibility coefficients from equation~\eqref{definition_1/D} are not dominated only by the self-gravitating amplification.

\begin{figure*}
\begin{center}
\epsfig{file=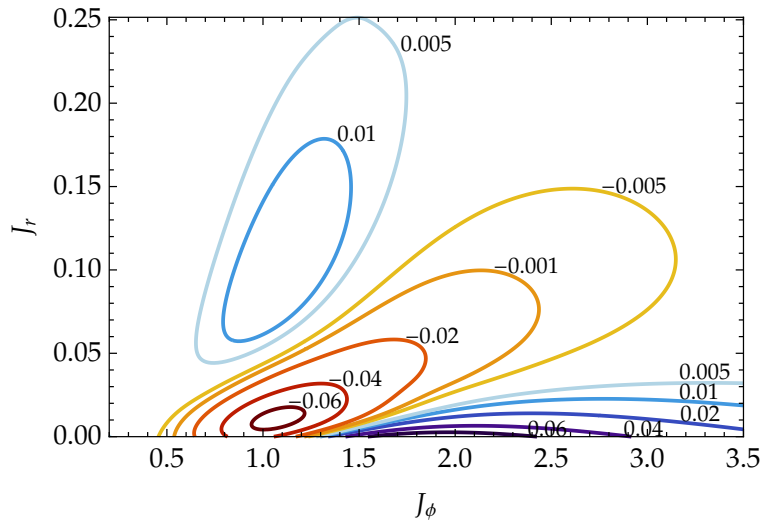,angle=-00,
width=1.1\columnwidth
}
\caption{\small{Map of the divergence of the total flux $\bm{\mathcal{F}}_{\rm tot} $ summed over the three resonances (ILR, COR and OLR). Red contours, for which ${ \text{div} \left( \bm{\mathcal{F}}_{\rm tot} \right) \!<\! 0 }$, correspond to regions from which the orbits will be depleted, whereas blue contours, for which ${ \text{div} \left( \bm{\mathcal{F}}_{\rm tot} \right) \!>\! 0}$, correspond to regions where secular diffusion will tend to increase the value of the DF.  The net fluxes  involve simultaneously  radial migration near  ${ (J_\phi,J_r) \!\sim\! (1.8,0) }$, and heating near  ${ (J_\phi,J_r) \!\sim\! (1,0.1) }$.}}
\label{figDiv_Flux_full}
\end{center}
\end{figure*}

\subsection{Computing the diffusion flux}
\label{sec:diffusionflux}

Given the knowledge of the  eigenvalues, it is now straightforward to compute the susceptibility coefficients within the improved approximation of the small denominators thanks to equation~\eqref{calculation_1/D^2_I}, where the integration on $k_{r}$ is performed for ${ k_{r} \!\in\! [k_{\rm inf} \, ; \, k_{\rm sup}] }$. One can then compute the associated drift and diffusion coefficients respectively given by equations~\eqref{WKB_drift} and~\eqref{WKB_diff}.    The diffusion flux ${ \bm{\mathcal{F}}_{\rm tot} }$ defined in equation~\eqref{definition_Ftot} immediately follows, where the sum on $\bm{m}$ is restricted only to the three resonances ILR, COR and OLR. 
In Appendices~\ref{sec:Schwarzschildplot} and~\ref{sec:frequencyselection}, we discuss two specific properties of such a truncated Mestel disc, namely the cancellation between the radial components of the diffusion and drift elements (the \textit{Schwarzshild conspiracy}, Appendix~\ref{sec:Schwarzschildplot}) and the natural and intrinsic presence of a temporal frequency bias (Appendix~\ref{sec:frequencyselection}), which both enlighten the subtle arbitrations between the different resonances. 

\begin{figure}
\begin{center}
\epsfig{file=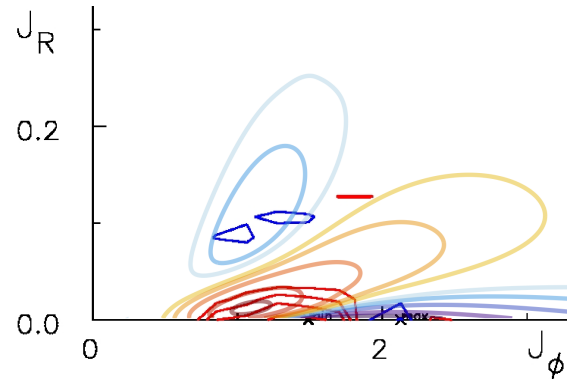,angle=-00,width=0.4\textwidth}
\caption{\small{Overlay of the WKB predictions for the divergence of the diffusion flux ${ \text{div} \left( \bm{\mathcal{F}}_{\rm tot} \right) }$ and the differences between the initial and the evolved DF in S12's simulation. The opaque contours correspond to the differences in the action-space for the DF in S12  between the time ${ t_{\rm S12} \!=\! 1000 }$ and ${ t_{\rm S12} \!= 0 }$ (see the upper panel of S12's figure 10). The red opaque contours correspond to negative differences, so that these regions are emptied from their orbits, whereas blue opaque contours correspond to positive differences, \textit{i.e} regions where the DF has increased through diffusion. The transparent contours correspond to the predicted values of ${ \text{div} \left( \bm{\mathcal{F}}_{\rm tot} \right) }$ using the same conventions as in figure~\ref{figDiv_Flux_full}.
Note the overlap of the predicted transparent red and blue contours with the measured solid ones. 
}}
\label{figDiv_Flux_S12_1000}
\end{center}
\end{figure}

\begin{figure}
\begin{center}
\epsfig{file=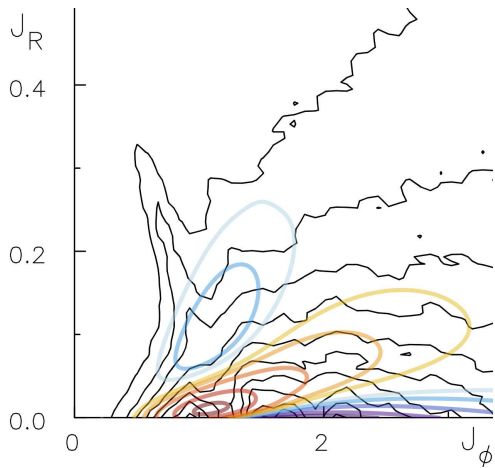,angle=-00,width=0.35\textwidth}
\caption{\small{Overlay of the WKB predictions for the divergence of the diffusion flux ${ \text{div} \left( \bm{\mathcal{F}}_{\rm tot} \right) }$ on top of the contours of the DF in action-space measured in the S12 simulation. The black background contours are the levels contours of the DF at time ${ t_{\rm S12} \!=\! 1400 }$ (see the lower panel of figure 7 of S12). These contours are spaced linearly between 95\% and 5\% of the DF maximum and exhibit clearly the appearance of a resonant ridge. The colored transparent contours correspond to the predicted values of ${ \text{div} \left( \bm{\mathcal{F}}_{\rm tot} \right) }$ using the same conventions as in figure~\ref{figDiv_Flux_full}. 
Note that the developed  late time ridge is consistent with the predicted depletion (red) and enrichment  (blue) of orbits.}}
\label{figDiv_Flux_S12_1400}
\end{center}
\end{figure}

Finally let  us compute the divergence of this flux, ${ \text{div} \left( \bm{\mathcal F}_{\rm tot} \right) }$ given by equation~\eqref{BL_divergence_flux} 
in order to compare quantitatively the WKB predictions with the results from S12's simulations.
Figure~\ref{figDiv_Flux_full} represents the contours of ${ \text{div} \left( \bm{\mathcal{F}}_{\rm tot} \right) }$. A comparison of our WKB predictions with the results from S12's simulation are illustrated in the figures~\ref{figDiv_Flux_S12_1000} and~\ref{figDiv_Flux_S12_1400}.
In figure~\ref{figDiv_Flux_full},  red contours correspond to regions for which ${ \text{div} \left( \bm{\mathcal{F}}_{\rm tot} \right) \!<\! 0 }$, so that thanks to equation~\eqref{BL_divergence_flux} they are associated to action-space regions where the WKB Balescu-Lenard equation predicts a decrement  of the DF during  secular evolution. In contrast, blue contours are associated to regions for which ${ \text{div} \left( \bm{\mathcal{F}}_{\rm tot} \right) \!>\! 0 }$, so that the DF will increase  there. The overall picture involves two competing processes: i) the beginning of a ridge forming towards ${ (J_\phi,J_r) \!\sim\! (1,0.1) }$, and  ii) the formation of an over density near  ${ (J_\phi,J_r) \!\sim\! (1.8,0) }$.  Point i) is in fact consistent with both
the early time measurement of S12 as shown in figure~\ref{figDiv_Flux_S12_1000}
and the late time measurement of S12 as shown in 
figure~\ref{figDiv_Flux_S12_1400}. These qualitative agreements are in fact surprisingly good,
given that the WKB theory is approximate and 
was only estimated for ${ t \!=\! 0^+ }$. 
Interestingly, the early time measurement from figure~\ref{figDiv_Flux_S12_1000} also displays a hint of an over-density on the ${ J_{r} \!=\! 0 }$ axis, in agreement with 
point ii), while the late time measurement suggests that the over density has split, with a hint of a second ridge forming. The time  evolution of equation~(\ref{initial_BL}) is likely to explain why this over density on the axis seems to split,
and why the ridge gets amplified.

From figure~\ref{figDiv_Flux_full}, we    explicitly compute ${ \text{div} \left( \bm{\mathcal{F}}_{\rm tot} \right) }$, so that we may now study the typical timescale of collisional relaxation predicted by the WKB Balescu-Lenard equation. This is the purpose of the next section.

\subsection{Physical timescales}
\label{sec:timescales}

Given estimates of the diffusion flux, one can explicitly compute the \textit{collisional timescale}, \textit{i.e.} the timescale for which the finite-$N$ effects become significant. Indeed, the larger the number $N$ of particles, the later these effects will come into play. One should note that our writing of the Balescu-Lenard equation~\eqref{initial_BL} is independent of the number $N$ of particles.  However, when correctly dimensionalized, this kinetic equation takes the form
\begin{equation}
\frac{\partial F}{\partial t} + L [F] = \frac{1}{N} \, C_{\rm BL} [F] \, ,
\label{BL_with_N}
\end{equation}
where $L$ is the operator of pure advection, and ${ C_{\rm BL} }$ is the Balescu-Lenard collisional operator, \textit{i.e.} the r.h.s of equation~\eqref{initial_BL}. Equation~\eqref{BL_with_N} underlines the fact that the collisional term is associated to a kinetic Taylor expansion in the parameter ${ \varepsilon \!=\! 1 / N \!\ll\! 1 }$. Within the angle-actions coordinates, the advection operator is immediately given by
\begin{equation}
L = \bm{\Omega} \!\cdot\! \frac{\partial }{\partial \bm{\theta}} \, .
\label{operator_L_angle_action}
\end{equation}
Because we have assumed that $F$ is always quasi-stationary, so that ${ F = F (\bm{J} , t) }$ (adiabatic approximation), one has ${ L [F] = 0 }$. We now introduce the time
\begin{equation}
\tau = \frac{t}{N} \, ,
\label{definition_mu}
\end{equation}
so that equation~\eqref{BL_with_N} immediately becomes
\begin{equation}
\frac{\partial F}{\partial \tau} = C_{\rm BL} [F] \, .
\label{BL_with_mu}
\end{equation}
Equation~\eqref{BL_with_mu} corresponds to a rewriting of the Balescu-Lenard equation, where $N$ is not present anymore. This will allow us to quantitatively compare the time during which the S12 simulation was run to the diffusion time predicted by our WKB Balescu-Lenard  formalism. In order to ease this comparison, we place ourselves in the same units system as the one used by S12. Figure 7 of S12 for which the ridge was observed was obtained with the parameters ${ N \!=\! 50 \!\times\! 10^{6} }$ and ${ \Delta t_{\rm S12} \!=\! 1400 }$. Using the rescaled time introduced in equation~\eqref{definition_mu}, one obtains that S12 observed the resonant ridge after a time ${ \Delta \tau_{\rm S12} \!=\! \Delta t_{\rm S12} / N \!\simeq\! 3 \!\times\! 10^{-5} }$. One can then compare this time, with the typical time required to obtain a resonant ridge within our WKB formalism. Given the map of ${ \text{div} \left( \bm{\mathcal{F}}_{\rm tot} \right) }$ described in section~\ref{sec:diffusionflux}, one can estimate the typical time for which this flux could lead to the features observed in S12. The contours presented in the figure 7 of S12 are separated by a value of ${ 0.1 \!\times\! F_{0}^{\rm max} }$, where $F_{0}^{\rm max} \!\simeq\! 0.12 $ corresponds to the maximum of the normalized DF from equation~\eqref{definition_DF_Schwarzschild}. As a consequence, to observe the resonant ridge, the DF should typically change by a value of the order of ${ \Delta F_{0} \!\simeq\! 0.1 \!\times\! F_{0}^{\rm max} }$. From figure~\ref{figDiv_Flux_full}, one can note that the maximum value of the divergence of the flux is given by ${ | \text{div} \left( \bm{\mathcal{F}}_{\rm tot} \right) | \!\simeq\! 0.06 }$. Finally, thanks to equation~\eqref{BL_divergence_flux}, one may write the relation ${ \Delta F_{0} \!\simeq\! \Delta \tau_{\rm WKB} \, |\text{div} \left( \bm{\mathcal{F}}_{\rm tot} \right) |}$, where ${ \Delta \tau_{\rm WKB} }$ is the \textit{minimal time} during which the WKB Balescu-Lenard equation has to be considered in order to develop a ridge. Thanks to the previous typical numerical values, one obtains that ${ \Delta \tau_{\rm WKB} \!\simeq\! 3 \!\times\! 10^{-1} }$. When comparing these two typical times, ${ \Delta \tau_{\rm S12} }$ the duration during which S12 simulation was performed and ${ \Delta \tau_{\rm WKB} }$ the duration required to observe secular diffusion in the WKB Balescu-Lenard formalism, one obtains the order of magnitude
\begin{equation}
\frac{\Delta \tau_{\rm S12}}{\Delta \tau_{\rm WKB}} \simeq 10^{-4} \, .
\label{comparison_Delta_mu}
\end{equation}
Hence the direct application of the WKB-limited Balescu-Lenard equation does not allow us to predict the observed timescale for the diffusion features in simulations. Indeed, the timescale of collisional diffusion predicted by this WKB formalism seems  much larger than the time during which the numerical simulation was performed. This discrepancy is also strengthened by the use of a \textit{softening length} in numerical simulations, which induces an effective thickening of the disc, so as to slow down the collisional relaxation. A possible explanation  for this timescale discrepancy is discussed in the next section.

\subsection{Interpretation}
\label{sec:understandingS12simulation}

In order to  interpret S12 simulation under the light of a collisional secular diffusion equation, such as the Balescu-Lenard equation and its WKB limit, one should first note the undisputed  presence of collisional effects  in S12's simulation. Indeed, figure 2 of S12 shows that when the number of particles of the simulation is increased, the strength of the density fluctuations are delayed, which in turn is likely to be related to the amplitude of the secular features. 
The larger the number of particles, the later the effect of  secular diffusion. Such dependence illustrates the fact that discreteness effects do play a role in the secular diffusion observed in S12. 

\cite{SellwoodKahn} have argued that a sequence of causally connected swing amplified transients could occur subject to  a (possibly non local) resonant condition between 
successive spirals waves.  
 The Balescu-Lenard formalism captures precisely such sequences -- in as much as it 
integrates over dressed {\sl correlated} potential fluctuations subject to relative resonant conditions, but does {\sl not} preserve causality nor resolve them on dynamical timescales.  The exact initial phases are not relevant in 
the Balescu-Lenard formalism: see Appendix~\ref{sec:derivation} for a sketch of a full derivation which makes this point clear.

The timescale discrepancy observed in equation~\eqref{comparison_Delta_mu} might be driven by the  \textit{incompleteness} of the WKB basis. Indeed, equation~\eqref{definition_WKB_basis}'s   basis -- thanks to which the susceptibility coefficients were evaluated in equation~\eqref{calculation_1/D^2_I} -- does not form a complete set, as it can only represent correctly tightly wound spirals.
It also enforces local resonances, and does not allow for remote orbits to resonate, or wave packets to propagate between such non local resonances. 
The seminal works from~\cite{GoldreichLyndenBell1965a,JulianToomre1966,Toomre1981} showed that any leading spiral wave during its unwinding to a trailing wave undergoes a significant amplification, coined swing amplification. Because it involves open spirals this mechanism is not captured by the WKB formalism. 
 This additional amplification is expected to increase the susceptibility of the disc and therefore accelerate  secular diffusion (both drift and diffusion), so that the timescales discrepancy from equation~\eqref{comparison_Delta_mu} will become less restrictive. Following the notations from~\cite{Toomre1981}, the truncated Mestel disc considered in the S12 simulation corresponds to ${ Q \!\simeq\! 1.5 }$ and ${ X \!=\! 2 }$, so that figure 7 from~\cite{Toomre1981} shows that significant swing amplification (of order $\!\sim\!\! 10$) may be expected.
It has also been claimed \citep{ToomreKalnajs1991} that swing amplified shot noise in the shearing sheet 
approximation would behave like significantly heavier macro-particles.
 Such an amplification would  keep a dependence of the secular response with the total number $N$ of particles, but would reduce significantly the {\sl effective} number of particles.
 
The Balescu-Lenard WKB limit seems to capture qualitatively the main features of the initial diffusion process in action space (as discussed in Section~\ref{sec:diffusionflux}), but falls short in predicting the relevant timescale.
The remaining questions are therefore: what is the exact impact of  swing amplification? Can it 
explain the timescale discrepancy?

\section{Conclusion}
\label{sec:conclusion}

We implemented the inhomogeneous Balescu-Lenard equation~\eqref{initial_BL} for an infinitely thin galactic disc using two main approximations. We first assumed the disc to be tepid. We could then use the epicyclic approximation which allowed for an explicit  mapping between the physical coordinates ${ (\bm{x} , \bm{v}) }$ and the angle-actions coordinates ${ (\bm{\theta} , \bm{J}) }$ via equation~\eqref{mapping_action}. Our second approximation relied on the introduction of the tightly wound basis elements from equation~\eqref{definition_WKB_basis}. Because of the
corresponding WKB approximation, we obtained in equation~\eqref{diagonal_M_tepid} a diagonal response matrix, so that gravity is effectively treated \textit{locally}. The associated scale-decoupling hypothesis yields a crucial restriction to only local resonances, as shown in equation~\eqref{restriction_local_resonances}. We then derived in equation~\eqref{calculation_1/D^2_I} a simple quadrature for the susceptibility coefficients, given by equations~\eqref{1/D_nocoll} and~\eqref{expression_Am1_m2_WKB} for the bare ones. Thanks to this restriction to local resonances, we were also  able to write the drift and diffusion coefficients as simple quadratures in equations~\eqref{WKB_drift} and~\eqref{WKB_diff}.

These simple expressions derived within the WKB formalism yield, to our knowledge, {\sl a first non trivial explicit expression}
 for the  Balescu-Lenard diffusion and drift coefficients.
 They are certainly useful to provide insight into the physical processes at work during the secular diffusion of a self-gravitating discrete disc. 
 Moreover, modulo the restriction to the three physically motivated resonances ILR, COR and OLR, our WKB formalism 
 can be used for quantitative comparisons to numerical experiments such as the one
 presented in~\cite{Sellwood2012}. 
 It considered a stable  { \sl isolated }  Mestel disc sampled by pointwise particles,
 whose secular evolution is induced by finite-${N}$ effect ideally  captured by the Balescu-Lenard equation. 

The straightforward calculation in the WKB limit of the divergence of the full diffusion flux, ${ \text{div} \left( \bm{\mathcal{F}}_{\rm tot} \right) }$, 
(illustrated in figures~\ref{figDiv_Flux_full},~\ref{figDiv_Flux_S12_1000} and~\ref{figDiv_Flux_S12_1400}), 
recovered most of the secular features observed in S12.
This qualitative agreement is impressive, given the level of approximation involved in the WKB limit.
The hints for the formation of a ridge -- depletion and enrichment of orbits along a preferred direction -- is qualitatively consistent with the findings of S12 and~\cite{FouvryPichon2015,FouvryBinneyPichon2015},
without postulating additional assumptions about the source of fluctuations\footnote{In contrast, the formalism of secular forcing presented in~\cite{FouvryPichon2015,FouvryBinneyPichon2015} postulated a partially ad-hoc shape of the perturbation power spectrum, see equation~\eqref{assumption_noise_secular_forcing}.}.
 
The  comparison of the collisional time predicted in the WKB limit (equation \eqref{comparison_Delta_mu}) to the diffusion time  of S12 simulation, highlights nonetheless   a significant quantitative overestimation. 
 We provided a possible explanation which relies on the intrinsic limitations of the WKB formalism, as it cannot account for  swing amplification, during which unwinding transient spirals are strongly amplified. This additional amplification, which involves explicitly non local wave absorption and emission,
 may be the missing contribution required to reconcile quantitavely our predictions and the simulation. One venue will be to compute numerically
 exactly equations~\eqref{initial_drift} and~\eqref{initial_diff} in action space -- without assuming tightly wound spirals or epicyclic orbits -- with a complete basis
 \footnote{An alternative middle ground would be to account for non local interferences of the WKB wave packets, given by equation~\eqref{definition_WKB_basis},
 which in turn would allow for non local resonances to come into play.}.
 This is the topic of an upcoming numerical investigation \citep{Fouvry2015}.
 
Should this complementary investigation explain the timescale mismatch, one would be in a stronger position to validate the accuracy of ${N-}$body schemes to correctly capture secular evolution of discrete self-gravitating cold discs over very long timescales. 
This would clearly be a worthy assessment of such schemes relying on the Balescu-Lenard theory.
Once the above described conundrum is resolved, 
we also will be able to {\sl evolve} over secular times the Balescu-Lenard equation  and 
predict the full cosmic time evolution of such discrete discs. This may also contribute to solving the timescale discrepancy.

In closing, beyond the application presented in section~\ref{sec:application},
the  above developed tightly wound  Balescu-Lenard formalism may
 for instance describe the secular diffusion of giant molecular clouds in galactic discs
(which in turn could play a role in migration driven metallicity gradients and disc thickening),
 the secular migration of planetesimals in partially self-gravitating proto-planetary discs, 
 or even the long-term evolution of  population of stars, gas blobs  and debris near the Galactic center.
 Such topics will be subject to further investigations.

\begin{acknowledgements}
JBF thank the Institute of Astronomy, Cambridge, for hospitality
while this investigation was completed.
JBF and CP also thank the theoretical physics sub-department, Oxford, for hospitality and the CNRS-Oxford 
exchange program for funding. We thank  Donald Lynden-Bell,  James Binney,
Simon Prunet, Walter Dehnen, John Magorrian and Mir Abbas Jalali for their feedback.
This work is partially supported by the Spin(e) grants ANR-13-BS05-0005 of the French {\sl Agence Nationale de la Recherche}
(\url{http://cosmicorigin.org})
and by the  LABEX Institut Lagrange de Paris (under reference ANR-10-LABX-63) which  is funded by  ANR-11-IDEX-0004-02.
\end{acknowledgements}


\bibliographystyle{aa}
\bibliography{references}

\appendix

\section{Sketch of Balescu-Lenard Derivation}
\label{sec:derivation}

Two derivations of the inhomogeneous Balescu-Lenard equation have been presented in the literature. The first one~\citep{Heyvaerts2010} is based on the appropriate truncation at the order ${1/N}$ of the BBGKY hierarchy. The second~\citep{Chavanis2012} relies on the Klimontovich equation, using a quasilinear approximation. 
We now briefly sketch the derivation presented in~\cite{Chavanis2012}. We consider an isolated system of $N$ particles in interaction, of mass ${ m \!=\! 1 }$, in a physical space of dimension $d$. Their dynamics is entirely described by Hamilton's equations
\begin{equation}
\frac{\mathrm{d} \bm{x}_{i}}{\mathrm{d} t} = \frac{\partial H}{\partial \bm{v}_{i}} \;\;\; ; \;\;\; \frac{\mathrm{d} \bm{v}_{i}}{\mathrm{d} t} = - \frac{\partial H}{\partial \bm{x}_{i}} \, ,
\label{Hamiltons_equations_derivation}
\end{equation}
where the Hamiltonian of the system is given by
\begin{equation}
H = \sum_{i = 1}^{N} \frac{1}{2} \bm{v}_{i}^{2} + \sum_{i < j} u (|\bm{x}_{i} \!-\! \bm{x}_{j}|) \, .
\label{Hamiltonian_derivation}
\end{equation}
Here ${ u (|\bm{x}_{i} \!-\! \bm{x}_{j}|) }$ is the binary potential of interaction. In the  gravitational case, it satisfies ${ u(\bm{x}) \!=\! - G / |\bm{x}| }$. One can now introduce the discrete distribution function ${ F_{\rm d} (\bm{x} , \bm{v} , t) }$ defined as
\begin{equation}
F_{\rm d} (\bm{x} , \bm{v} , t) = \sum_{i = 1}^{N} \delta_{\rm D} (\bm{x} \!-\! \bm{x}_{i} (t)) \, \delta_{\rm D} (\bm{v} \!-\! \bm{v}_{i} (t)) \, ,
\label{discrete_DF_derivation}
\end{equation}
along with the corresponding potential
\begin{equation}
\psi_{\rm d} (\bm{x}, t) = \!\! \int \!\! \mathrm{d} \bm{x}' \mathrm{d} \bm{v}' \, u (|\bm{x} \!-\! \bm{x}'|) \, F_{\rm d} (\bm{x} ' , \bm{v} ' , t) \, .
\label{discrete_potential_derivation}
\end{equation}
One can show that $F_{\rm d}$ satisfies the Klimontovich equation~\citep{Klimontovich1967}
\begin{equation}
\frac{\partial F_{\rm d}}{\partial t} \!+\! \frac{\partial H_{\rm d}}{\partial \bm{v}} \!\cdot\! \frac{\partial F_{\rm d}}{\partial \bm{x}} \!-\! \frac{\partial H_{d}}{\partial \bm{x}} \!\cdot\! \frac{\partial F_{\rm d}}{\partial \bm{v}} = 0 \, ,
\label{Klimontovich_equation_derivation}
\end{equation}
where we have defined the Hamiltonian $H_{\rm d}$ as
\begin{equation}
H_{\rm d} = \frac{1}{2} \, \bm{v}^{2} \!+\! \psi_{\rm d} (\bm{x} , t) \, .
\label{definition_Hamiltonian_discrete_derivation}
\end{equation}
At this stage, it is important to note that the Klimontovich equation~\eqref{Klimontovich_equation_derivation} contains exactly the same information as the Hamilton equation~\eqref{Hamiltons_equations_derivation}. We now introduce the smooth distribution function ${ F (\bm{x} , \bm{v} , t) \!=\! \langle F_{\rm d} (\bm{x} , \bm{v} , t) \rangle }$, corresponding to an average of $F_{\rm d}$ over a large number of initial conditions. One can then write ${ F_{\rm d} \!=\! F \!+\! \delta F}$, where ${ \delta F }$ denotes fluctuations around the smooth distribution. In a similar way, we introduce ${ \psi (\bm{x} , \bm{v} , t) \!=\! \langle \psi_{\rm d} (\bm{x} , \bm{v} , t) \rangle }$, so that ${ \psi_{\rm d} \!=\! \psi \!+\! \delta \psi }$. We have therefore decomposed the discrete distribution function $F_{\rm d}$ into a smooth component $F$ that evolves slowly with time, whereas the fluctuating component ${ \delta F }$ evolves more rapidly. As a consequence, when considering the evolution of the fluctuations, one can assume the smooth distribution to be \textit{frozen}. Using this timescale-decoupling approach, one can use the angle-actions coordinates ${ (\bm{\theta}_{1} , \bm{J}_{1}) }$ associated with the quasi-stationary smooth potential $\psi$ to describe the fast evolution of the fluctuations. Using these decompositions and this change of coordinates, equation~\eqref{Klimontovich_equation_derivation} takes the form of two evolution equations
\begin{equation}
\frac{\partial \delta F}{\partial t} \!+\! \bm{\Omega}_{1} \!\cdot\! \frac{\partial \delta F}{\partial \bm{\theta}_{1}} \!-\! \frac{\partial \delta \psi}{\partial \bm{\theta}_{1}} \!\cdot\! \frac{\partial F}{\partial \bm{J}_{1}} = 0 \, , 
\label{decoupled_evolution_equation_fast}
\end{equation}
and
\begin{equation}
\frac{\partial F}{\partial t} = \frac{\partial }{\partial \bm{J}_{1}} \!\cdot\! \left\langle \delta F \, \frac{\partial \delta \psi}{\partial \bm{\theta}_{1}} \right\rangle \, ,
\label{decoupled_evolution_equation_slow}
\end{equation}
where we have introduced the intrinsic frequencies of the system ${ \bm{\Omega}_{1} \!=\! \bm{\Omega} (\bm{J}_{1}) }$, as in equation~\eqref{definition_Omega}, and where ${ \langle \, . \, \rangle }$ denotes an angle average. Because of our timescale-decoupling approach, we  may neglect the time variation of ${ F (\bm{J}_{1} , t) }$ in the calculation of the collision term (adiabatic approximation). It implies that $F$ may be treated as a constant in equation~\eqref{decoupled_evolution_equation_fast}, because $F$ evolves on a (relaxation) timescale much larger than the (dynamical) time corresponding to the evolution of ${ \delta F }$. In order to be valid, this approximation requires to have ${ N \!\gg\! 1 }$. Finally, we also assume that the distribution $F$ remains Vlasov stable, so that its evolution is only governed by correlations and not by dynamical instabilities.

The first step of the derivation of the Balescu-Lenard equation is then to study the short timescale evolution equation~\eqref{decoupled_evolution_equation_fast}. One defines the Fourier-Laplace transform of the fluctuation ${ \delta F }$ as
\begin{equation}
\delta \tilde{F}_{\bm{m}_{1}} (\bm{J}_{1} , \omega_{1}) = \!\! \int \!\! \frac{\mathrm{d} \bm{\theta}_{1}}{(2 \pi)^{d}} \!\! \int_{0}^{+ \infty} \!\!\!\!\!\! \mathrm{d} t \, e^{- i (\bm{m}_{1} \cdot \bm{\theta}_{1} - \omega_{1} t)} \, \delta F (\bm{\theta}_{1} , \bm{J}_{1} , t) \, ,
\label{definition_Fourier_Laplace_derivation}
\end{equation}
valid for ${ \text{Im} (\omega_{1}) }$ sufficiently large. Similarly to equation~\eqref{definition_Fourier_angles}, one also defines the spatial Fourier tranform of the initial value as
\begin{equation}
\delta \hat{F}_{\bm{m}_{1}} (\bm{J}_{1} , 0) = \!\! \int \!\! \frac{\mathrm{d} \bm{\theta}_{1}}{(2 \pi)^{d}} \, e^{ - i \bm{m}_{1} \cdot \bm{\theta}_{1}} \, \delta F (\bm{\theta}_{1} , \bm{J}_{1} , 0) \, .
\label{definition_Fourier_initial_derivation}
\end{equation}
Thanks to these transformations, equation~\eqref{decoupled_evolution_equation_fast} may be rewritten under the form
\begin{equation}
\delta \tilde{F}_{\bm{m}_{1}} ( \bm{J}_{1} , \omega_{1}) = \frac{\bm{m}_{1} \!\cdot\! \partial F / \partial \bm{J}_{1}}{\bm{m}_{1} \!\cdot\! \bm{\Omega}_{1} \!-\! \omega_{1}} \delta \tilde{\psi}_{\bm{m}} (\bm{J} , \omega_{1}) \!+\! \frac{\delta \hat{F}_{\bm{m}_{1}} (\bm{J}_{1} , 0)}{i (\bm{m}_{1} \!\cdot\! \bm{\Omega}_{1} \!-\! \omega_{1})} \, .
\label{calculation_evolution_fast_I_derivation}
\end{equation}
We now use the basis elements introduced in equation~\eqref{definition_basis}, so that we may decompose the potential fluctuations under the form
\begin{equation}
\delta \psi (\bm{\theta}_{1} , \bm{J}_{1} , t) = \sum_{p} a_{p} (t) \, \psi^{(p)} (\bm{\theta}_{1} , \bm{J}_{1}) \, .
\label{decomposition_potential_fluctuations_derivations}
\end{equation}
We introduce the Laplace transform of ${ a_{p} (t) }$ as
\begin{equation}
\tilde{a}_{p} (\omega_{1}) = \!\! \int_{0}^{+ \infty} \!\!\!\!\!\! \mathrm{d} t \, a_{p} (t) \, e^{i \omega_{1} t} \, ,
\label{definition_Laplace_transform_derivation}
\end{equation}
Let us then  take the inverse Fourier transform of equation~\eqref{calculation_evolution_fast_I_derivation}, multiply by ${ \psi^{(q)}_{\bm{m}_{2}} (\bm{\theta}_{1} , \bm{J}_{1}) }$ and integrate over $\bm{\theta}_{1}$ and $\bm{J}_{1}$ (using the property that ${ \mathrm{d} \bm{x} \, \mathrm{d} \bm{v} = \mathrm{d} \bm{\theta}_{1} \, \mathrm{d} \bm{J}_{1} }$). One gets
\begin{equation}
\tilde{a}_{p} (\omega_{1}) \!=\! - (2 \pi)^{d} \!\! \sum_{q} \! [ \bm{I} \!-\! \widehat{\bm{M}} (\omega_{1}) ]_{pq}^{-1} \! \sum_{\bm{m}_{2}} \!\! \int \!\! \mathrm{d} \bm{J}_{2} \frac{\delta \hat{F}_{\bm{m}_{2}} (\bm{J}_{2} , 0)}{i (\bm{m}_{2} \!\cdot\! \bm{\Omega}_{2} \!-\! \omega_{1})} \, \psi^{(q) *}_{\bm{m}_{2}} (\bm{J}_{2}) \, ,
\label{calculation_evolution_fast_II_derivation}
\end{equation}
where the response matrix $\widehat{\bm{M}}$ is given by equation~\eqref{Fourier_M}. Equation~\eqref{calculation_evolution_fast_II_derivation} can be rewritten under the form
\begin{equation}
\delta \tilde{\psi}_{\bm{m}_{1}} (\bm{J}_{1} , \omega_{1}) \!=\! - (2 \pi)^{d} \!\! \sum_{\bm{m}_{2}} \!\! \int \!\! \mathrm{d} \bm{J}_{2}  \frac{1}{\mathcal{D}_{\bm{m}_{1} , \bm{m}_{2}} (\bm{J}_{1} , \bm{J}_{2} , \omega_{1})} \frac{\delta \hat{F}_{\bm{m}_{2}} (\bm{J}_{2} , 0)}{i (\bm{m}_{2} \!\cdot\! \bm{\Omega}_{2} \!-\! \omega_{1})} \, ,
\label{calculation_evolution_fast_III_derivation}
\end{equation}
where the susceptibility coefficients have been introduced in equation~\eqref{definition_1/D}.  One can now compute the collision term appearing in the r.h.s of equation~\eqref{decoupled_evolution_equation_slow}. It requires us to evaluate
\begin{align}
\left\langle \delta F \, \frac{\partial \delta \psi}{\partial \bm{\theta}_{1}} \right\rangle = \!\! \sum_{\bm{m}_{1} , \bm{m}_{2}} \!\!  \int  & \frac{\mathrm{d} \omega_{1}}{2 \pi} \frac{\mathrm{d \omega_{2} }}{2 \pi} \, i \bm{m}_{2} \, e^{i (\bm{m}_{1} \cdot \bm{\theta}_{1} - \omega_{1} t)} e^{i (\bm{m}_{2} \cdot \bm{\theta}_{1} - \omega_{2} t)} \nonumber
\\
& \times \langle \delta \tilde{F}_{\bm{m}_{1}} (\bm{J}_{1} , \omega_{1}) \,  \delta \tilde{\psi}_{ \bm{m}_{2}} (\bm{J}_{1} , \omega_{2} ) \rangle \, .
\label{calculation_evolution_slow_I_derivation}
\end{align}
Using equation~\eqref{calculation_evolution_fast_I_derivation}, one immediately obtains that
\begin{align}
\langle \delta \tilde{F}_{\bm{m}_{1}} (\bm{J}_{1} , \omega_{1}) \, & \delta \tilde{\psi}_{\bm{m}_{2}} (\bm{J}_{1} , \omega_{2}) \rangle  = \nonumber
\\
&  \frac{\bm{m}_{1} \!\cdot\! \partial F / \partial \bm{J}_{1}}{\bm{m}_{1} \!\cdot\! \bm{\Omega}_{1} \!-\! \omega_{1}} \, \langle \delta \tilde{\psi}_{\bm{m}_{1}} (\bm{J}_{1} , \omega_{1}) \, \delta \tilde{\psi}_{\bm{m}_{2}} (\bm{J}_{1} , \omega_{2}) \rangle  \nonumber
 \\
& + \frac{\langle \delta \hat{F}_{\bm{m}_{1}} (\bm{J}_{1} , 0) \, \delta \tilde{\psi}_{\bm{m}_{2}} (\bm{J}_{1} , \omega_{2}) \rangle}{i (\bm{m}_{1} \!\cdot\! \bm{\Omega}_{1} \!-\! \omega_{1})} \, .
\label{calculation_evolution_slow_II_derivation}
\end{align}
In equation~\eqref{calculation_evolution_slow_II_derivation}, the first term corresponds to the self-correlation of the potential whereas the second term corresponds to the correlations between the potential fluctuations and the distribution function at time ${ t \!=\! 0 }$. Each of these terms must then be considered one at a time.
Assuming that there is {\sl  no correlation  in the initial phases,} one can show~\citep[see Appendix C from][]{Chavanis2012} that
\begin{equation}
\langle \delta \hat{F}_{\bm{m}_{1}} (\bm{J}_{1} , 0) \, \delta \hat{F}_{\bm{m}_{2}} (\bm{J}_{2} , 0) \rangle \!=\! \frac{1}{(2 \pi)^{d}} \delta_{\bm{m}_{1}}^{- \bm{m}_{2}} \, \delta_{\rm D} (\bm{J}_{1} \!-\! \bm{J}_{2}) \, F (\bm{J}_{1}) \, .
\label{initial_correlation_derivation}
\end{equation}
Hence, given equation~\eqref{calculation_evolution_fast_III_derivation}, one can rewrite the first term of equation~\eqref{calculation_evolution_slow_II_derivation} under the form
\begin{align}
\langle \delta & \tilde{\psi}_{\bm{m}_{1}} (\bm{J}_{1} , \omega_{1}) \, \delta \tilde{\psi}_{\bm{m}_{2}} (\bm{J}_{1} , \omega_{2}) \rangle  = (2 \pi)^{d} \!\! \sum_{\bm{m}_{3}} \!\! \int \!\! \mathrm{d} \bm{J}_{3} \, \frac{1}{\mathcal{D}_{\bm{m}_{1} , \bm{m}_{3}} (\bm{J}_{1} , \bm{J}_{3} , \omega_{1})}  \nonumber
\\
& \;\;\;\;\;\;\;\;  \frac{1}{\mathcal{D}_{\bm{m}_{2} , - \bm{m}_{3}} (\bm{J}_{1} , \bm{J}_{3} , \omega_{2})} \frac{F (\bm{J}_{3})}{(\bm{m}_{3} \!\cdot\! \bm{\Omega}_{3} \!-\! \omega_{1}) (\bm{m}_{3} \!\cdot\! \bm{\Omega}_{3} \!+\! \omega_{2})} \, .
\label{first_term_slow_derivation}
\end{align}
If we consider only the contributions that do not decay in time, one can perform the substitution
\begin{equation}
\frac{1}{(\bm{m}_{3} \!\cdot\! \bm{\Omega}_{3} \!-\! \omega_{1}) (\bm{m}_{3} \!\cdot\! \bm{\Omega}_{3} \!+\! \omega_{2})} \!\to\! (2 \pi)^{2} \delta_{\rm D} (\omega_{1} \!+\! \omega_{2}) \, \delta_{\rm D} (\bm{m}_{3} \!\cdot\! \bm{\Omega}_{3} \!-\! \omega_{1}) \, .
\label{substitution_Landau_derivation} \nonumber
\end{equation}
Starting from equation~\eqref{first_term_slow_derivation}, thanks to the previous substitution and using the fact that ${ \mathcal{D}_{- \bm{m}_{1} , \bm{m}_{3}} (\bm{J} , \bm{J}_{3} , - \omega_{1}) = \mathcal{D}_{\bm{m}_{1} , \bm{m}_{3}} (\bm{J}_{1} , \bm{J}_{3} , \omega_{1})^{*} }$, one can show that the first contribution from equation~\eqref{calculation_evolution_slow_I_derivation} takes the form
\begin{align}
\left\langle \delta F \, \frac{\partial \delta \psi}{\partial \bm{\theta}_{1}} \right\rangle_{\rm I} = - i (2 \pi)^{d + 1} \!\!\!\! \sum_{\bm{m}_{1} , \bm{m}_{2}} \!\! \int & \!  \frac{\mathrm{d} \omega_{1}}{2 \pi} \!\! \int \!\! \mathrm{d} \bm{J}_{2} \, \bm{m}_{1} \frac{\bm{m}_{1} \!\cdot\! \partial F / \partial \bm{J}_{1}}{\bm{m}_{1} \!\cdot\! \bm{\Omega}_{1} \!-\! \omega_{1}} \nonumber
\\
& \hskip -0.5cm \frac{\delta_{\rm D} (\bm{m}_{2} \!\cdot\! \bm{\Omega}_{2} \!-\! \omega_{1})}{| \mathcal{D}_{\bm{m}_{1} , \bm{m}_{2}} (\bm{J}_{1} , \bm{J}_{2} , \omega_{1})  |^{2}} F (\bm{J}_{2}) \, .
\label{calculation_first_term_slow_derivation}
\end{align}
The last step of the calculation is to use the Landau prescription ${ \omega \!\to\! \omega \!+\! i 0^{+} }$ along with Plemelj formula
\begin{equation}
\frac{1}{x \pm i 0^{+}} = \mathcal{P} \left(\! \frac{1}{x} \!\right) \mp i \pi \delta_{\rm D} (x) \, ,
\label{Plemelj_formula_derivation}
\end{equation}
where $\mathcal{P}$ is the principal value. One can then rewrite equation~\eqref{calculation_first_term_slow_derivation} under the form
\begin{align}
\left\langle \delta F \, \frac{\partial \delta \psi}{\partial \bm{\theta}_{1}} \right\rangle_{\rm I} = \pi \, (2 \pi)^{d} \!\!\! \sum_{\bm{m}_{1} , \bm{m}_{2}} \!\! \int \! & \mathrm{d} \bm{J}_{2} \, \bm{m}_{1} \frac{\delta_{\rm D} (\bm{m}_{1} \!\cdot\! \bm{\Omega}_{1} \!-\! \bm{m}_{2} \!\cdot\! \bm{\Omega}_{2})}{\mathcal{D}_{\bm{m}_{1} , \bm{m}_{2}} (\bm{J}_{1} , \bm{J}_{2} , \bm{m}_{1} \!\cdot\! \bm{\Omega}_{1})} \nonumber
\\
& \times \left(\! \bm{m}_{1} \!\cdot\! \frac{\partial F}{\partial \bm{J}_{1}} \!\right) F (\bm{J}_{2}) \, .
\label{calculation_first_term_slow_final_derivation}
\end{align}
Similarly, one can rewrite the second term of equation~\eqref{calculation_evolution_slow_II_derivation} under the form
\begin{align}
& \frac{\langle \delta \hat{F}_{\bm{m}_{1}} (\bm{J}_{1} , 0) \, \delta \hat{\psi}_{\bm{m}_{2}} (\bm{J}_{1} , \omega_{2}) \rangle}{i (\bm{m}_{1} \!\cdot\! \bm{\Omega}_{1} \!-\! \omega_{1}) } =  \nonumber
\\
& \;\;\;\;\; \;\;\;\;\; \;\;\;\;\; \;\;\; - (2 \pi)^{2} \frac{\delta_{\rm D} (\omega_{1} \!+\! \omega_{2}) \, \delta_{\rm D} (\bm{m}_{1} \!\cdot\! \bm{\Omega}_{1} \!-\! \omega_{1})}{\mathcal{D}_{\bm{m}_{2} , - \bm{m}_{1}} (\bm{J}_{1} , \bm{J}_{1} , - \omega_{1})} F (\bm{J}_{1}) \, .
\label{second_term_slow_derivation}
\end{align}
Using symmetries of the matrix ${ [\bm{I} \!-\! \widehat{\bm{M}}]^{-1} }$, starting from equation~\eqref{second_term_slow_derivation}, one can show that the second contribution from equation~\eqref{calculation_evolution_slow_I_derivation} finally takes the form
\begin{align}
\left\langle \delta F \, \frac{\partial \delta \psi}{\partial \bm{\theta}_{1}} \right\rangle_{\rm II} = - \pi \, (2 \pi)^{d} \!\!\! \sum_{\bm{m}_{1} , \bm{m}_{2}} \!\! \int \! & \mathrm{d} \bm{J}_{2} \, \bm{m}_{1} \frac{\delta_{\rm D} (\bm{m}_{1} \!\cdot\! \bm{\Omega}_{1} \!-\! \bm{m}_{2} \!\cdot\! \bm{\Omega}_{2})}{\mathcal{D}_{\bm{m}_{1} , \bm{m}_{2}} (\bm{J}_{1} , \bm{J}_{2} , \bm{m}_{1} \!\cdot\! \bm{\Omega}_{1})} \nonumber
\\
& \times \left(\! \bm{m}_{2} \!\cdot\! \frac{\partial F}{\partial \bm{J}_{2}} \!\right) F (\bm{J}_{1}) \, .
\label{calculation_second_term_slow_final_derivation}
\end{align}
Combining the two contributions obtained in equations~\eqref{calculation_first_term_slow_final_derivation} and~\eqref{calculation_second_term_slow_final_derivation}, one can rewrite equation~\eqref{calculation_evolution_slow_I_derivation} under the form
\begin{align}
\left\langle \delta F \, \frac{\partial \delta \psi}{\partial \bm{\theta}_{1}} \right\rangle = & \pi \, (2 \pi)^{d} \!\! \sum_{\bm{m}_{1} , \bm{m}_{2}} \!\! \int \!\! \mathrm{d} \bm{J}_{2} \, \bm{m}_{1} \, \frac{\delta_{\rm D} (\bm{m}_{1} \!\cdot\! \bm{\Omega}_{1} \!-\! \bm{m}_{2} \!\cdot\! \bm{\Omega}_{2})}{|\mathcal{D}_{\bm{m}_{1} , \bm{m}_{2}} (\bm{J}_{1} , \bm{J}_{2} , \bm{m}_{1} \!\cdot\! \bm{\Omega}_{1} ) |^{2}} \nonumber
\\
& \;\;\;\;\;\;\; \left( \bm{m}_{1} \!\cdot\! \frac{\partial }{\partial \bm{J}_{1}} \!-\! \bm{m}_{2} \!\cdot\! \frac{\partial }{\partial \bm{J}_{2}} \right) F (\bm{J}_{1} , t) \, F (\bm{J}_{2} , t) \, .
\label{calculation_evolution_slow_III_derivation}
\end{align}
Hence, using the slow evolution equation~\eqref{decoupled_evolution_equation_slow}, we recover the Balescu-Lenard equation introduced in equation~\eqref{initial_BL}. As a final remark, one must note that on the short dynamical timescale, the evolution is governed by equation~\eqref{decoupled_evolution_equation_fast}, which involves the fluctuating components ${ \delta F }$ and ${ \delta \psi }$ of the distribution function and the potential. In contrast, on the long secular timescale, the evolution is governed by equation~\eqref{decoupled_evolution_equation_slow}, which after an angle-average only involves the mean distribution function $F$. Indeed, 
thanks to the ensemble average, 
all the cross-correlations between the fluctuations ${ \delta F }$ and ${ \delta \psi }$, as in equations~\eqref{initial_correlation_derivation},~\eqref{first_term_slow_derivation} and~\eqref{second_term_slow_derivation} can be expressed in terms of the underlying smooth distribution function $F$ only, so that the fluctuating components are absent from the secular Balescu-Lenard collision operator from equation~\eqref{calculation_evolution_slow_III_derivation}.

\section{Turning off  collective effects}
\label{sec:noselfgravity}

The reference \cite{Chavanis2013} (see also Appendix A of \cite{Chavanis2012}) considers the inhomogeneous Balescu-Lenard equation without collective effects. This collisional kinetic equation is the equivalent of the Landau equation for inhomogeneous systems. It can be straightforwardly obtained as an approximation of the Balescu-Lenard equation~\eqref{initial_BL}. Indeed, one has to make the substitution from the \textit{dressed} susceptibility coefficients ${ 1 / \mathcal{D}_{\bm{m}_{1} , \bm{m}_{2}} (\bm{J}_{1} , \bm{J}_{2} , \omega) }$ to the \textit{bare} ones given by ${ A_{\bm{m}_{1} , \bm{m}_{2}} (\bm{J}_{1} , \bm{J}_{2}) }$, so that the inhomogeneous Balescu-Lenard equation without collective effects (\textit{i.e.} the inhomogeneous Landau equation) is given by
\begin{align}
& \frac{\partial F}{\partial t} = \pi (2 \pi)^{d} \frac{\partial }{\partial \bm{J}_{1}} \!\cdot\! \bigg[ \sum_{\bm{m}_{1} , \bm{m}_{2}} \!\! \bm{m}_{1} \!\! \int \!\! \mathrm{d} \bm{J}_{2} \, \delta_{\rm D} (\bm{m}_{1} \!\cdot\! \bm{\Omega}_{1} \!-\! \bm{m}_{2} \!\cdot\! \bm{\Omega}_{2}) \nonumber
\\
& \times  | A_{\bm{m}_{1} , \bm{m}_{2}} (\bm{J}_{1} , \bm{J}_{2}) |^{2} \!\left( \bm{m}_{1} \!\cdot\! \frac{\partial }{\partial \bm{J}_{1}} \!-\! \bm{m}_{2} \!\cdot\! \frac{\partial }{\partial \bm{J}_{2}} \right) F (\bm{J}_{1} , t) \, F (\bm{J}_{2} , t) \bigg] \, , \label{initial_BL_bare}
\end{align}
where the coefficients ${ A_{\bm{m}_{1} , \bm{m}_{2}} }$ are associated to the Fourier transform in angles of the interaction potential~\citep{Pichon1994,Chavanis2013} and read
\begin{align}
& A_{\bm{m}_{1} , \bm{m}_{2}} (\bm{J}_{1} , \bm{J}_{2}) =  \nonumber
\\
& \;\;\;\;\;\; \frac{1}{(2 \pi)^{4}} \!\! \int \!\! \mathrm{d} \bm{\theta}_{1} \mathrm{d} \bm{\theta}_{2} \, u (| \bm{x} (\bm{\theta}_{1} , \bm{J}_{1}) \!-\! \bm{x} (\bm{\theta}_{2} , \bm{J}_{2}) |) e^{- i (\bm{m}_{1} \cdot \bm{\theta}_{1} \!-\! \bm{m}_{2} \cdot \bm{\theta}_{2})} \, , \label{definition_Am1_m2}
\end{align}
where ${ u (\bm{x}) }$ is the binary interaction given by ${ u (\bm{x}) \!=\! - G / |\bm{x}| }$ in the gravitational case. A key remark at this stage is that the expression~\eqref{definition_Am1_m2} does not require to introduce biorthogonal basis elements as in equation~\eqref{definition_basis}, whereas in order to estimate the dressed susceptibility coefficients from equation~\eqref{definition_1/D}, one must necessarily rely on Kalnajs' matrix method~\citep{Kalnajs2}. It is however possible to express ${ A_{\bm{m}_{1} , \bm{m}_{2}} }$ using the potential basis. Indeed, for a fixed value of ${ \bm{x}_{2} }$, we consider the function ${ \bm{x}_{1} \!\mapsto\! u (\bm{x}_{1} \!-\! \bm{x}_{2}) }$. One can then decompose this function on the basis elements ${ \psi^{(p)} (\bm{x}_{1}) }$, so that we may write
\begin{equation}
u(\bm{x}_{1} \!-\! \bm{x}_{2}) = \sum_{p} a_{p} (\bm{x}_{2}) \, \psi^{(p)} (\bm{x}_{1}) \, ,
\label{decomposition_u_basis}
\end{equation}
where it is important to note that the basis coefficients ${ a_{p} (\bm{x}_{2}) }$ are functions of ${ \bm{x}_{2} }$. Thanks to the biorthogonality property of the basis detailed in equation~\eqref{definition_basis}, one can obtain the expression of the coefficients ${ a_{p} (\bm{x}_{2}) }$ which reads
\begin{align}
a_{p} (\bm{x}_{2}) & = - \!\! \int \!\! \mathrm{d} \bm{x}_{1} \, u (\bm{x}_{1} \!-\! \bm{x}_{2}) \, \rho^{(p) *} (\bm{x}_{1}) \nonumber
\\
& = - \left[ \int \!\! \mathrm{d} \bm{x}_{1} \, u (\bm{x}_{1} \!-\! \bm{x}_{2}) \, \rho^{(p)} (\bm{x}_{1})  \right]^{*} \nonumber
\\
& = - \psi^{(p) *} (\bm{r}_{2}) \, , \label{expression_ap_x2}
\end{align}
where we used the fact that $u$ is a real function. Hence we finally obtain
\begin{equation}
u (\bm{x}_{1} \!-\! \bm{x}_{2}) = - \sum_{p} \psi^{(p)} (\bm{x}_{1}) \, \psi^{(p) *} (\bm{x}_{2}) \, .
\label{Expression_u_basis}
\end{equation}
Taking appropriately a Fourier transform with respect to the angles as in equation~\eqref{definition_Am1_m2}, we obtain
\begin{equation}
A_{\bm{m}_{1} , \bm{m}_{2}} (\bm{J}_{1} , \bm{J}_{2}) = - \sum_{p} \psi^{(p)}_{\bm{m}_{1}} (\bm{J}_{1}) \, \psi^{(p) *}_{\bm{m}_{2}} (\bm{J}_{2}) \, .
\label{Am1_m2_basis}
\end{equation}
This fairly simple relation allows us to express the bare susceptibility coefficients ${ A_{\bm{m}_{1} , \bm{m}_{2}} }$ using the potential basis. One does not need anymore to perform a Fourier transform in angles of the interaction potential, because the resolution of Poisson's equation has been \textit{implicitly hidden} in the effective construction of the basis elements.\footnote{This also explains  why the factorization assumption ${ A_{\bm{m}_{1} , \bm{m}_{2}} (\bm{J}_{1} , \bm{J}_{2}) = A_{\bm{m}_{1}} (\bm{J}_{1}) \, A_{\bm{m}_{2}} (\bm{J}_{2}) }$ used in~\cite{Chavanis2007} does not hold.} Neglecting the collective effects in the expression~\eqref{definition_1/D} of the dressed susceptibility coefficients amounts to taking ${ \widehat{\mathbf{M}} (\omega) \!=\! 0 }$, so that we obtain
\begin{equation}
\frac{1}{\mathcal{D}_{\bm{m}_{1} , \bm{m}_{2}} (\bm{J}_{1} , \bm{J}_{2} , \omega)} \bigg|_{\rm w \!/\! o \, coll.} \!\!\!\!= - A_{\bm{m}_{1} , \bm{m}_{2}} (\bm{J}_{1} , \bm{J}_{2}) \, .
\label{1/D_nocoll}
\end{equation}
The negative sign in equation~\eqref{1/D_nocoll} plays no significant role in the kinetic equations, since one has to make the substitution of the square modulus ${ 1/|\mathcal{D}|^{2} \!\mapsto\! |A|^{2} }$ in the Balescu-Lenard equation~\eqref{initial_BL}, to obtain the inhomogeneous Landau equation~\eqref{initial_BL_bare}. Using our WKB basis, one can proceed similarly as in the dressed case, by limiting oneself only to local resonances. Equation~\eqref{calculation_1/D_III} therefore becomes in the bare case
\begin{align}
A_{\bm{m}_{1} , \bm{m}_{1}} & (R_{1} ,  J_{r}^{1} , R_{1} , J_{r}^{2}) = - \frac{1}{2 \pi} \frac{G}{R_{1}} \nonumber
\\
& \times \, \int_{1\!/\! \sigma_{k}}^{\infty} \!\!\!\!\! \mathrm{d} k_{r} \, \frac{1}{k_{r}} \, \mathcal{J}_{m_{1}^{r}} \!\left[\!\! \sqrt{\tfrac{2 J_{r}^{1}}{\kappa_{1}}} k_{r} \!\right] \, \mathcal{J}_{m_{2}^{r}} \!\left[\!\! \sqrt{\tfrac{2 J_{r}^{2}}{\kappa_{1}}} k_{r} \!\right] \, . \label{expression_Am1_m2_WKB}
\end{align}
This expression of the bare susceptibility coefficients can be straightforwardly obtained from the dressed ones by imposing ${ \lambda \!=\! 0 }$. One should note that because of the absence of the amplification term ${ 1 / (1 \!-\! \lambda_{k_{r}}) }$, the approximation of the small denominators cannot be used. Hence a physically motivated evaluation of the expression~\eqref{expression_Am1_m2_WKB} becomes more subtle to perform, especially because of the possibly important role that  the Coulomb logarithm ${ 1/ k_{r} }$ might play. Finally, one can note that even for exactly local resonances, \textit{i.e} ${ \bm{m}_{1} \!=\! \bm{m}_{2} }$ and ${ R_{1} \!=\! R_{2} }$, the use of our WKB formalism allowed us to obtain non-diverging bare susceptibility coefficients. On the other hand, one could try to estimate the bare susceptibility coefficients starting from equation~\eqref{definition_Am1_m2}, \textit{i.e.} without using any potential basis. Using the polar coordinates ${ (R , \phi) }$, one has to compute
\begin{align}
A_{\bm{m}_{1} , \bm{m}_{2}} (\bm{J}_{1} , \bm{J}_{2}) = & - \frac{G}{(2 \pi)^{4}} \!\! \int \!\! \mathrm{d} \theta_{1}^{r} \mathrm{d} \theta_{1}^{\phi} \mathrm{d} \theta_{2}^{r} \mathrm{d} \theta_{2}^{\phi} \nonumber
\\
& \times \, \frac{e^{i \bm{m}_{1} \cdot \bm{\theta}_{1}} e^{i \bm{m}_{2} \cdot \bm{\theta}_{2}}}{\sqrt{R_{1}^{2} \!+\! R_{2}^{2} \!-\! 2 R_{1} R_{2} \cos (\phi_{1} \!-\! \phi_{2})}} \, . \label{calculation_Am1_m2_TF}
\end{align}
By azimuthal symmetry, it is straightforward to show~\citep{Pichon1997} that
\begin{equation}
A_{\bm{m}_{1} , \bm{m}_{2}} \propto \delta_{m_{1}^{\phi}}^{m_{2}^{\phi}} \, .
\label{azimuthal_Am1_m2_TF}
\end{equation}
Hence the different ${ m_{\phi}-}$modes of the ${ A_{\bm{m}_{1} , \bm{m}_{2}} }$ coefficients are independent. This result is identical to what was obtained in equation~\eqref{equality_mphi_kphi} for the dressed case. In order to illustrate the \textit{regularizing role} of the WKB basis from equation~\eqref{definition_WKB_basis}, we will place ourselves in the context of an extremely tepid disc, and therefore assume ${ J_{r}^{1} \!=\! J_{r}^{2} \!=\! 0 }$. Thanks to the epicyclic mapping from equation~\eqref{mapping_action}, one can drop all the dependences on $\theta_{R}$ appearing in $R$ and $\phi$. Equation~\eqref{calculation_Am1_m2_TF} then immediately implies that
\begin{equation}
m_{1}^{r} = m_{2}^{r} = 0 \, ,
\label{mr_nul_calculation_Am1_m2}
\end{equation}
and the bare susceptibility coefficients are given by
\begin{align}
A_{\bm{m}_{1} , \bm{m}_{2}} (R_{1} , 0, R_{2} , 0) = - \frac{G}{(2 \pi)^{2}} \delta_{0}^{m_{1}^{r}} \delta_{0}^{m_{2}^{r}} \delta_{m_{1}^{\phi}}^{m_{2}^{\phi}} \!\!\ \int \!\! \mathrm{d} \theta_{1}^{\phi} \mathrm{d} \theta_{2}^{\phi} \nonumber
\\
\times \, \frac{e^{- i m_{1}^{\phi} \theta_{1}^{\phi}} e^{i m_{2}^{\phi} \theta_{2}^{\phi}} }{\sqrt{R_{1}^{2} \!+\! R_{2}^{2} \!-\! 2 R_{1} R_{2} \cos (\theta_{1}^{\phi} \!-\! \theta_{2}^{\phi})}} \, . \label{calculation_Am1_m2_TF_II}
\end{align}
In this illustrative limit, we have by construction no contributions from the ILR and OLR resonances for which ${ m_{r} \!\neq\! 0 }$. After an immediate change of variables, using ${ \Delta \!=\! \theta_{1}^{\phi} \!-\! \theta_{2}^{\phi} }$, it becomes
\begin{align}
A_{0 , m_{\phi} , 0 , m_{\phi}} & (R_{1} , 0 , R_{2} , 0) =  \nonumber
\\
& - \frac{G}{2 \pi} \!\! \int_{- \pi}^{\pi} \!\!\!\! \mathrm{d} \Delta \, \frac{e^{- i m_{\phi} \Delta}}{\sqrt{(R_{1} \!-\! R_{2})^{2} \!+\! 2 R_{1} R_{2} ( 1 \!-\! \cos (\Delta) ) }} \, . \label{calculation_Am1_m2_TF_III}
\end{align}
The resonance condition from equation~\eqref{definition_f_R2r} takes the form ${ m_{\phi} \Omega_{\phi} (R_{1}) \!=\! m_{\phi} \Omega_{\phi} (R_{2}) }$, because we restricted ourselves in equation~\eqref{mr_nul_calculation_Am1_m2} to the case ${ m_{r} \!=\! 0 }$. Hence assuming that the function ${ R_{g} \mapsto \Omega_{\phi} (R_{g}) }$ is a monotonic function, one has to satisfy the constraint ${ R_{1} \!=\! R_{2} }$, so that we are restricting ourselves only to local resonances as in equation~\eqref{restriction_local_resonances}. For such resonances, equation~\eqref{calculation_Am1_m2_TF_III} becomes
\begin{equation}
A_{0 , m_{\phi}, 0 , m_{\phi}} (R_{1} , 0 , R_{1} , 0) = - \frac{G}{2 \pi \! \sqrt{2} R_{1}} \!\! \int_{- \pi}^{\pi} \!\!\!\! \mathrm{d} \Delta \, \frac{ e^{- i m_{\phi} \Delta} }{\sqrt{1 \!-\! \cos(\Delta)}} \, .
\label{calculation_Am1_m2_TF_IV}
\end{equation}
At this stage, one must note that this integral is divergent. Indeed, for ${
\Delta \!\to\! 0 }$, one has ${ 1\!/\!\sqrt{1 \!-\! \cos(\Delta)} \!\sim\! \!\sqrt{2} /
\Delta }$. Hence the expression of the bare susceptibility coefficients derived
from the Fourier transform of the interaction potential as in
equation~\eqref{definition_Am1_m2}, when restricted to exactly local resonances
becomes logarithmically divergent. This divergence, which is observed in the
case of local resonances, is induced by the interaction of singular orbits. It
is important to note that this divergence was not observed in
equation~\eqref{expression_Am1_m2_WKB} when computing the bare susceptibility
coefficients using the WKB basis from equation~\eqref{definition_WKB_basis}.
This implies that the WKB basis is not complete.
For a complete biorthogonal basis, the
expression~\eqref{Am1_m2_basis} of the ${ A_{\bm{m}_{1} , \bm{m}_{2}} }$
coefficients is an exact expression. However, our calculation shows that the \textit{scale-decoupled} WKB
basis we used, possesses a subtle \textit{regularizing} incompleteness which
allows to get rid of the diverging contributions to the coefficients
${ A_{\bm{m}_{1} , \bm{m}_{2}} }$ in the limit of exactly local resonances.

\section{The  Schwarzschild DF case}
\label{sec:explicitappendix}

When considering a Schwarzchild distribution function as in equation~\eqref{definition_DF_Schwarzschild}, while relying on the approximation of the small denominators from equation~\eqref{expression_1/D^2_ASD}, one can explicitly perform the remaining integration on the radial action $J_{r}^{2}$ in the expressions~\eqref{WKB_drift} and~\eqref{WKB_diff} of the drift and diffusion coefficients. We now detail this explicit calculation. For such a Schwarzschild distribution function, it is straightforward to check that the gradients of the distribution function with respect to the actions are given by
\begin{equation}
\frac{\partial F}{\partial J_{r}} = - \frac{\kappa}{\sigma_{r}^{2}} F \;\;\; ; \;\;\; \frac{\partial F}{\partial J_{\phi}} = F \frac{\partial }{\partial J_{\phi}} \!\left[ \ln \!\left(\! \frac{\Omega \Sigma}{\pi \kappa \sigma_{r}^{2}} \!\right) \!-\! \frac{\kappa J_{r}}{\sigma_{r}^{2}} \!\right] \, .
\label{explicit_gradients_DF}
\end{equation}
Using the expression of the susceptibility coefficients from equation~\eqref{expression_1/D^2_ASD}, after some simple algebra, one can rewrite the drift coefficients from equation~\eqref{WKB_drift} under the form
\begin{align}
A_{\bm{m}_{1}} & (\bm{J}_{1}) = - g_{\bm{m}_{1}} (J_{\phi}^{1}  , J_{r}^{1}) \!\! \int \!\! \mathrm{d} J_{r}^{2} \, \exp \!\left[\! - \frac{\kappa}{\sigma_{r}^{2}} J_{r}^{2} \!\right] \mathcal{J}_{m_{1}^{r}}^{2} \!\left[\!\! \sqrt{\tfrac{2 J_{r}^{2}}{\kappa_{1}}} k_{\rm max} \!\right] \, \nonumber
\\
&\times \, \left[ m_{1}^{\phi} \!\left\{ \frac{\partial }{\partial J_{\phi}} \!\left[ \ln \!\left(\! \frac{\Omega \Sigma}{\pi \kappa \sigma_{r}^{2}} \!\right) \right] \!-\! J_{r}^{2} \frac{\partial }{\partial J_{\phi}} \!\left[\! \frac{\kappa}{\sigma_{r}^{2}} \!\right] \right\} \!-\! m_{1}^{r} \frac{\kappa}{\sigma_{r}^{2}} \right] \, .
\label{explicit_WKB_drift}
\end{align}
Similarly, the diffusion coefficients from equation~\eqref{WKB_diff} take the form
\begin{equation}
D_{\bm{m}_{1}} (\bm{J}_{1}) \!=\! g_{\bm{m}_{1}} (J_{\phi}^{1} , J_{r}^{1}) \!\!\! \int \!\!\! \mathrm{d} J_{r}^{2} \, \exp \!\left[\! - \frac{\kappa}{\sigma_{r}^{2}} J_{r}^{2} \!\right] \mathcal{J}_{m_{1}^{r}}^{2} \!\left[\!\! \sqrt{\tfrac{2 J_{r}^{2}}{\kappa_{1}}} k_{\rm max} \!\right]  .
\label{explicit_WKB_diff}
\end{equation}
In equations~\eqref{explicit_WKB_drift} and~\eqref{explicit_WKB_diff}, in order to shorten the notations, we introduced the function ${ g_{\bm{m}_{1}} (J_{\phi}^{1} , J_{r}^{1}) }$ defined as
\begin{equation}
g_{\bm{m}_{1}} (J_{\phi}^{1} , J_{r}^{1}) \!=\! \frac{1}{(\bm{m}_{1} \!\cdot\! \bm{\Omega}_{1})'} \frac{G^{2}}{R_{1}^{2}} \frac{\Omega\Sigma}{\kappa \sigma_{r}^{2}} \frac{(\Delta k_{\lambda})^{2}}{k_{\rm max}^{2}} \!\left[\! \frac{1}{1 \!-\! \lambda_{\rm max}} \!\right]^{2} \! \mathcal{J}_{m_{1}^{r}}^{2} \!\left[\!\! \sqrt{\tfrac{2 J_{r}^{1}}{\kappa_{1}}} k_{\rm max} \!\right] \, ,
\label{explicit_definition_gm1} \nonumber
\end{equation}
where we used the same shortened notation for the resonant factor ${ 1/(\bm{m}_{1} \!\cdot\! \bm{\Omega}_{1})' }$ as in equation~\eqref{short_resonance_factor}. In addition to the integration formula~\eqref{integration_formula_Bessel}, we may also rely on the additional identity
\begin{align}
\!\! \int_{0}^{+ \infty} \!\!\!\!\!\!\! \mathrm{d} J_{r} \, & J_{r} \, e^{- \alpha J_{r}} \mathcal{J}_{m_{r}}^{2} \!\left[ \beta \sqrt{J_{r}} \right] =   \nonumber
\\
& \!\!\! \frac{e^{- \beta^{2} / 2 \alpha}}{\alpha^{2}} \!\left\{ \left[ - \frac{\beta^{2}}{2 \alpha} \!+\! 1 \!+\! |m_{r}| \right] \mathcal{I}_{m_{r}} \!\left[\! \frac{\beta^{2}}{2 \alpha} \!\right] \!+\! \frac{\beta^{2}}{2 \alpha} \mathcal{I}_{|m_{r}| \!+\! 1} \!\left[\! \frac{\beta^{2}}{2 \alpha} \!\right] \right\} \, , \label{integration_formula_Bessel_complex}
\end{align}
where ${ \alpha \!>\! 0 }$, ${ \beta \!>\! 0 }$, and ${ m_{r} \!\in\! \mathbb{Z} }$. In analogy with the definition from equation~\eqref{definition_chi}, we also introduce $\chi_{\rm max}$ as
\begin{equation}
\chi_{\rm max} = \frac{\sigma_{r}^{2} \, k_{\rm max}^{2}}{\kappa^{2}} \, .
\label{definition_chi_max}
\end{equation}
One can then immediately perform the integration on $J_{r}^{2}$ from equation~\eqref{explicit_WKB_diff}, so that the diffusion coefficients are given by
\begin{equation}
D_{\bm{m}_{1}} (\bm{J}_{1}) = h_{\bm{m}_{1}}^{D} (J_{\phi}^{1}) \, \mathcal{J}_{m_{1}^{r}}^{2} \!\left[\!\! \sqrt{\tfrac{2 J_{r}^{1}}{\kappa_{1}}} k_{\rm max} \!\right] \, ,
\label{explicit_WKB_diff_II}
\end{equation}
where the function ${ h_{\bm{m}_{1}}^{D} (J_{\phi}^{1}) }$ is defined as
\begin{equation}
h_{\bm{m}_{1}}^{D} (J_{\phi}^{1}) = \frac{1}{(\bm{m}_{1} \!\cdot\! \bm{\Omega}_{1})'} \frac{G^{2}}{R_{1}^{2}} \frac{\Omega \Sigma}{\kappa^{2}} \frac{(\Delta k_{\lambda})^{2}}{k_{\rm max}^{2}} \!\left[\! \frac{1}{1 \!-\! \lambda_{\rm max}} \!\right]^{2} \! e^{- \chi_{\rm max}} \mathcal{I}_{m_{1}^{r}} [ \chi_{\rm max} ] \, .
\label{explicit_definition_h_D} \nonumber
\end{equation}
After some algebra, the drift coefficients from equation~\eqref{explicit_WKB_drift} are given by
\begin{equation}
A_{\bm{m}_{1}} (\bm{J}_{1}) = -\,  h_{\bm{m}_{1}}^{A} (J_{\phi}^{1}) \, \mathcal{J}_{m_{1}^{r}}^{2} \!\left[\!\! \sqrt{\tfrac{2 J_{r}^{1}}{\kappa_{1}}} k_{\rm max} \!\right] \, ,
\label{explicit_WKB_drift_II}
\end{equation}
where the function ${ h_{\bm{m}_{1}}^{A} (J_{\phi}^{1}) }$ is defined as
\begin{align}
h_{\bm{m}_{1}}^{A} & (J_{\phi}^{1}) =  \, h_{\bm{m}_{1}}^{D} (J_{\phi}^{1}) \, \bigg\{ \!-\! m_{1}^{r} \frac{\kappa}{\sigma_{r}^{2}} \!+\! m_{1}^{\phi} \frac{\partial }{\partial J_{\phi}} \bigg[\! \ln \bigg(\! \frac{\Omega \Sigma}{\pi \kappa \sigma_{r}^{2}} \!\bigg) \bigg] \nonumber
\\
& + m_{1}^{\phi} \frac{\kappa}{\sigma_{r}^{2}} \frac{\partial}{\partial J_{\phi}} \bigg[\! \frac{\sigma_{r}^{2}}{\kappa} \bigg] \bigg[ 1 \!+\! |m_{1}^{r}| \!-\! \chi_{\rm max} \!+\! \frac{\mathcal{I}_{|m_{1}^{r}| + 1} [\chi_{\rm max}] }{\mathcal{I}_{m_{1}^{r}} [\chi_{\rm max}]} \bigg]  \bigg\} \, . \label{explicit_definition_h_A}
\end{align}
These explicit expressions of the diffusion and drift coefficients obtained in equations~\eqref{explicit_WKB_diff_II} and~\eqref{explicit_WKB_drift_II} allow to estimate in a simple way the secular flux in the entire action space ${ \bm{J} \!=\! (J_{\phi} , J_{r}) }$, once we assume that the distribution function is a Schwarzschild DF given by equation~\eqref{definition_DF_Schwarzschild} and that the susceptibility coefficients can be approximated by equation~\eqref{expression_1/D^2_ASD}.

\section{Relation to other kinetic equations}
\label{sec:comp}

The kinetic equation governing the collisional evolution of a system 
of $N$ stars at the order ${ 1/N }$ is the inhomogeneous Balescu-Lenard equation~\eqref{initial_BL}. This equation conserves the total number of stars  and the
energy, and monotonically increases the Boltzmann entropy (H-Theorem). We note
that the collisional evolution of the system is due to a condition of resonance
encapsulated in the term ${ \delta_{\rm D} (\bm{m}_{1} \!\cdot\! \bm{\Omega}_{1} \!-\! \bm{m}_{2} \!\cdot\! \bm{\Omega}_{2}) }$. In general, this condition can allow for
local and non local resonances. In the case of tepid discs considered in the
present paper, assuming that only tightly wound spirals are sustained by the disc, 
we justified in equation~\eqref{restriction_local_resonances} the fact that the resonances are purely local,
so that ${ \bm{m}_{1} \!=\! \bm{m}_{2} }$ and ${ J_{\phi}^{1} \!=\! J_{\phi}^{2} }$.
Furthermore, because of the epicyclic approximation, the
intrinsic frequencies of the system, given by equation~\eqref{definition_Omega_kappa}, depend only on $J_{\phi}$, so that
${ \bm{\Omega} \!=\! \bm{\Omega} (J_{\phi}) }$. Under these conditions, the Balescu-Lenard equation
giving the collisional evolution of ${ F \!=\! F(J_{\phi},J_{r},t) }$ may be rewritten as
\begin{align}
&\frac{\partial F}{\partial t}=  \, 4\pi^3 \frac{\partial}{\partial \bm{J}} \!\cdot\! \Biggl\lbrace \sum_{\bm{m}}\bm{m}\frac{1}{\left| \frac{\partial }{\partial J_{\phi}} [\bm{m} \!\cdot\! \bm{\Omega}] \right|_{J_{\phi}}} \!\! \int \!\! \mathrm{d} J_{r}' \mathrm{d}J_{\phi}'    \nonumber
\\
& \hskip -0.15cm  \times  \frac{\delta_{\rm D} (J_{\phi} \!-\! J_{\phi}')}{|\mathcal{D}_{\bm{m}, \bm{m}} (J_{\phi}, J_{r}, J_{\phi}', J_{r}', \bm{m}\!\cdot\! \bm{\Omega}) |^{2}}\nonumber\\
& \hskip -0.15cm \times  \bm{m} \!\cdot\! \Biggl\lbrack \! F (J_{\phi}' , J_{r}',t) \frac{\partial F}{\partial \bm{J}}  (J_{\phi}, J_{r},t) \!-\! F (J_{\phi} , J_{r},t) \frac{\partial F}{\partial \bm{J}}  (J_{\phi}', J_{r}',t)\Biggr\rbrack\Biggr\rbrace \, . \label{comp1} 
\end{align}
The integration on ${ J_{\phi}' }$ is straightforward because of the 
$\delta_{\rm D}$-function (local resonance), and we are left with
\begin{align}
\frac{\partial F}{\partial t}= & \, 4\pi^3 \frac{\partial}{\partial \bm{J}} \!\cdot\! \Biggl\lbrace \sum_{\bm{m}}\bm{m}  \frac{1}{\left| \frac{\partial }{\partial J_{\phi}} [\bm{m} \!\cdot\! \bm{\Omega}] \right|_{J_{\phi}}} \!\! \int \!\! \mathrm{d} J_{r}' \nonumber
\\
& \times \frac{1}{|\mathcal{D}_{\bm{m}, \bm{m}} (J_{\phi}, J_{r}, J_{\phi}, J_{r}', \bm{m}\!\cdot\! \bm{\Omega}) |^{2}} \nonumber
\\
& \times  \bm{m} \!\cdot\! \Biggl\lbrack \! F (J_{\phi}, J_{r}',t) \frac{\partial F}{\partial \bm{J}}  (J_{\phi}, J_{r},t) \!-\! F (J_{\phi} , J_{r},t) \frac{\partial F}{\partial \bm{J}}  (J_{\phi}, J_{r}',t)\Biggr\rbrack\Biggr\rbrace \, , \label{comp2}
\end{align}
where the susceptibility coefficients are generally given by equation~\eqref{calculation_1/D^2_I}.

The kinetic equation \eqref{comp2} is an integro-differential 
equation that governs the evolution of the system {\it as a whole}. It
describes the effects of encounters between any test particle characterized
by the angle-action coordinates  ${ (J_{\phi}, J_{r}) }$ and the field particles
characterized by the (running) angle-action coordinates ${ (J_{\phi}', J_{r}') }$.
Actually, there is no distinction between test and field particles, so that they are
characterized by the same distribution function ${ F(\cdot,t) }$ that evolves in a
self-consistent manner, hence the integro-differential character of the kinetic
equation. This is a characteristic of the Balescu-Lenard equation describing the
evolution of the system as a whole. 

\subsection{Fokker-Planck limit}

We can also use this formalism to directly obtain the 
Fokker-Planck  equation governing the relaxation of a test star in a
{\it bath} of field stars, assumed to be in a steady state with a
distribution function ${ F_0 (J_{\phi}', J_{r}') }$. Proceeding as in
\cite{Chavanis2012},  we just have to replace in equation \eqref{comp2} the
distribution function of the field particles ${ F(J_{\phi}, J_{r}',t) }$ by the
static distribution ${ F_0(J_{\phi}, J_{r}') }$, while the time evolving
distribution function of the test particle is rewritten as ${ P(J_{\phi},
J_{r},t) }$ for clarity. This heuristic procedure is justified in
\cite{Chavanis2012} by an explicit calculation of the diffusion and drift
coefficients of the Fokker-Planck equation. It transforms the integro-differential equation~\eqref{comp2} into a differential equation
\begin{align}
\frac{\partial P}{\partial t}= & \, 4\pi^3 \frac{\partial}{\partial \bm{J}} \!\cdot\! \Biggl\lbrace \sum_{\bm{m}}\bm{m}  \frac{1}{\left| \frac{\partial }{\partial J_{\phi}} [\bm{m} \!\cdot\! \bm{\Omega}] \right|_{J_{\phi}}} \!\! \int \!\! \mathrm{d} J_{r}'  \, \nonumber
\\
& \times \frac{1}{|\mathcal{D}_{\bm{m}, \bm{m}} (J_{\phi}, J_{r}, J_{\phi}, J_{r}', \bm{m}\!\cdot\! \bm{\Omega}) |^{2}} \nonumber
\\
& \times  \bm{m} \!\cdot\! \Biggl\lbrack F_0 (J_{\phi}, J_{r}') \frac{\partial P}{\partial \bm{J}}  (J_{\phi}, J_{r},t) \!-\! P (J_{\phi} , J_{r},t) \frac{\partial F_0}{\partial \bm{J}}  (J_{\phi}, J_{r}')\Biggr\rbrack\Biggr\rbrace \, , \label{comp3}
\end{align}
which can be interpreted as a Fokker-Planck equation.
If we assume that the field particles are at statistical 
equilibrium (thermal bath), described by the Boltzmann distribution 
\begin{equation}
F_0( \bm{J}) = C e^{-\beta H( \bm{J})} \, ,
\label{comp4}
\end{equation}
then, using the relation ${ \partial F_0/\partial \bm{J} \!=\! -\beta F_0 \bm{\Omega} }$
(see the definition of $\bm{\Omega}$ in equation~\eqref{definition_Omega}), we
can reduce the Fokker-Planck equation~\eqref{comp3} to the form
\begin{equation}
\frac{\partial P}{\partial t}= \frac{\partial}{\partial \bm{J}} \!\cdot\! \Biggl\lbrace \sum_{\bm{m}}\bm{m} \, D_{\bm{m}}(\bm{J}) \, \bm{m} \!\cdot\! \Biggl\lbrack  \frac{\partial P}{\partial \bm{J}} + \beta \bm{\Omega}(J_{\phi})\,P\Biggr\rbrack\Biggr\rbrace \, , 
\label{comp5}
\end{equation}
with the diffusion coefficient
\begin{equation}
D_{\bm{m}}(\bm{J})=\frac{4\pi^3}{\left| \frac{\partial }{\partial J_{\phi}} [\bm{m} \!\cdot\! \bm{\Omega}] \right|_{J_{\phi}}} \!\!\int \!\! \mathrm{d} J_{r}'  \, \frac{F_0 (J_{\phi}, J_{r}')}{|\mathcal{D}_{\bm{m}, \bm{m}} (J_{\phi}, J_{r}, J_{\phi}, J_{r}', \bm{m}\!\cdot\! \bm{\Omega}) |^{2}} \, .
\label{comp6}
\end{equation}
We note that the friction term in the Fokker-Planck equation~\eqref{comp5} is proportional
and opposite to the intrinsic frequency $\bm{\Omega}$, and that the friction
coefficient $\xi$ satisfies a
(generalized) Einstein relation ${ \xi \!=\! D \beta }$ (for each resonance)
\citep{Chavanis2012}.

This Fokker-Planck formalism may have other applications. For example, if we consider a
system with two species of particles (\textit{e.g.} characterized by different masses
$m_{A}$ and $m_{B}$), and if species B has reached an equilibrium state with a
distribution ${ F_0(J_{\phi}', J_{r}') }$, we can use equation~\eqref{comp3} to
describe the relaxation of particles of species A due to the encounters with
particles of species B (but neglecting the encounters between particles of
species A). This approach is further discussed in Appendix F of
\cite{Chavanis2013}. On the other hand, for a single species system, we may
describe the early dynamics of the system as a whole with a good approximation
by replacing  ${ F(J_{\phi}, J_{r}',t) }$  in the Balescu-Lenard equation~\eqref{comp2} by the
{\it initial} distribution function ${ F_0(J_{\phi}, J_{r}') }$, leading again to an
equation of the form~\eqref{comp3}. 

\subsection{Other kinetic equations}

We now compare the previous results to other related kinetic equations.  For
spatially homogeneous systems with long-range interactions, the Balescu-Lenard equation
reads (see \cite{Chavanis2012b}):
\begin{align}
\frac{\partial F}{\partial t}= & \, \pi (2\pi)^d m  \frac{\partial}{\partial \bm{v}} \!\cdot\!  \Biggl\lbrace \! \int \!\! \mathrm{d} \bm{k} \mathrm{d}\bm{v}' \, \bm{k}  \frac{\hat{u}(\bm{k})^2}{\left| \epsilon(\bm{k},\bm{k} \!\cdot\! \bm{v})\right|^2} \, \delta_{\rm D} [\bm{k} \!\cdot\! (\bm{v} \!-\! \bm{v}')]  \nonumber
\\
& \times  \bm{k} \!\cdot\! \Biggl\lbrack F (\bm{v}',t) \frac{\partial F}{\partial \bm{v}}  (\bm{v},t) \!-\! F (\bm{v},t) \frac{\partial F}{\partial \bm{v}}  (\bm{v}'\!,t)\Biggr\rbrack\Biggr\rbrace \, , \label{comp7}
\end{align}
where
\begin{equation}
\epsilon(\bm{k},\omega)=1 \!-\! (2\pi)^d \hat{u}(\bm{k}) \!\! \int \!\!  \mathrm{d} \bm{v} \,  \frac{\bm{k} \!\cdot\! \partial F / \partial v  }{\bm{k} \!\cdot\! \bm{v} \!-\! \omega} \,  \label{comp8}
\end{equation}
is the dielectric function. For one dimensional systems, it reduces to the trivial form
\begin{align}
\frac{\partial F}{\partial t}= & \, 2\pi^2 m  \frac{\partial}{\partial {v}} \!\! \int \!\! \mathrm{d} {k} \mathrm{d}{v}' \, {|k|}  \frac{\hat{u}({k})^2}{\left| \epsilon({k},{k} {v}) \right|^2} \, \delta_{\rm D} ({v} \!-\! {v}') \nonumber
\\
& \times   \Biggl\lbrack F ({v}'\! ,t) \frac{\partial F}{\partial {v}}  ({v},t) \!-\! F ({v},t) \frac{\partial F}{\partial {v}}  ({v}'\!,t)\Biggr\rbrack=0 \, . \label{comp9}
\end{align}
In ${ d \!>\! 1 }$, there are always resonances between particles with different
velocities, implying that the Balescu-Lenard equation relaxes towards the Boltzmann
distribution on a timescale of the order ${ N t_D }$, where $t_D$ is the dynamical
time. By contrast, in ${ d \!=\! 1 }$, the resonances become local (in velocity space) and
since the term in brackets is anti-symmetric with respect to the interchange of
$v$ and ${ v' }$, the Balescu-Lenard diffusion current vanishes exactly. This implies that the
relaxation time is larger than ${ N t_D }$, presumably of the order ${ N^2 t_D }$,
corresponding to the next order term in the expansion of the dynamics in
powers of ${ 1/N }$. The Fokker-Planck equation describing the evolution of a test particle in
a bath of field particles is discussed in \cite{Chavanis2012b,Chavanis2013b}.

The kinetic equation governing the collisional evolution of a system of $N$
point vortices  in two dimensional hydrodynamics  at the order ${ 1/N }$ can be
written, in the axisymmetric case, as (see \cite{Chavanis2012c}):
\begin{align}
\frac{\partial \omega}{\partial t}= & \, 2\pi^2 \gamma\frac{1}{r}  \frac{\partial}{\partial {r}} \!\! \int_0^{+\infty} \!\!\!\!\!\! \mathrm{d}{r}'  \, {r}'   \chi({r},{r}'\!,\Omega(r,t)) \, \delta_{\rm D} [\Omega(r,t) \!-\! \Omega(r'\!,t)]   \nonumber
\\
& \times   \Biggl\lbrack \omega ({r}'\!,t) \frac{1}{r}\frac{\partial \omega}{\partial {r}}  ({r},t) \!-\! \omega({r},t) \frac{1}{r'}\frac{\partial \omega}{\partial {r}}  ({r}'\!,t)\Biggr\rbrack \, \,  , \label{comp10}
\end{align}
where ${ \omega(r,t) }$ is the profile of vorticity, ${ \Omega(r,t) }$ the profile of
angular velocity, and ${ \chi({r},{r}',\Omega(r,t)) }$ is related to the dressed
potential of interaction between the point vortices (see \cite{Chavanis2012c}
for more details). This equation conserves the total number of point vortices
and the energy, and monotonically increases the Boltzmann entropy (H-Theorem). 
If the profile of angular velocity is monotonic, the kinetic equation reduces to
the form
\begin{align}
\frac{\partial \omega}{\partial t}= & 2\pi^2 \gamma\frac{1}{r}  \frac{\partial}{\partial {r}} \int_0^{+\infty}  \!\!\!\!\!\! \mathrm{d}{r}' \, {r}'   \chi({r},{r}'\!,\Omega(r,t)) \frac{1}{\left| \Omega'(r,t)\right|} \, \delta_{\rm D} (r \!-\! r') \nonumber
\\
& \times   \Biggl\lbrack \omega ({r}'\!,t) \frac{1}{r}\frac{\partial \omega}{\partial {r}}  ({r},t) \!-\! \omega({r},t) \frac{1}{r'}\frac{\partial \omega}{\partial {r}}  ({r}'\!,t)\Biggr\rbrack=0 \, . \label{comp11}
\end{align}
For non-monotonic profile of angular velocity, one can have non-local
resonances (\textit{i.e.} distant collisions between point vortices), as studied in
\cite{ChavanisLemou2007}. This produces a diffusion current. If the profile of
angular velocity is, or becomes, monotonic, the resonances are purely local and,
since the term in brackets is anti-symmetric with respect to the interchange of
$r$ and ${ r' }$, the diffusion current also vanishes. This implies that the relaxation
time is larger than ${ N t_D }$ as discussed above. The Fokker-Planck equation describing the evolution of a test vortex in
a sea of field vortices is discussed in \cite{Chavanis2012c}.

If we focus on purely local resonances, we note that the inhomogeneous  Balescu-Lenard
equation~\eqref{comp1} is different from equations~\eqref{comp9} and~\eqref{comp11} because it is two-dimensional in $J_r$ and $J_{\phi}$, and the
resonances act only on $J_{\phi}$. Therefore, purely local resonances do not
yield a zero flux, contrary to equations~\eqref{comp9} and~\eqref{comp11}. This
really is an effect of the two-dimensionality of the system. Indeed, for local
resonances, the ${ 1D }$ inhomogeneous  equation also yields a zero flux:
\begin{align}
\frac{\partial F}{\partial t}= & 2\pi^2 \frac{\partial}{\partial {J}}
\Biggl\lbrace \sum_{{m}}\left |\frac{m}{\Omega'(J)}\right| \! \int
\!\! \mathrm{d} J' \, \frac{ \delta_{\rm D} (J\!-\!J')}{|\mathcal{D}_{{m}, {m}} (J,J,{m}{\Omega}) |^{2}} \nonumber
\\
& \times    \Biggl\lbrack F (J',t) \frac{\partial
F}{\partial {J}}  (J,t) \!-\! F (J,t) \frac{\partial
F}{\partial {J}}  (J',t)\Biggr\rbrack\Biggr\rbrace=0 \, . \label{comp12}
\end{align}

\section{The Schwarzschild conspiracy}
\label{sec:Schwarzschildplot}

\begin{figure}
\begin{center}
\epsfig{file=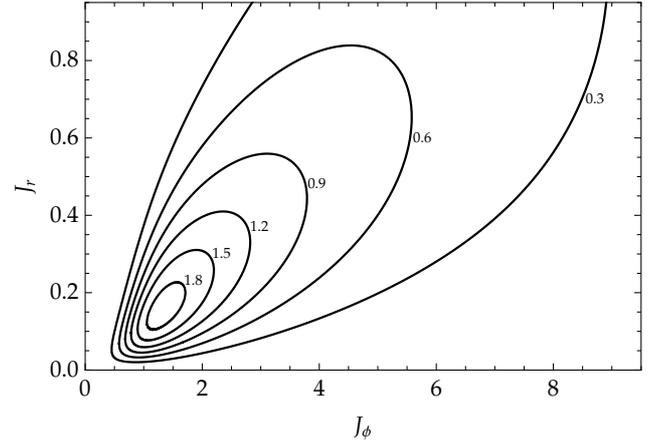,angle=-00,width=0.45\textwidth}
\caption{\small{
Contours in action-space ${ (J_{\phi} , J_{r}) }$ of the difference between the sampled anisotropic DF of S12 simulation ${ F_{\rm ani} \!\propto\! J_{\phi}^{q} \exp [- E / \sigma_{r}^{2}]}$ and its Schwarzschild epicyclic approximation ${ F_{\rm Sch} }$ from equation~\eqref{definition_DF_Schwarzschild}. The plotted quantity is ${ | F_{\rm ani} \!-\! F_{\rm Sch} | / F_{\rm Sch}^{\rm max} }$, and contours labels are expressed in percentages.
}}
\label{figContourDiffDF}
\end{center}
\end{figure}

The Schwarzschild distribution function introduced in equation~\eqref{definition_DF_Schwarzschild} and considered in S12 simulation has the specificity to be exponential in the ${J_{r}-}$direction, so that $F$ is \textit{Boltzmannian} with respect to the $J_{r}$ variable. It is  known \citep[\textit{e.g.}][]{Chavanis2012} that the (complete) Boltzmann distribution is the steady state of the Balescu-Lenard equation~\eqref{initial_BL}. Therefore, we can expect that the (partial) exponential behavior of the  Schwarzschild distribution will induce simplifications that we now detail. 
In analogy with equation~\eqref{definition_Ftot}, the flux associated to a given resonance $\bm{m}$ is defined as
\begin{equation}
\bm{\mathcal{F}}_{\bm{m}}  = \bm{m} \left[ A_{\bm{m}} (\bm{J}) \, F (\bm{J}) + D_{\bm{m}} (\bm{J}) \, \bm{m} \!\cdot\! \frac{\partial F}{\partial \bm{J}} \right]  
 \equiv \bm{m} \, \mathcal{F}_{\bm{m}} \, , \label{definition_Flux_m}
\end{equation}
where the non-bold $\mathcal{F}_{\bm{m}}$ is a scalar. Using shortened notations and forgetting numerical prefactors, we may rewrite the drift and diffusion coefficients from equations~\eqref{WKB_drift} and~\eqref{WKB_diff} under the form
\begin{equation}
\begin{cases}
\displaystyle A_{\bm{m}} (\bm{J}_{1}) = - \frac{1}{(\bm{m} \!\cdot\! \bm{\Omega})'} \!\! \int \!\! \mathrm{d} J_{r}^{2} \, \bm{m}\!\cdot\! \frac{\partial F}{\partial \bm{J}_{2}} \frac{1}{|\mathcal{D}_{\bm{m}}|^{2}} \, ,
\\
\displaystyle D_{\bm{m}} (\bm{J}_{1}) = \frac{1}{(\bm{m} \!\cdot\! \bm{\Omega})'} \!\! \int \!\! \mathrm{d} J_{r}^{2} \, F \frac{1}{| \mathcal{D}_{\bm{m}} |^{2}} \, ,
\end{cases}
\label{shortened_drift_diffusion}
\end{equation}
where we used the shortened notation
\begin{equation}
\frac{1}{(\bm{m} \!\cdot\! \bm{\Omega})'} = \frac{1}{ \left| \frac{\partial }{\partial J_{\phi}} [\bm{m} \!\cdot\! \bm{\Omega}] \right|_{J_{\phi}^{1}}} \, 
\label{short_resonance_factor}
\end{equation}
for the term appearing as a prefactor in the expressions~\eqref{WKB_drift} and~\eqref{WKB_diff} of the drift and diffusion coefficients. The flux can then be decomposed as $\mathcal{F}_{\bm{m}} \!=\! \mathcal{F}_{\bm{m}}^{r} \!+\! \mathcal{F}_{\bm{m}}^{\phi}$ with
\begin{equation}
\mathcal{F}_{\bm{m}}^{r}\!=\!  \frac{m_{r}}{(\bm{m} \!\cdot\! \bm{\Omega})'} \!\! \int \!\! \mathrm{d} J_{r}^{2} \frac{1}{|\mathcal{D}_{\bm{m}}|^{2}}  
\left [ F (J_{r}^{2}) \, \frac{\partial F}{\partial J_{r}} (J_{r}^{1}) \!-\! F (J_{r}^{1}) \, \frac{\partial F}{\partial J_{r}} (J_{r}^{2}) \right], \label{expression_Fm_0}
\end{equation}
and
\begin{equation}
\mathcal{F}_{\bm{m}}^{\phi}\!=\!  \frac{m_{\phi}}{(\bm{m} \!\cdot\! \bm{\Omega})'} \!\! \int \!\! \mathrm{d} J_{r}^{2} \frac{1}{|\mathcal{D}_{\bm{m}}|^{2}}  
\left [ F (J_{r}^{2}) \, \frac{\partial F}{\partial J_{\phi}} (J_{r}^{1}) \!-\! F (J_{r}^{1}) \, \frac{\partial F}{\partial J_{\phi}} (J_{r}^{2}) \right]. 
\hskip -0.3cm \label{expression_Fm}
\end{equation}
We are interested in the value of the flux at the initial time, where the distribution function is given by the Schwarzschild distribution. Because of the exponential dependence in $J_{r}$ of the Schwarzschild distribution function, one has ${ \partial F / \partial J_{r} \!=\!  - (\kappa / \sigma_{r}^{2}) \, F }$. As a result,  the radial component~\eqref{expression_Fm_0} of the flux cancels out and the flux  is simply given by equation~\eqref{expression_Fm}. This coincidence could be called  the {\it Schwarzschild conspiracy} and has important consequences on the properties of the collisional diffusion. Indeed, for a tepid disc, one has ${ | \partial F / \partial J_{r} | \!\gg\! | \partial F / \partial J_{\phi} | }$. Hence one would expect the gradients with respect to $J_{r}$ to be the major contributors to the diffusion. When considering independently the drift and  diffusion coefficients, the gradients in ${ \partial F / \partial J_{r} }$ dominate the diffusion current. Thus for $m_{r} \!\neq\! 0$, the diffusion-only flux can be approximated by
\begin{align}
\mathcal{F}_{\bm{m}}^{\rm Diff} \simeq & \frac{m_{r}}{(\bm{m} \!\cdot\! \bm{\Omega})'} \!\! \int \!\! \mathrm{d} J_{r}^{2} \frac{1}{|\mathcal{D}_{\bm{m}}|^{2}} F (J_{r}^{2}) \, \frac{\partial F}{\partial J_{r}} (J_{r}^{1}) \, . 
\label{expression_Fm_0d}
\end{align}
As the ILR and the OLR  have a non-zero $m_{r}$ compared to the COR,  these resonances should dominate independently the drift and diffusion components. However, when considering the full flux made of the contributions from the drift and diffusion coefficients, because of the Schwarzschild conspiracy, there is a simplification of the dominant terms in ${ \partial F / \partial J_{r} }$, so that one recovers as in equation~\eqref{expression_Fm} that only the smaller gradients ${ \partial F / \partial J_{\phi} }$ remain present. The Schwarzschild conspiracy between drift and diffusion will therefore tend to slightly reduce the magnitude of the full diffusion flux, so as to moderately slow down the collisional relaxation. More importantly, the Schwarzschild conspiracy will favor the COR resonance (radial migration) over the ILR resonance (${J_{r}-}$heating).
One should note that the DF which is effectively sampled in S12 is of the form ${ F \!=\! F (E,J_{\phi}) \!\propto\! J_{\phi}^{q} \exp [- E / \sigma_{r}^{2}] }$, with ${ q \!=\! V_{0}^{2} / \sigma_{r}^{2} \!-\! 1 }$ \citep{Toomre1977,BinneyTremaine2008}. It is only within the epicyclic approximation that this DF takes the form of the Schwarschild DF from equation~\eqref{definition_DF_Schwarzschild}. As a consequence, in S12 simulation, the Schwarzschild conspiracy is not exactly satisfied as observed in equation~\eqref{expression_Fm_0}, but the residual 
difference driving the secular diffusion is likely to be subdominant, as illustrated in figure~\ref{figContourDiffDF}.

\section{Temporal frequency selection}
\label{sec:frequencyselection}

An important feature of the diffusion equation~\eqref{initial_BL_rewrite} is that the diffusion takes place along specific resonance directions associated to the vectors $\bm{m}$ as discussed in equation~\eqref{Flux_m_fast_slow}. Hence being able to determine the dominant resonance is crucial in order to estimate the direction of the secular diffusion in action space. The temporal frequency associated to a given resonance $\bm{m}$ in a location $\bm{J}$ of action-space is given by ${ \omega \!=\! \bm{m} \!\cdot\! \bm{\Omega}  }$. Thanks to the expression~\eqref{intrinsic_frequencies_Mestel}, one immediately notes that for a Mestel disc, one has
\begin{equation}
0 < \omega_{\rm ILR} < \omega_{\rm COR} < \omega_{\rm OLR} \, .
\label{comparison_frequencies_resonances}
\end{equation}
In~\cite{FouvryPichon2015,FouvryBinneyPichon2015}, we studied the same S12 simulation using the WKB limit of the secular diffusion equation, which intends to describe the secular forcing of a collisionless self-gravitating system perturbed by an external source. An essential assumption of this approach was to consider the external perturbation as originating from numerical Poisson shot noise, and therefore assume it to be proportional to the local active surface density. The autocorrelation of the external perturbation $\psi^{\rm ext}$ was taken to be equal to
\begin{equation}
\left< \left| \psi^{\rm ext} \right|^{2} \right> (\omega , k_{r} , J_{\phi}) \propto \Sigma_{\rm t} (J_{\phi}) \, .
\label{assumption_noise_secular_forcing}
\end{equation}
One should note that this crude assumption on the noise properties has no $\omega$ dependence, so that all resonances are equally favored by the Poisson shot noise and are perturbed similarly whatever their associated intrinsic frequencies ${ \omega \!=\! \bm{m} \!\cdot\! \bm{\Omega} }$. This \textit{ad hoc} and simple noise approximation is one of the limitations of the formalism presented in~\cite{FouvryPichon2015,FouvryBinneyPichon2015}.
In contrast, in the WKB Balescu-Lenard equation described in this paper, this preferential selection of the resonances based on their intrinsic frequency is naturally present. Indeed, one can note in the expression~\eqref{expression_Fm} of the flux associated to a resonance $\bm{m}$, the presence of the prefactor ${ 1 / (\bm{m} \!\cdot\! \bm{\Omega})' }$ which arose in equation~\eqref{rewriting_Dirac_delta} when handling the resonance condition. For  the Mestel disc, whose intrinsic frequencies are given by equation~\eqref{intrinsic_frequencies_Mestel}, this term can be straightforwardly computed and reads
\begin{equation}
\frac{1}{(\bm{m} \!\cdot\! \bm{\Omega})'} = \frac{1}{| m_{\phi} \!+\! \sqrt{2} \, m_{r} |} \frac{1}{ | \partial \Omega_{\phi} / \partial J_{\phi} |} \, .
\label{frequency_selection}
\end{equation}
Comparing the ILR resonance to the OLR and COR, one immediately obtains that
\begin{equation}
\frac{(\bm{m}_{\rm OLR} \!\cdot\! \bm{\Omega})'}{(\bm{m}_{\rm ILR} \!\cdot\! \bm{\Omega})'} = \frac{2 \!+\!\! \sqrt{2}}{2 \!-\!\! \sqrt{2}} \simeq 5.8 \, ,
\,\,\,
\frac{(\bm{m}_{\rm COR} \!\cdot\! \bm{\Omega})'}{(\bm{m}_{\rm ILR} \!\cdot\! \bm{\Omega})'} = \frac{2}{2 \!-\!\! \sqrt{2}} \simeq 3.4 \, .
\label{comparison_frequency_selection}
\end{equation}
Hence because the ILR resonance is associated to lower intrinsic temporal frequency ${ \omega \!=\! \bm{m} \!\cdot\! \bm{\Omega} }$, the resonant factor ${ 1 / (\bm{m} \!\cdot\! \bm{\Omega})' }$ naturally tends to favor the ILR resonance with respect to the OLR and COR, and therefore performs natively a temporal frequency biasing which was absent from the \textit{ad hoc} assumption of equation~\eqref{assumption_noise_secular_forcing} describing the external forcing considered in~\cite{FouvryPichon2015,FouvryBinneyPichon2015}.

One can even be more specific when comparing the ILR and OLR resonances. The only difference between these two resonances is the sign of $m_{r}$. The expression~\eqref{diagonal_M_tepid} of the amplification eigenvalues shows that its value only depends on $s^{2}$, so that ${ \lambda_{\rm ILR} \!=\! \lambda_{\rm OLR} }$. The expression of the susceptibility coefficients from equation~\eqref{calculation_1/D^2_I} is also independent of the sign of $m_{r}$ so that ${ 1 / | \mathcal{D}_{\rm ILR} |^{2} \!=\! 1 / | \mathcal{D}_{\rm OLR} |^{2} }$. Hence, when considering the flux $\bm{\mathcal{F}}_{\bm{m}} $ given by equation~\eqref{expression_Fm}, one notes that between the ILR and OLR resonances, $\bm{\mathcal{F}}_{\bm{m}}$ only changes through the factor ${ 1 / (\bm{m} \!\cdot\! \bm{\Omega})' }$. Thanks to equation~\eqref{comparison_frequency_selection}, one immediately obtains
\begin{equation}
\frac{\mathcal{F}_{\rm ILR}}{\mathcal{F}_{\rm OLR}} \simeq 5.8 \, .
\label{comparison_flux_ILR_OLR}
\end{equation}
The secular diffusion flux associated to the OLR resonance is therefore always much smaller than the one associated to the ILR resonance, because of this effect of temporal frequency biasing.

As a conclusion, the temporal frequency selection effect described in equation~\eqref{comparison_frequency_selection} will tend to favor the ILR resonance because it is associated to a smaller intrinsic frequency. However, one can note from figure~\ref{figLambdaJphi} that the COR resonance is always more amplified than the ILR resonance. Finally, the crucial remark is to note that the susceptibility coefficients from equation~\eqref{calculation_1/D^2_I} involve Bessel functions $\mathcal{J}_{m_{r}}$, which are such that ${ \lim_{x \to 0} \mathcal{J}_{m_{r}} (x) \!=\! 1}$ if ${m_{r} \!=\! 0}$, or ${ \!=\! 0 }$ otherwise. As a consequence, close to the ${ J_{r} \!=\! 0 }$ axis, the COR resonance will always tend to become the dominant resonance. There is therefore a non trivial arbitration between these opposite effects when considering the respective contributions of the various resonances to the full secular diffusion flux.

\end{document}